\documentclass[superscriptaddress,prd,a4paper,nofootinbib,11pt]{revtex4-1}

\usepackage[utf8]{inputenc}
\usepackage[T2A]{fontenc}

\usepackage{amsmath}
\usepackage{hyperref} 
\usepackage{ulem}

\usepackage{graphicx}
\usepackage{mdframed}
\usepackage[export]{adjustbox}

 \usepackage{caption}
 \usepackage{subcaption}
 
\newcommand{\bea}{\begin{eqnarray}}  
\newcommand{\eea}{\end{eqnarray}}  

\usepackage{xcolor}

\def\fb{b{\bar{b}}b{\bar{b}}}
\usepackage{wrapfig}
\usepackage{graphicx}
\usepackage{caption}
\begin{document}



\title{Probing the Rare Four-Bottom Higgs Decay $H\to b\bar b b\bar b$ at the HL-LHC and ILC}

 \author{Eduard Boos}
 \email{boos@theory.sinp.msu.ru}
 \affiliation{Skobeltsyn Inst. of Nuclear Physics, M.V.Lomonosov Moscow State Univ., Moscow 119992, Russia}
\affiliation{Faculty of Physics, Lomonosov Moscow State University, Leninskie Gory,
119991, Moscow, Russia}

  \author{Vyacheslav Bunichev}
 \email{bunichev@theory.sinp.msu.ru}
 \affiliation{Skobeltsyn Inst. of Nuclear Physics, M.V.Lomonosov Moscow State Univ., Moscow 119992, Russia}

 \author{Guliya Nurbakova}
\email{g.nurbakova@gmail.com}
\affiliation{Al-Farabi Kazakh National University, 71 Al-Farabi avenue, Almaty, Kazakhstan}
\affiliation{Scientific Research Institutes of Experimental and theoretical physics, 96a Tole bi street, Almaty, Kazakhstan}
 
 \author{Saniya Rustembayeva}
\email{rustembayeva@gmail.com}
\affiliation{Al-Farabi Kazakh National University, 71 Al-Farabi avenue, Almaty, Kazakhstan} 
\affiliation{Scientific Research Institutes of Experimental and theoretical physics, 96a Tole bi street, Almaty, Kazakhstan}

\begin{abstract}
We propose the rare SM Higgs decay $H\to b\bar b b\bar b$ as a probe of the structure of Higgs interactions with bottom quarks and gauge bosons, and as a baseline for searches for new physics producing four-bottom final states in Higgs decays. We compute the leading contributions to this decay, including the dominant $H\to b\bar b g\to b\bar b b\bar b$ topology, the sizeable $H\to ZZ^\ast\to b\bar b b\bar b$ channel, and the loop-induced $H\to gg\to b\bar b b\bar b$ contribution. We find a branching ratio of order $1.6\times10^{-3}$ and show that destructive interference among the leading amplitudes is phenomenologically relevant.

We demonstrate that this decay can be probed in associated Higgs production at both the HL-LHC and the ILC. For $pp\to WH\to Wb\bar b b\bar b$ at $\sqrt{s}=14$ TeV, we use a multivariate analysis based on boosted decision trees to exploit correlations among the four-$b$ kinematic observables. At $3000~{\rm fb}^{-1}$, the statistical significance reaches about $3.5$, while a tighter high-purity working point gives $S/B\simeq5\%$ with significance close to $3\sigma$. A combined high-luminosity LHC dataset could therefore make this rare decay observable.

For $e^+e^-\to ZH\to Zb\bar b b\bar b$ at the ILC with $\sqrt{s}=250$ GeV, we demonstrate that the cleaner collider environment gives a high-purity signal sample. In the nominal setup, the multivariate analysis gives a significance above $5\sigma$ already at $300~{\rm fb}^{-1}$. At integrated luminosities of order $1-3~{\rm ab}^{-1}$, the branching ratio can be measured with several-percent precision.
\end{abstract}


\maketitle
\tableofcontents

\section{Introduction}
The discovery of a new scalar boson at the LHC \cite{Aad:2012tfa, Chatrchyan:2012xdj}, followed by confirmation,
that its properties are in agreement with those expected for the Higgs Boson of 
the Standard Model (SM), completed the pattern of the SM 
particles.
The decay modes of the Higgs  boson such as $\gamma\gamma$ \cite{Khachatryan:2014ira, Aad:2014eha},
$ZZ^{*}$ \cite{Chatrchyan:2013mxa, Aad:2014eva}, $WW^{*}$ \cite{Chatrchyan:2013iaa, ATLAS:2014aga, Aad:2015ona, Sirunyan:2018egh}, $\tau\tau$ \cite{Chatrchyan:2014nva, Aad:2015vsa, Sirunyan:2017khh} have been observed and measured sometime ago  while the dominant decay mode of the Higgs boson to b-quark pairs
has been registered relatevely recently \cite{Aaboud:2018zhk, CMS:2018nsn}. All experimental results obtained to date are consistent with SM predictions  evaluated at  NNLO QCD and NLO EW level (see, \cite{LHCHiggsCrossSectionWorkingGroup:2016ypw}). It should be emphasized that it is important to measure as many Higgs boson decay modes as possible, either to further confirm the expectations of the SM, or to detect possible deviations from the predictions of the SM.

In this study, we propose to explore the Higgs-boson decay into four
$b$ quarks, $H \to b \bar{b} b \bar{b}$ (H4B), as a probe of Higgs-boson
properties. This decay contributes to the NNLO correction to the total
Higgs decay width through hard-emission tree-level $1\to4$ diagrams.
There are several motivations to study the H4B decay mode.
\begin{itemize}
\item

The partial width of the H4B decay is not negligible, and its leading
contribution is proportional to the bottom-quark Yukawa coupling.
\item

Higgs decay modes into four light fermions, such as $\tau$ leptons or
$c$ quarks, are typically dominated by the
$H\to ZZ^\ast,WW^\ast\to4f$ channels.
By contrast, final states containing a $b\bar b$ pair together with two
light fermions are dominated by the
$H\to b\bar b Z^\ast/\gamma^\ast$ topology, whose rate is enhanced by the
comparatively large bottom-quark Yukawa coupling.
However, the modes with two light quarks, although less suppressed than
the corresponding leptonic modes, are practically impossible to extract
from the QCD background, while those with $b\bar b$ plus two leptons are
experimentally much cleaner but too rare.
Therefore, the $H\to b\bar b b\bar b$ decay plays a special role among
four-fermion Higgs final states.
\item

The $H\to b\bar b b\bar b$ decay leads to a final state with four
$b$-jets and provides direct access to Higgs interactions with down-type
quarks, as well as to the internal structure of the decay through its
non-trivial kinematics.
\item

The H4B decay mode can also serve as a probe of Beyond Standard Model
(BSM) sectors, especially those involving light resonances coupled to
$b$ quarks, as motivated for example by models with partial Higgs
compositeness~\cite{Kaplan:1991dc} and their subsequent developments.
\end{itemize}

Beyond its SM interpretation, the $H\to4b$ final state also provides a sensitive probe of new physics that couples to the Higgs sector and preferentially produces bottom quarks. 
 Examples include light scalar or pseudoscalar states in exotic Higgs
decays, modified Higgs couplings, and models with partial compositeness. Establishing the SM prediction, its interference structure, and the realistic collider sensitivity is therefore essential: it defines the irreducible baseline against which any BSM enhancement or distortion of the four-$b$ final state must be tested.

In our study, we explore the $VH$ production channel, which is the same channel where the Higgs boson decay to $b\bar{b}$ was observed~\cite{CMS:2018nsn}. The choice of this channel is motivated by the challenges associated with exploring the gluon fusion ($GGH$) production channel for the $H\to 4b$ decay mode due to the large QCD background. As shown in~\cite{CMS:2017bcq}, there is some sensitivity to $H \to b\bar{b}$ in inclusive searches for a highly boosted Higgs boson, but due to low statistics, the significance of the signal in this CMS search at 35.9 fb$^{-1}$ (primarily coming from $GGH$) remains below $2\sigma$.

A similar situation arises for the vector boson fusion ($VBF$) production channel, where the signal significance is also low due to limited statistics. In this case, a significance of 2.4$\sigma$ was observed at 90.8 fb$^{-1}$~\cite{CMS:2023tfj}. After the $HW/Z$ process, the next most important channel for exploring the $H \to b\bar{b}$ decay is the $ttH$ production channel (see \cite{CMS:2023tfj} and references therein). The $ttH$ channel may also hold potential for studying the $H4B$ decay mode, though likely only at future colliders such as the FCC-hh, and it will be analysed separately.

Even for the most promising $VH$ production channel for the $H4B$ study, an advanced signal-to-background analysis is required. In our study, we applied multivariate data analysis, which enabled us to achieve significantly higher signal significance.
Besides the signal versus background analysis, in our work, we present several new results specific to the $H4B$ decay. These include a precise and thorough analysis of the interference between different classes of diagrams contributing to this decay, as well as an advanced kinematical study. Furthermore, we introduce new kinematical variables that play a critical role in improving the signal significance and deepening our understanding of the signal's properties.

It is important to note that while the $H4B$ decay was studied at next-to-leading-order (NLO) in~\cite{Gao:2019yp}, that work focused solely on the Higgs boson decay itself and did not include a signal-to-background analysis. Moreover, our study provides an accurate and correct treatment of the
interference effects between different diagrams contributing to $H4B$, which was not covered in~\cite{Gao:2019yp}. This interference analysis, combined with our advanced kinematical variables, represents an important  step towards observation and understanding of this decay
mode.

The paper is organised as follows.
In Sec.~\ref{sec:Decay} we establish the theoretical and phenomenological structure of the
$H\to b\bar b b\bar b$ decay. We compute the leading contributions to the decay width,
quantify the interference effects among them, and identify the kinematic variables that
distinguish the dominant $H\to b\bar b g\to b\bar b b\bar b$ topology from the
$H\to ZZ^\ast\to b\bar b b\bar b$ and $H\to gg\to b\bar b b\bar b$ contributions.
In Sec.~\ref{section:s-vs-b} we demonstrate that the decay can be probed at the HL-LHC
in associated $WH$ production, despite the large irreducible $W+4b$ background, by
exploiting correlations among four-$b$ kinematic observables in a multivariate analysis.
In Sec.~\ref{section:ilc} we show that the ILC provides the clean environment needed to
turn this rare decay from an observable signal into a precision Higgs measurement.
Our conclusions are presented in Sec.~\ref{sec:Conclusions}.

\section{The structure and the properties of the $H\to b\bar{b}b\bar{b}$ decay} 
\label{sec:Decay}
\subsection{$H4B$ decay width and structure}

The first step in our study is to explore  properties of $4b$ final state from  Higgs decay and the respective kinematical variables which will help us to 
separate signal from background at the later stage of analysis.
Let us take a closer look at $H\to b\bar{b}b\bar{b}$. On the one hand, this process contributes to the NNLO correction of the $H \to b\bar{b}$ decay, and its width is expected to be approximately two orders of magnitude smaller.

On the other hand, the five non-trivial kinematic variables describing the
phase space of the $H\to b\bar b b\bar b$ decay, compared with none for
$H\to b\bar b$, provide additional observables for optimising the
signal-to-background ratio.
 
\begin{figure}[htbp]
	\centering
		\includegraphics[width=\textwidth]{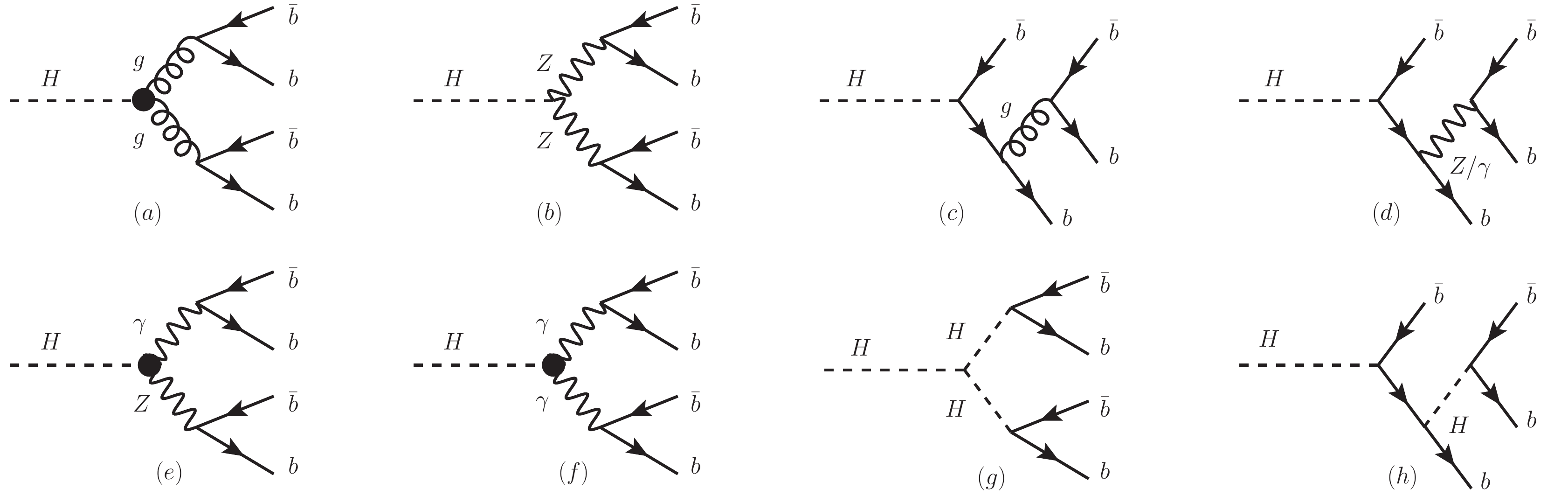}
	\caption{ \label{fig:h-decay} Feynman diagrams contributing to the Higgs boson decay $H\to b\bar{b}b\bar{b}$. }
\end{figure}

In order to maximise significance of the  $H\to b\bar{b}b\bar{b}$ signature from $pp\to WH$ process we 
 explore first the kinematical properties and the structure of the Higgs boson decay.
In Fig.~\ref{fig:h-decay} we present Feynman diagrams 
 contributing to  $H\to b\bar{b}b\bar{b}$ decay.
 These diagrams with different topologies define its principal kinematical properties. The loop-induced 
 vertices such as $Hgg, H\gamma\gamma$ and $H\gamma Z$ 
 are denoted by the  black blobs indicating their loop structure.  

The contribution from Fig.~\ref{fig:h-decay} diagrams to the  $H\to b\bar{b}b\bar{b}$ decay
amplitude, $A_{H\to4b}$, can be written as:
\begin{equation}
A_{H\to4b} = 
\kappa_b (A_g + A_{Z/\gamma} + \kappa_b^2 A_H )
+\kappa_g A_{gg} + \kappa_{Z} A_{ZZ} 
+\kappa_{Z\gamma}A_{Z\gamma} +\kappa_{\gamma}A_{\gamma\gamma}+\kappa_H \kappa_b^2 A_{HH}
\end{equation} 
with $A_{i}$ ($i=g,Z,\gamma,H, gg, ZZ, Z\gamma, \gamma\gamma, HH$) corresponding  to SM amplitudes  from diagram
with  one or two propagators with gluons, photons, $Z$ and Higgs bosons, while  $\kappa_b, \kappa_g, \kappa_Z, \kappa_\gamma, \kappa_{\gamma Z}$ and $\kappa_H$ stand for  the Higgs boson 
couplings to fermions or bosons normalized to the Standard Model ones. These normalised couplings  are convenient to use as modifiers to parameterise the new physics contribution to $H4B$ decay, assuming that the new particles are heavy enough to be produced on-shell.
In Table~\ref{tab:h-decay} (second column) we present the  $H\to b\bar{b}b\bar{b}$ width together with the contributions from individual diagrams 
to it from Fig.~\ref{fig:h-decay}.

The third column gives the relative contribution of each listed channel to
the full $H\to b\bar b b\bar b$ width, while the fourth column presents
the corresponding branching ratios.
Also, in the forth column we provide the relative contributions from  diagrams of Fig.~\ref{fig:h-decay} to the  $H\to b\bar{b}b\bar{b}$ ($\Gamma_{H4B}$).
First of all one can see that the main contribution to $H4B$ comes from only three leading diagrams
$H\to b\bar{b}g$, $H\to ZZ$ and $H\to gg$,
while the relative contribution from other diagrams is below per mille level. Secondly, one can notice that the total sum of the individual diagrams is larger than the total $H4B$, indicating the overall negative interference between 
diagrams, which we discuss below. Details of this interference are presented in Table~\ref{tab:h-interf}.
\begin{table}[htbp]
	\begin{center}
		\begin{tabular}{| l | c | c | c |} \hline
  &&&\\
			\hspace*{1cm}Process & Partial width (GeV) & $R = \Gamma_i/\Gamma_{H4B}$ & $Br = \Gamma_i/\Gamma_H^{\rm tot}$ \\
            &&&\\
   \hline\hline
			$H\to b\bar{b}g\to b\bar{b}b\bar{b}$ & $4.93\cdot10^{-6}$ & $6.85\cdot10^{-1}$ & $1.21\cdot10^{-3}$ \\
			$H\to ZZ\to b\bar{b}b\bar{b}$ & $2.15\cdot10^{-6}$ & $2.99\cdot10^{-1}$ & $5.26\cdot10^{-4}$ \\
			$H\to gg\to b\bar{b}b\bar{b}$ & $1.23\cdot10^{-7}$ & $1.71\cdot10^{-2}$ & $3.01\cdot10^{-5}$ \\
			$H\to b\bar{b}(Z/\gamma)\to b\bar{b}b\bar{b}$ & $1.67\cdot10^{-9}$ & $2.32\cdot10^{-4}$ & $4.09\cdot10^{-7}$ \\
			$H\to Z\gamma \to b\bar{b}b\bar{b}$ & $1.51\cdot10^{-9}$ & $2.10\cdot10^{-4}$ & $3.69\cdot10^{-7}$ \\
			$H\to \gamma\gamma \to b\bar{b}b\bar{b}$ & $6.20\cdot10^{-12}$ & $8.61\cdot10^{-7}$ & $1.59\cdot10^{-9}$ \\
			$H\to HH \to b\bar{b}b\bar{b}$ & $3.27\cdot10^{-13}$ & $4.54\cdot10^{-8}$ & $8.00\cdot10^{-11}$ \\
			$H\to Hbb \to b\bar{b}b\bar{b}$ & $3.68\cdot10^{-17}$ & $5.11\cdot10^{-12}$ & $9.00\cdot10^{-15}$ \\
\hline
Sum of listed contributions & $7.21\cdot10^{-6}$ & $1.06$ & $1.76\cdot10^{-3}$ \\
\hline
$H\to b\bar{b}b\bar{b}$ (full result) & $6.77\cdot10^{-6}$ & $1.00$ & $1.66\cdot10^{-3}$ \\
\hline		
\end{tabular}
\end{center}
	\caption{\label{tab:h-decay}
Partial widths in GeV for the listed contributions to $H\to b\bar{b}b\bar{b}$.
The third column shows the relative contribution of each listed channel to the full
$H\to b\bar{b}b\bar{b}$ width, $\Gamma_{H4B}$, given in the last row.
The fourth column gives the corresponding branching ratios, evaluated using
the SM total Higgs width
$\Gamma_H^{\rm tot}=4.088\cdot10^{-3}$~GeV from the LHC Higgs Cross Section
Working Group~\cite{LHCHiggsCrossSectionWorkingGroup:2016ypw}. The row ``Sum of listed contributions'' gives the sum of the individual contributions shown above, while the last row gives the full $H\to b\bar{b}b\bar{b}$ result including interference among all diagrams.}
\end{table}
\begin{table}[htbp]
	\begin{center}
		\begin{tabular}{|l|c|} \hline
			Subprocesses &  Interference, GeV\\ \hline
			$ \Gamma_{H4B}
    -\Gamma_{H4B}^{b\bar{b}g}
    -\Gamma_{H4B}^{ZZ}
    -\Gamma_{H4B}^{gg} $
    & 
    $-4.35\cdot10^{-7}$\\ \hline
	$b\bar{b}g \times  ZZ$ interference & $-1.29\cdot10^{-7}$\\ 
	$b\bar{b}g \times  gg$ interference& $-3.66\cdot10^{-7}$\\ 		 
	$gg \times  ZZ$ interference& $+6.00\cdot10^{-8}$\\ 		 
	\hline
		\end{tabular}
	\end{center}
	\caption{\label{tab:h-interf}  Interference between three leading  diagrams contributing $H\to b\bar{b}b\bar{b}$ decay. } 
\end{table}

Taking these results into account,
 the partial width,  $H\to b\bar{b}b\bar{b}$, can be accurately approximated  by the sum of  the three leading 
contributions
$H\to gg\to b\bar{b}b\bar{b}$ (Fig.~\ref{fig:h-decay}(a)),
$H\to ZZ\to b\bar{b}b\bar{b}$ (Fig.~\ref{fig:h-decay}(b)),
$H\to  b\bar{b}g\to b\bar{b}b\bar{b}$ (Fig.~\ref{fig:h-decay}(c))
and the interference between them 
as follows:
\begin{eqnarray}
\Gamma_{H\to b\bar{b}b\bar{b}} \approx  
 \kappa_g^2 \Gamma_{H\to gg\to b\bar{b}b\bar{b}} +
 \kappa_Z^2 \Gamma_{H\to ZZ\to b\bar{b}b\bar{b}}
 +
\kappa_{b}^2 
\Gamma_{H\to b\bar{b}g\to b\bar{b}b\bar{b}} 
+ \mbox{[Interference]}
\label{eq:h-decay} 
\end{eqnarray}

The main contribution to $H4B$ comes from the $H \to b\bar{b}g \to b\bar{b}b\bar{b}$ process, accounting for approximately 68\%, while the next largest contribution comes from the $H \to ZZ^* \to b\bar{b}b\bar{b}$ process, which contributes around 30\%. The contribution from $H \to gg \to b\bar{b}b\bar{b}$ is about 2\%.

\begin{figure}[htb]
	\includegraphics[width=0.45\textwidth]{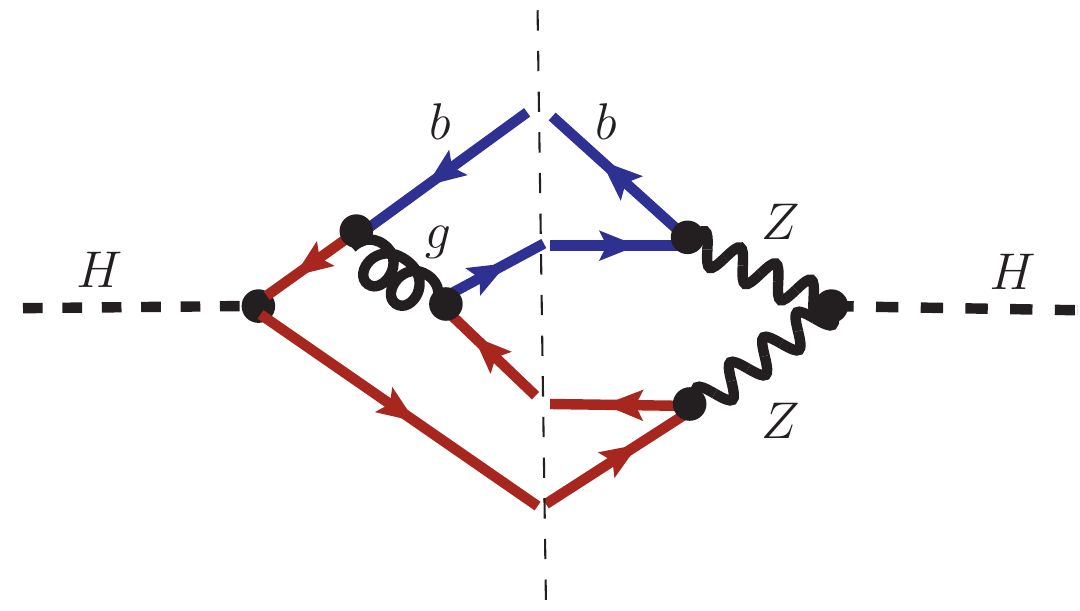}
	\hfill
	\includegraphics[width=0.45\textwidth]{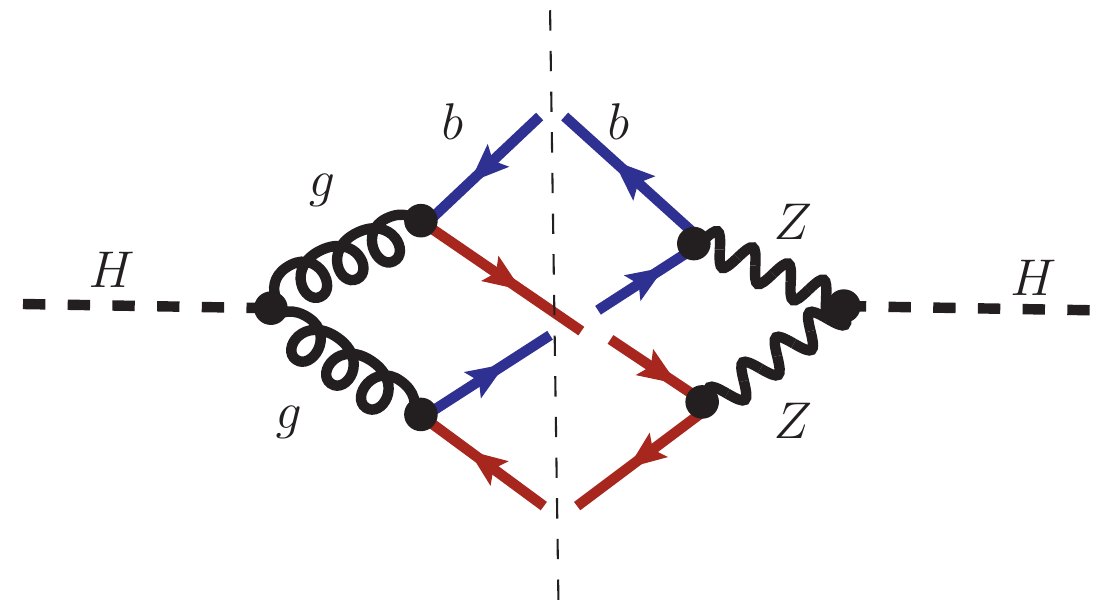}%
	\caption{\label{fig:chebur}
	Squared diagrams for non-zero interference between the $H \to b\bar{b}g \to b\bar{b}b\bar{b}$ and $H \to ZZ^* \to b\bar{b}b\bar{b}$ (left), and the $H \to gg \to b\bar{b}b\bar{b}$ and $H \to ZZ^* \to b\bar{b}b\bar{b}$ (right) topologies. Changes in b-quark line colour schematically indicate the respective changes in QCD colour flow.}
\end{figure}

As mentioned earlier, there is non-zero destructive interference between the $H \to b\bar{b}g \to b\bar{b}b\bar{b}$ and $H \to gg^* \to b\bar{b}b\bar{b}$ ($bbg \times gg$) topologies, with an interference effect of $-5.4\%$. Additionally, interference occurs between the $H \to b\bar{b}g \to b\bar{b}b\bar{b}$ and $H \to ZZ^* \to b\bar{b}b\bar{b}$ ($bbg \times ZZ$) topologies, which is also destructive ($-1.9\%$). Meanwhile, the interference between $H \to gg \to b\bar{b}b\bar{b}$ and $H \to ZZ^* \to b\bar{b}b\bar{b}$ ($gg \times ZZ$) topologies is constructive, contributing $+0.87\%$.
The squared diagrams for the last two interference cases are shown in Fig.~\ref{fig:chebur} with  the left diagram representing $bbg \times ZZ$ and the right diagram representing $gg \times ZZ$. The colour of the b-quark lines indicates the respective changes in QCD colour flow that lead to destructive or constructive interference effects.
Overall, the interference effect is destructive, amounting to approximately $-6.4\%$.

We would like to emphasize that the sign of the $bbg \times gg$ interference is reported as positive in Ref.~\cite{Gao:2019yp}, contrary to our findings, which were explicitly evaluated using CompHEP/CalcHEP. Furthermore, we have provided interference values for all leading topologies mentioned earlier, offering an additional, accurate, and correct treatment of interference effects between different diagrams contributing to the $H4B$ decay process.
Finally, we found that the $H \to ZZ^* \to b\bar{b}b\bar{b}$ channel, which contributes 30\% to the $H4B$ decays, is not as subdominant as stated in Ref.\cite{Gao:2019yp}. However, our findings regarding the $H \to gg \to b\bar{b}b\bar{b}$ channel, which contributes only about 2\% to the $H4B$ decay width, are in close agreement with those of Ref.\cite{Gao:2019yp}.

\subsection{Kinematical properties of $H4B$ decay.}

The study of the $H \to b\bar{b}b\bar{b}$ decay provides additional information, complementing current studies, on Higgs boson couplings to $b$-quarks and gauge bosons.

In particular, to achieve optimal sensitivity to the Yukawa coupling of the Higgs boson to the $b$-quark, it is crucial to separate the contribution of the $H \to b\bar{b}g \to b\bar{b}b\bar{b}$ process from those of $H \to ZZ^* \to b\bar{b}b\bar{b}$ and $H \to gg \to b\bar{b}b\bar{b}$. This section explores various observables and their distributions for this purpose.

Understanding the kinematical properties of the different signal subprocesses and the overall signal aids in distinguishing it from background processes, as discussed in Section~\ref{section:s-vs-b}.

\begin{figure}[htb]
 \includegraphics[trim={0 1cm 0 1cm}, clip,width=\textwidth]{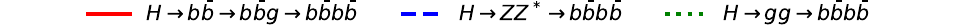}\\
    \includegraphics[trim={0 0cm 0 1cm}, clip,width=0.5\textwidth]{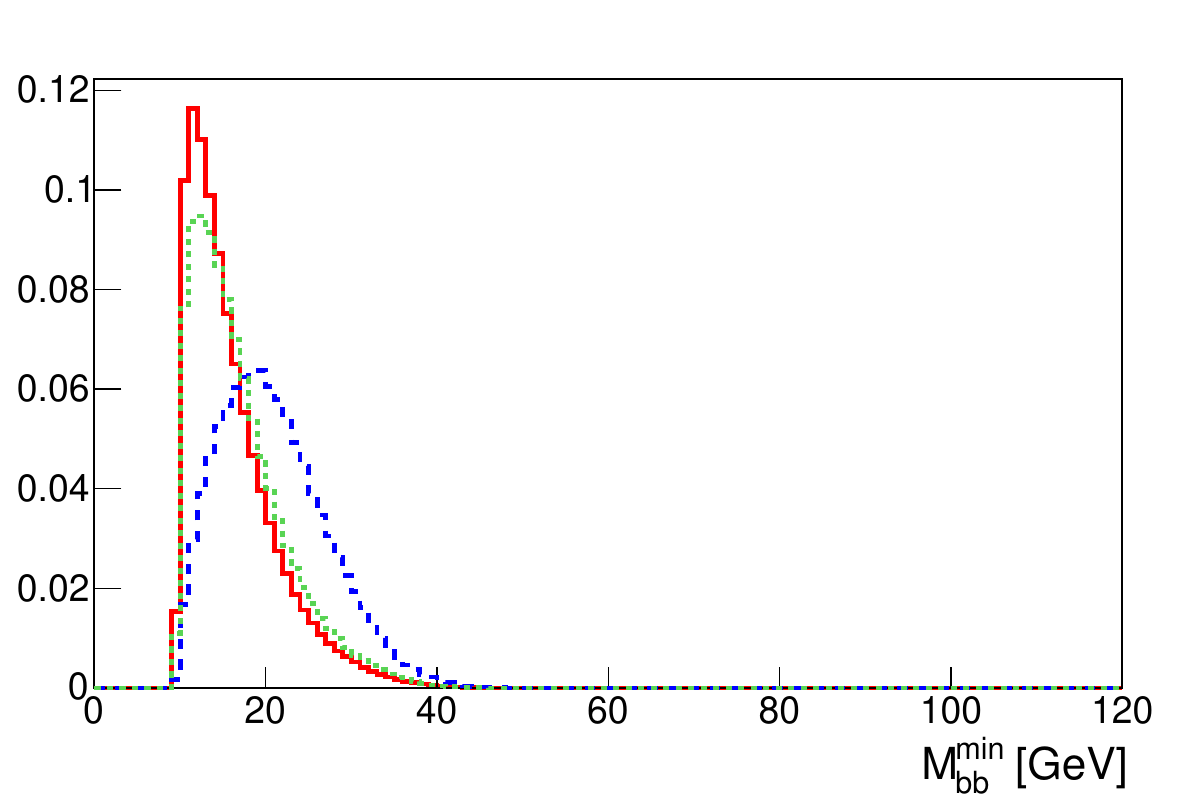}%
    \includegraphics[trim={0 0cm 0 1cm}, clip,width=0.5\textwidth]{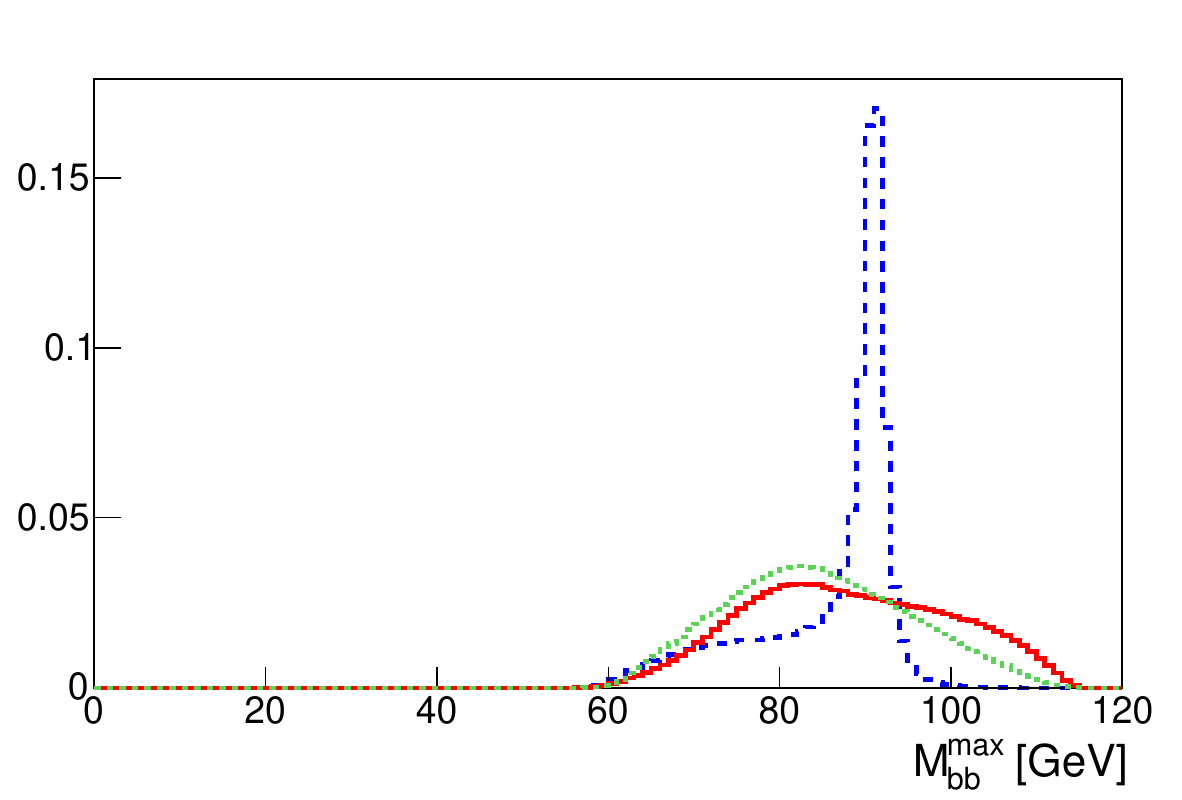}
    \caption{\label{fig:dec-1d-2b-tot} Normalized invariant mass distribution for subprocesses: $H \to b\bar{b} \to b\bar{b}g \to b\bar{b}b\bar{b}$ (red solid histogram), $H \to ZZ^* \to b\bar{b}b\bar{b}$ (blue dashed histogram), $H \to gg \to b\bar{b}b\bar{b}$ (green dotted histogram). \textbf{Left:} $b$-quark pair minimal invariant mass. \textbf{Right:} Maximal invariant mass of the $b$-quark pairs.}
\end{figure}
First, we examine one-dimensional plots of the invariant mass distributions for $b\bar{b}$ pairs. In Fig.~\ref{fig:dec-1d-2b-tot} (left), we present the normalized distributions for the minimal $b\bar{b}$ invariant mass. The distribution looks similar for different subprocesses and does not allow us to clearly separate their contributions. In Fig.~\ref{fig:dec-1d-2b-tot} (right), we show distributions for the maximal invariant mass of two $b$-quarks. The latter allows for the isolation of the \(H \to ZZ^*\) component.
\begin{figure}[htbp]
 \includegraphics[trim={0 1cm 0 1cm}, clip,width=\textwidth]{figs/legend1.pdf}\\    \includegraphics[trim={0 0cm 0 1cm}, clip,width=0.5\textwidth]{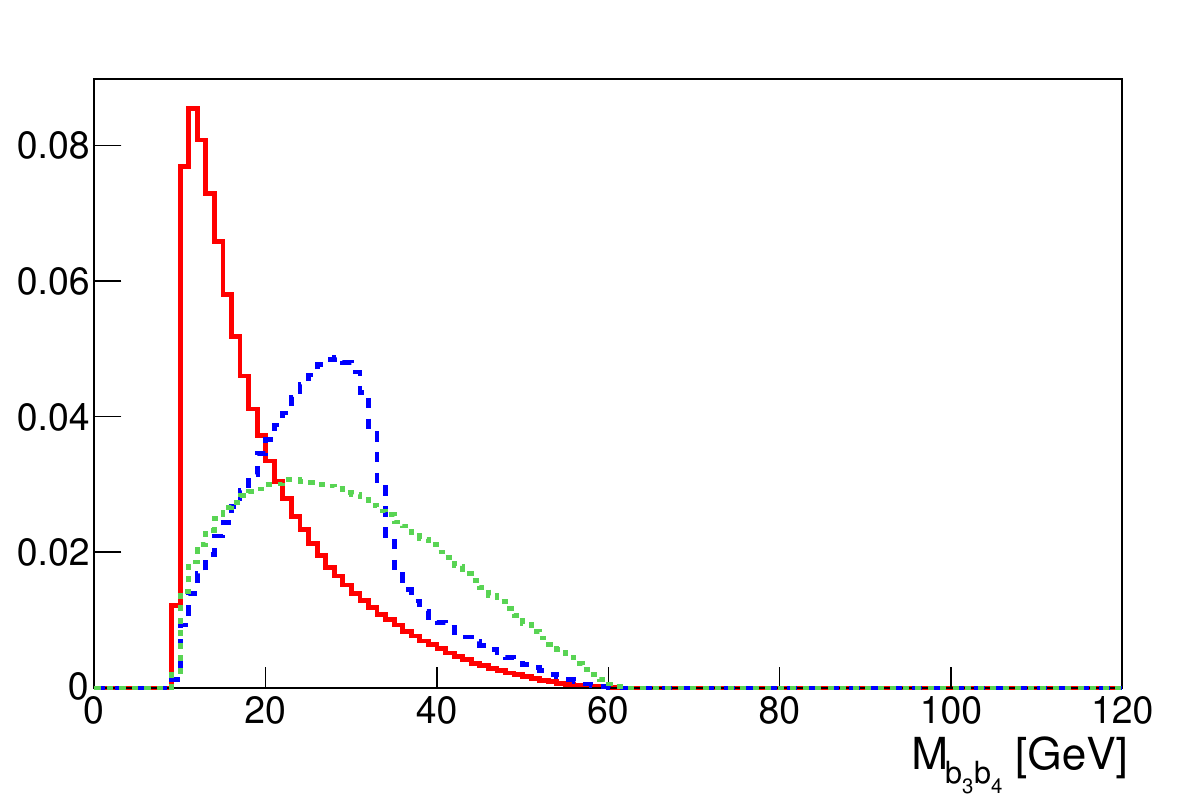}%
    \includegraphics[trim={0 0cm 0 1cm}, clip,width=0.5\textwidth]{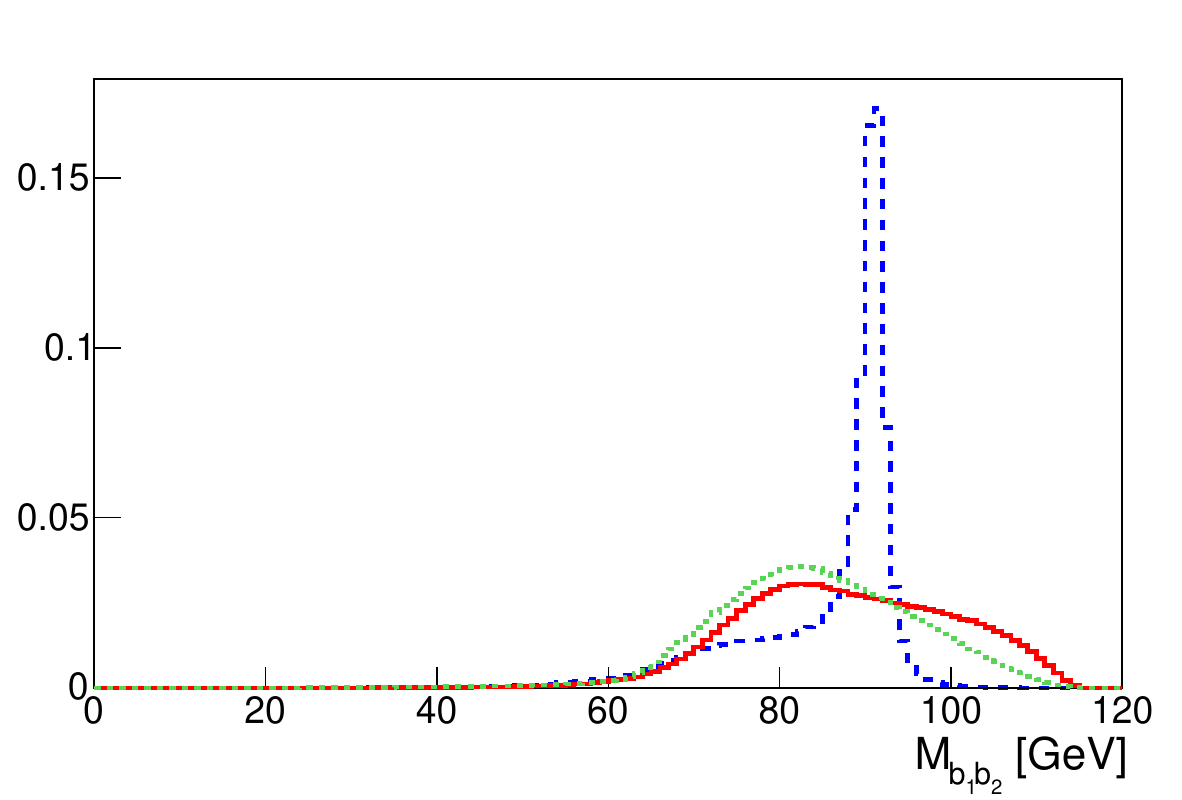}
    \caption{\label{fig:dec-1d-2b-contr} Normalized invariant mass distribution for subprocesses: $H \to b\bar{b} \to b\bar{b}g \to b\bar{b}b\bar{b}$ (red solid histogram), $H \to ZZ^* \to b\bar{b}b\bar{b}$ (blue dashed histogram), $H \to gg \to b\bar{b}b\bar{b}$ (green dotted histogram). \textbf{Left:} Invariant mass of the two lowest-energy $b$-quarks ($b_3, b_4$). \textbf{Right:} Invariant mass of the two highest-energy $b$-quarks ($b_1, b_2$).}
\end{figure}
\begin{figure}[htbp]
    \begin{center}
        {\hspace*{1cm} $H \to b\bar{b} \to b\bar{b}g \to b\bar{b}b\bar{b}$}
        \hfill
        {\it $H \to ZZ \to b\bar{b}b\bar{b}$}
        \hfill
        {\it $H \to gg \to b\bar{b}b\bar{b}$\hspace*{2cm}}
    \end{center}
    \includegraphics[width=0.33\textwidth]{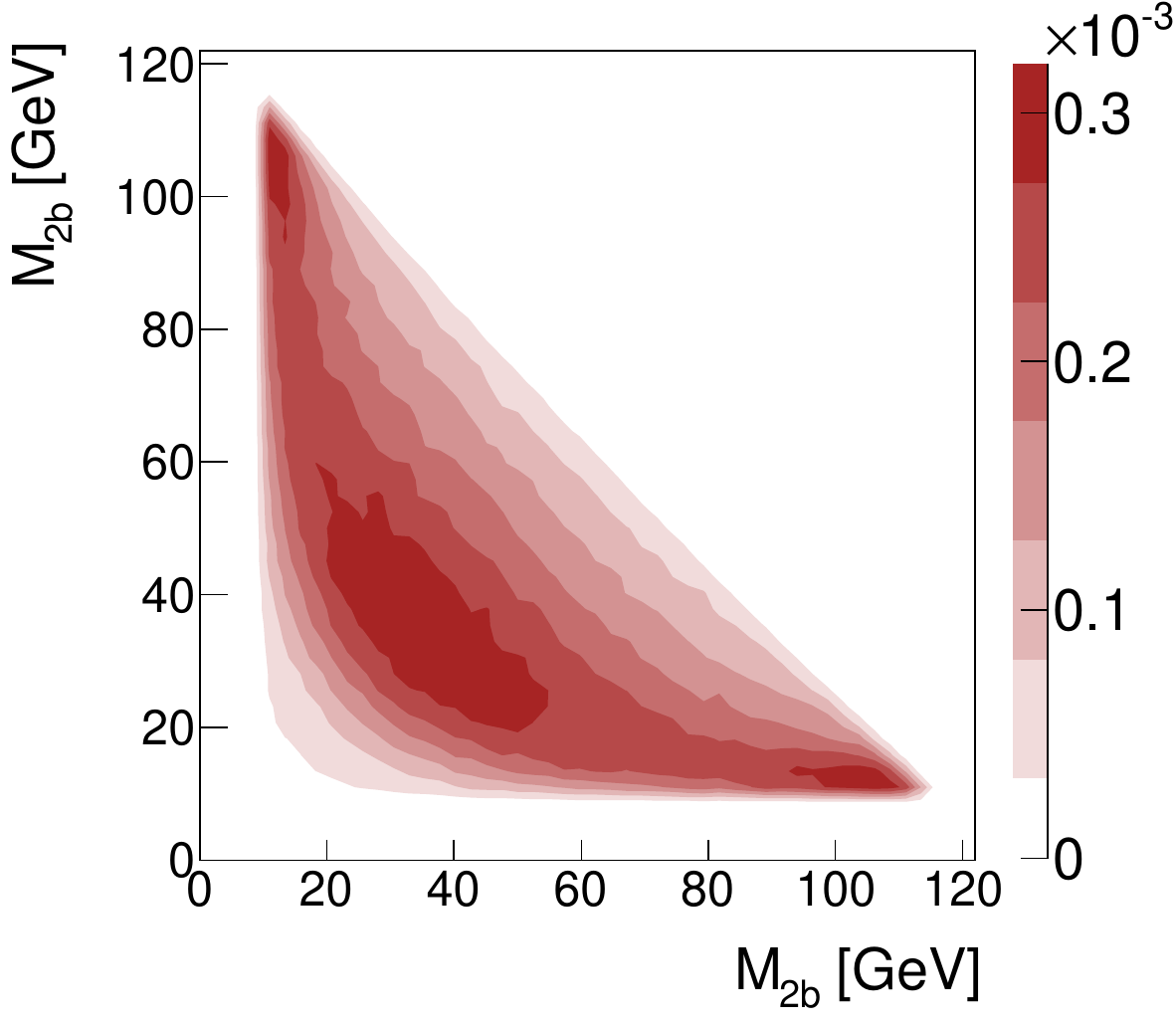}%
    \includegraphics[width=0.33\textwidth]{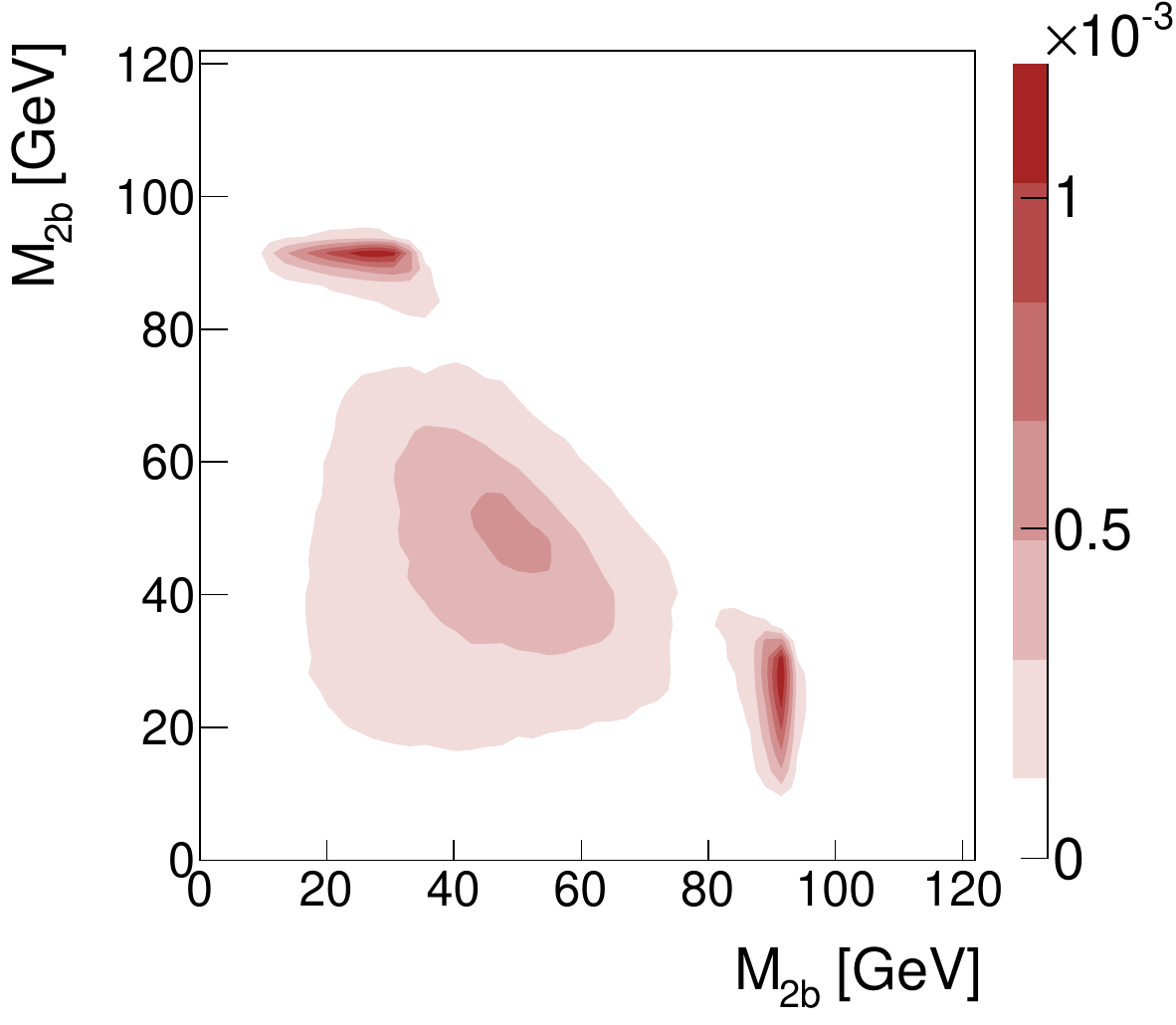}%
    \includegraphics[width=0.33\textwidth]{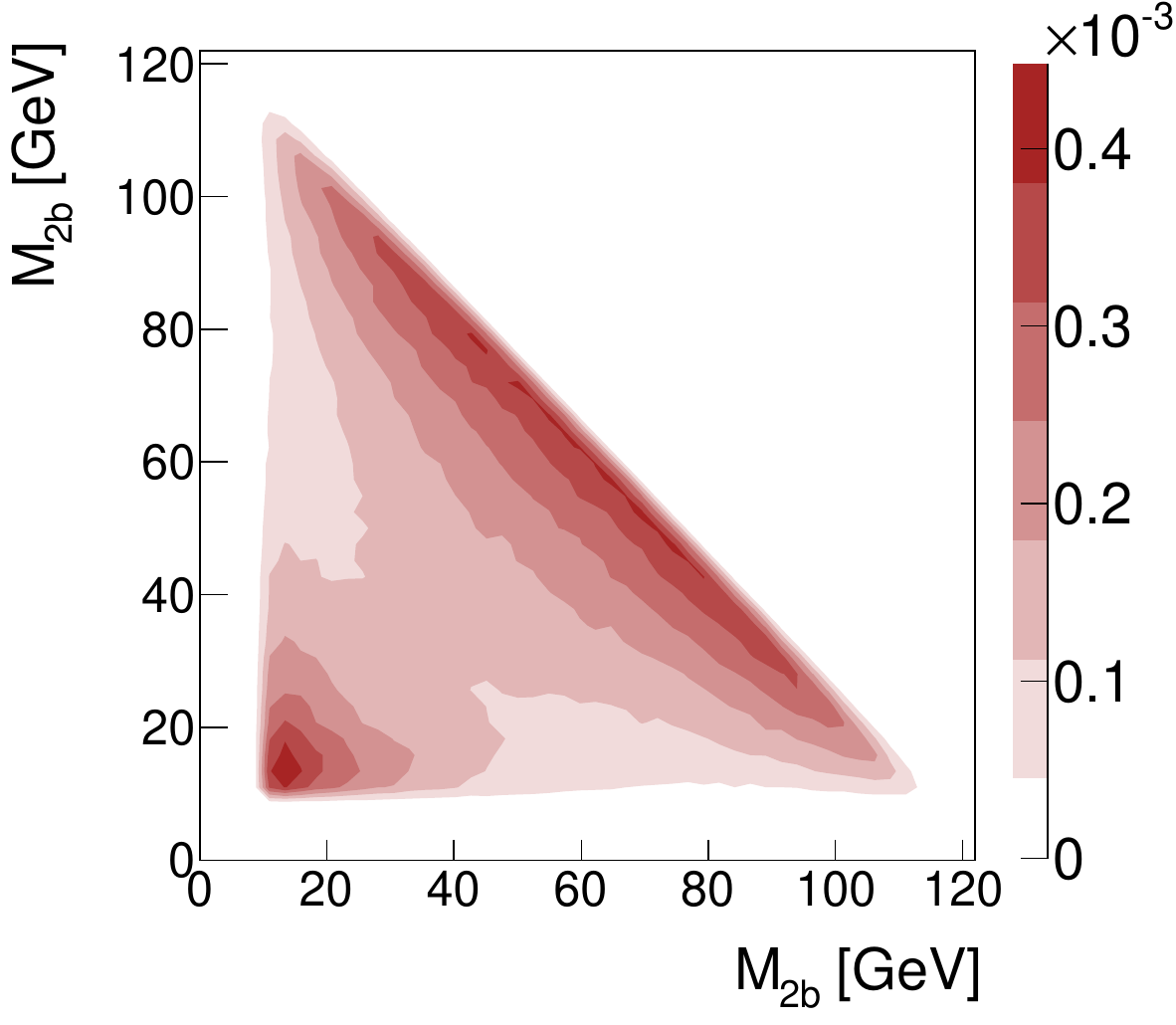}%
    \\
    \includegraphics[width=0.33\textwidth]{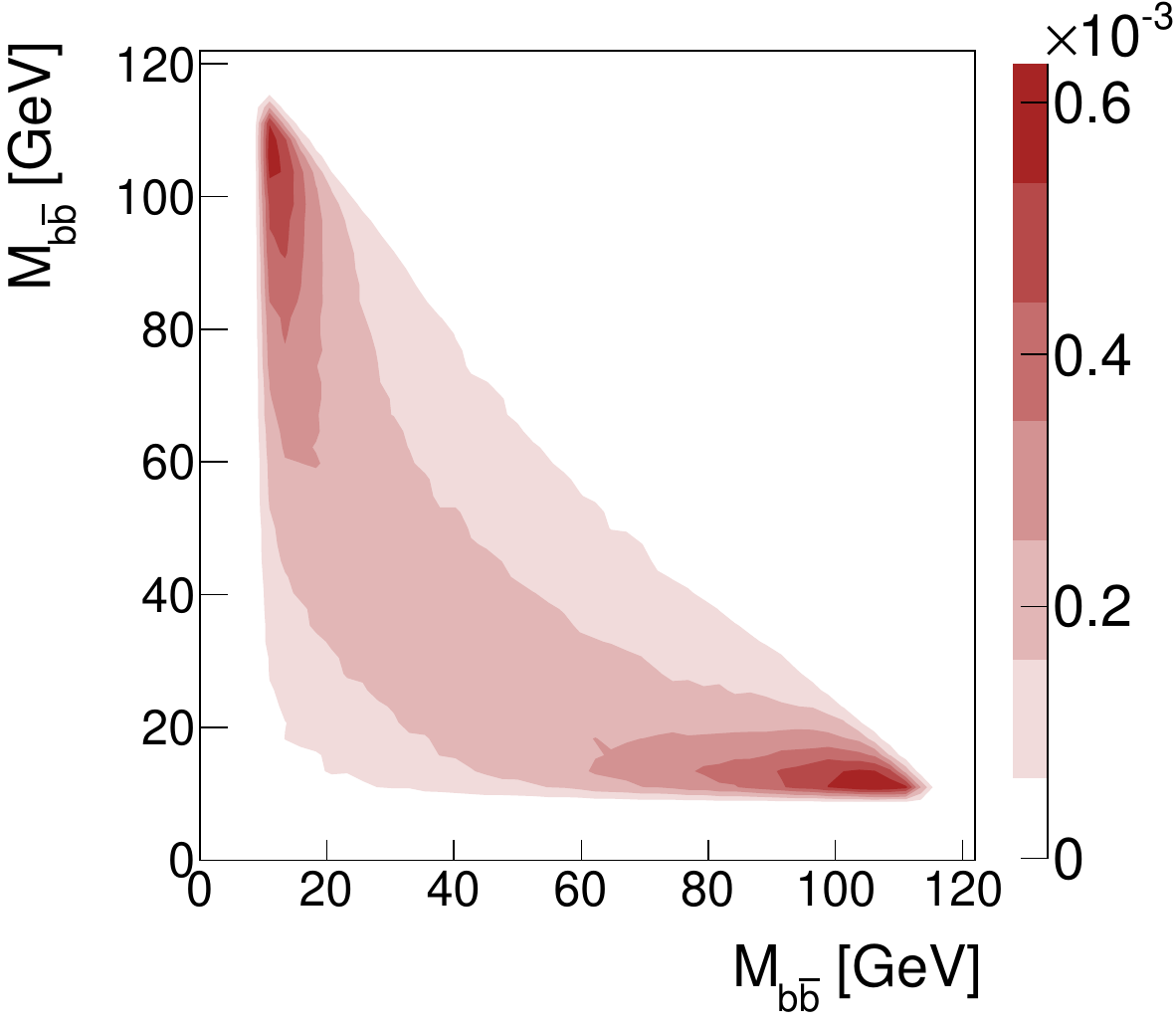}%
    \includegraphics[width=0.33\textwidth]{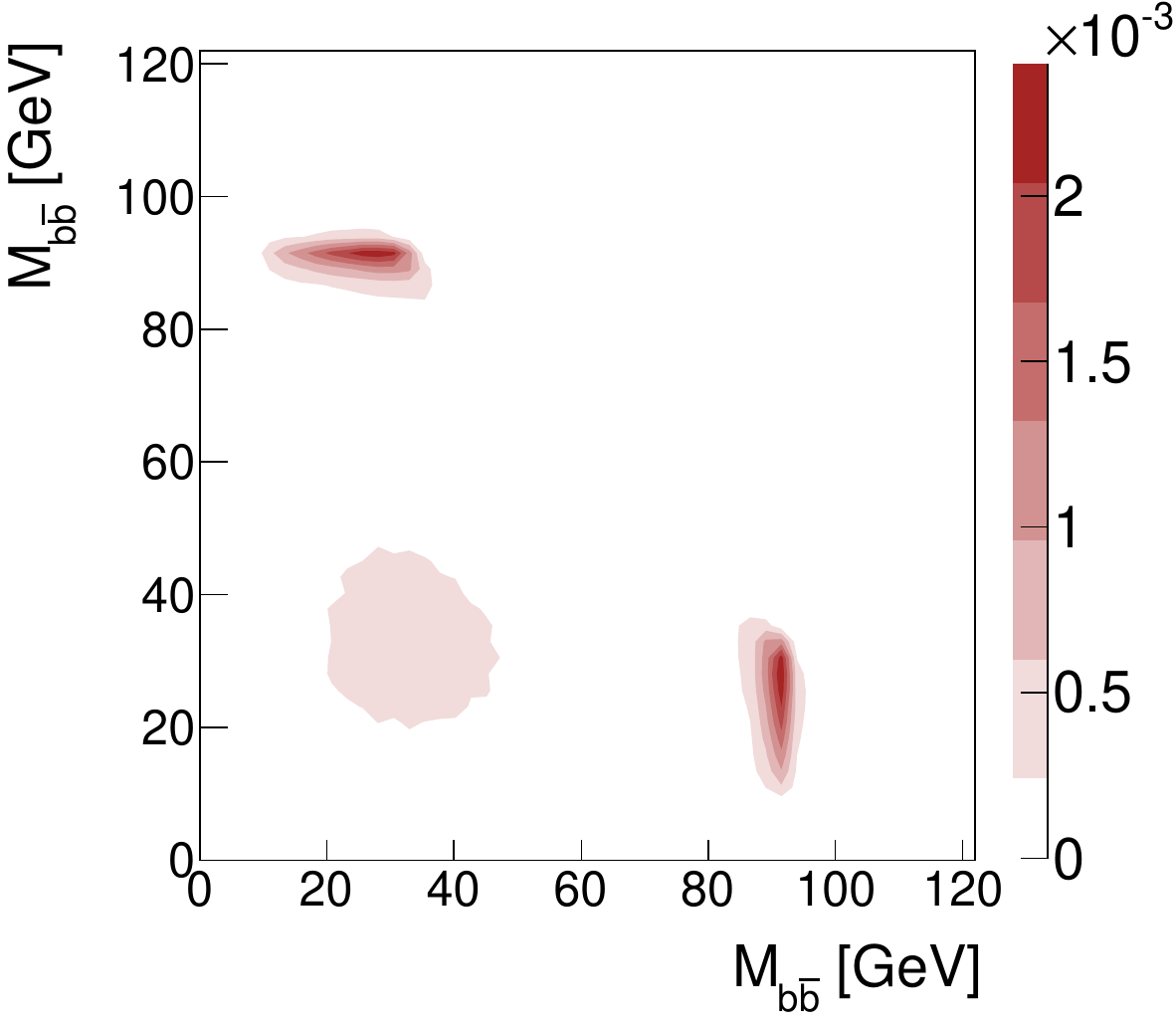}%
    \includegraphics[width=0.33\textwidth]{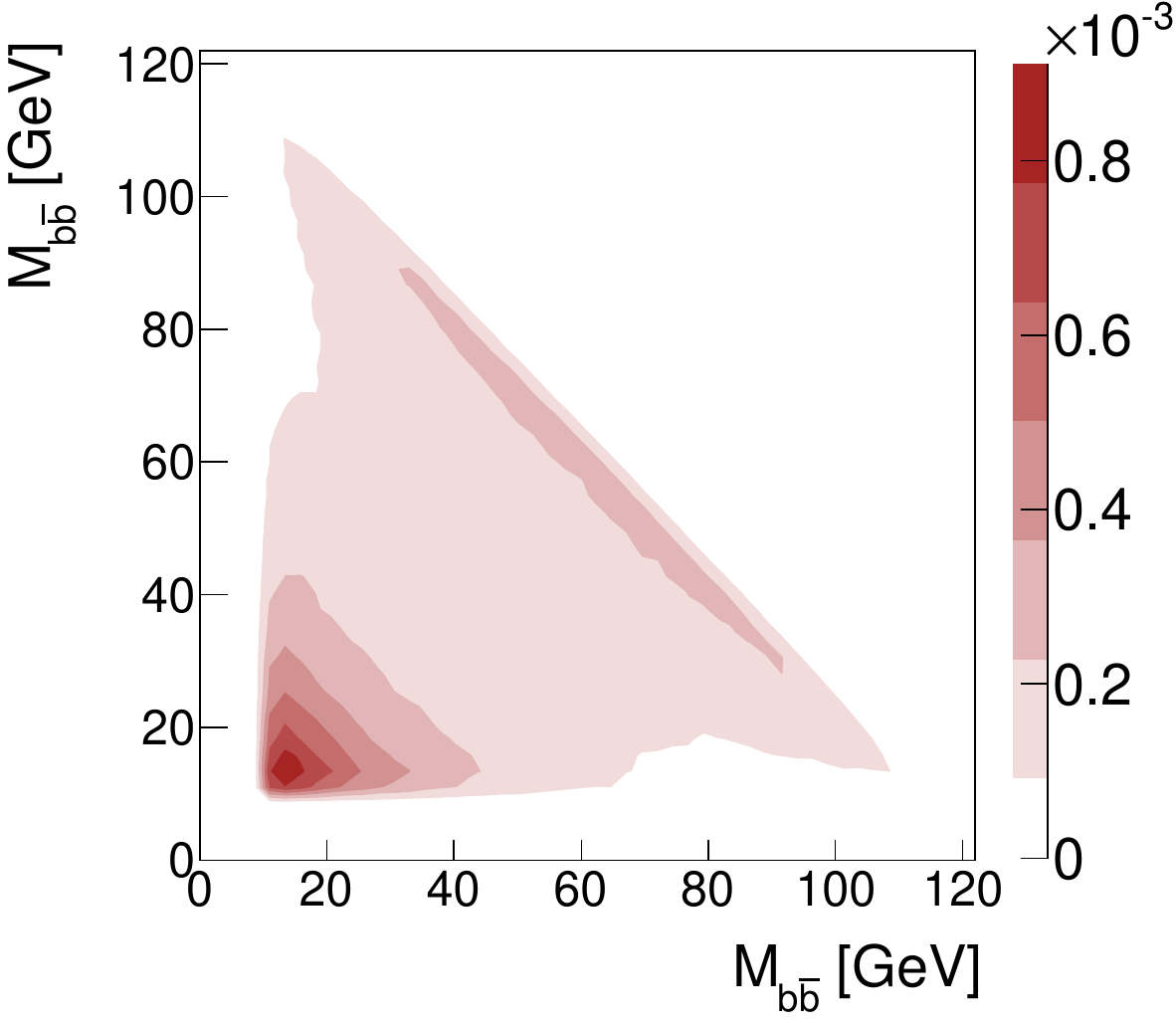}%
   \noindent
    \caption{\label{fig:decay-2d-m2b-random}Normalized double-differential distributions for 
    $H \to b\bar{b} \to b\bar{b}g \to b\bar{b}b\bar{b}$ (left), 
    $H \to ZZ^* \to b\bar{b}b\bar{b}$ (middle), and 
    $H \to gg \to b\bar{b}b\bar{b}$ (right). 
    \textbf{Top:} invariant mass of two randomly selected $b$-quarks versus that of the remaining pair, simulating the experimental case where quarks and antiquarks cannot be distinguished. 
    \textbf{Bottom:} invariant mass of one $b\bar{b}$ pair versus that of the other, corresponding to the hypothetical case where $b$-quarks and $\bar{b}$-quarks can be identified.}
\end{figure}

In Fig.~\ref{fig:dec-1d-2b-contr}, we present the invariant mass $M_{b_3b_4}$ of
the two lowest-energy $b$ quarks (left) and the invariant mass
$M_{b_1b_2}$ of the two highest-energy $b$ quarks (right) for
$H\to b\bar b\to b\bar b g\to b\bar b b\bar b$ (red solid histogram),
$H\to ZZ^\ast\to b\bar b b\bar b$ (blue dashed histogram), and
$H\to gg\to b\bar b b\bar b$ (green dotted histogram).
The $b$-quark energies are considered in the Higgs boson rest frame. It is evident that $M_{b_3,b_4}$ can distinguish all three subprocesses from each other, while $M_{b_1,b_2}$ clearly separates the $H \to ZZ^* \to b\bar{b}b\bar{b}$ subprocess, which peaks near the $Z$-boson resonance.

\begin{figure}[htb]
    \begin{center}
        {\hspace*{1cm} $H \to b\bar{b} \to b\bar{b}g \to b\bar{b}b\bar{b}$}
        \hfill
        {\it $H \to ZZ \to b\bar{b}b\bar{b}$}
        \hfill
        {\it $H \to gg \to b\bar{b}b\bar{b}$\hspace*{2cm}}
    \end{center}
    \includegraphics[width=0.33\textwidth]{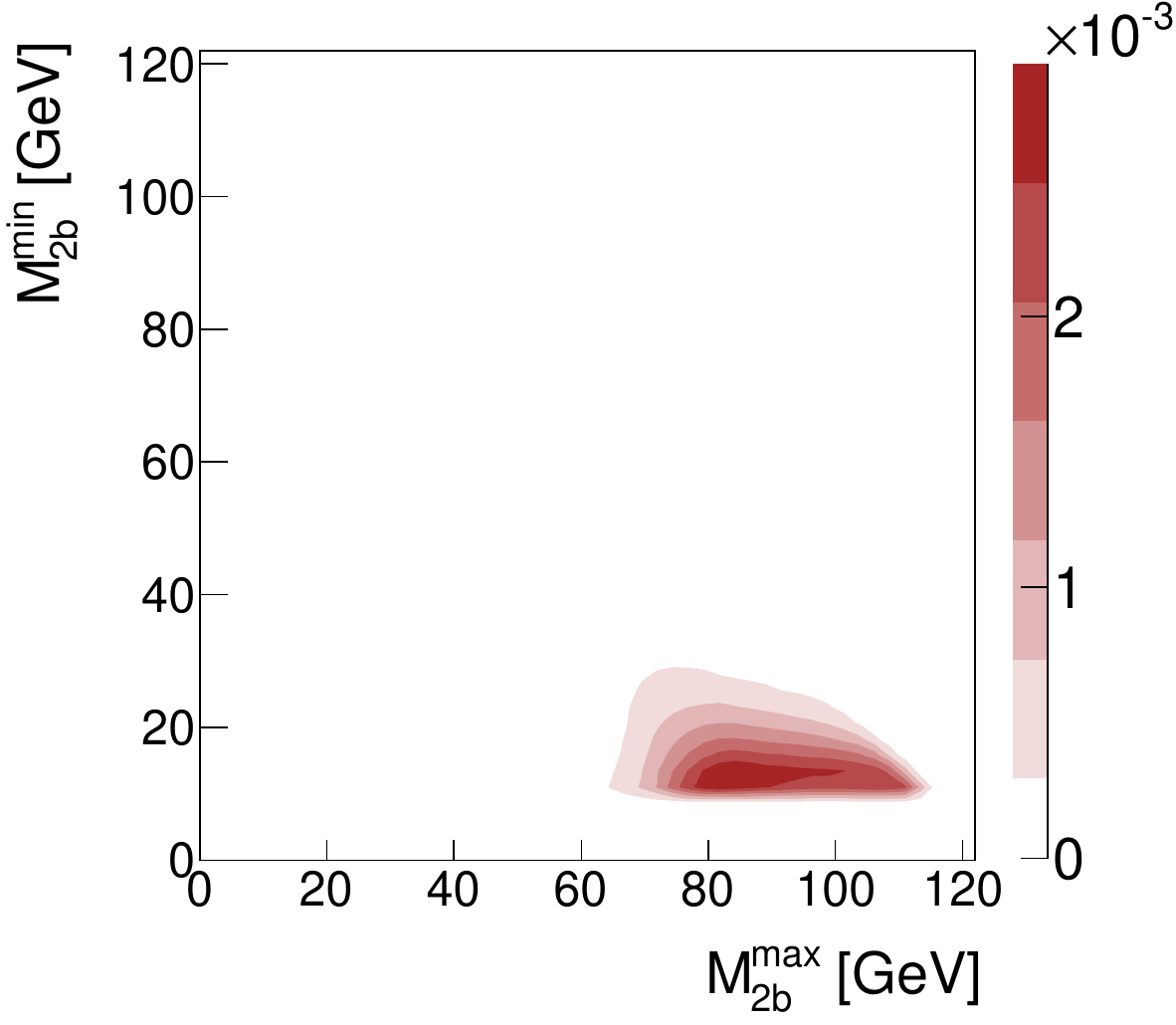}%
    \includegraphics[width=0.33\textwidth]{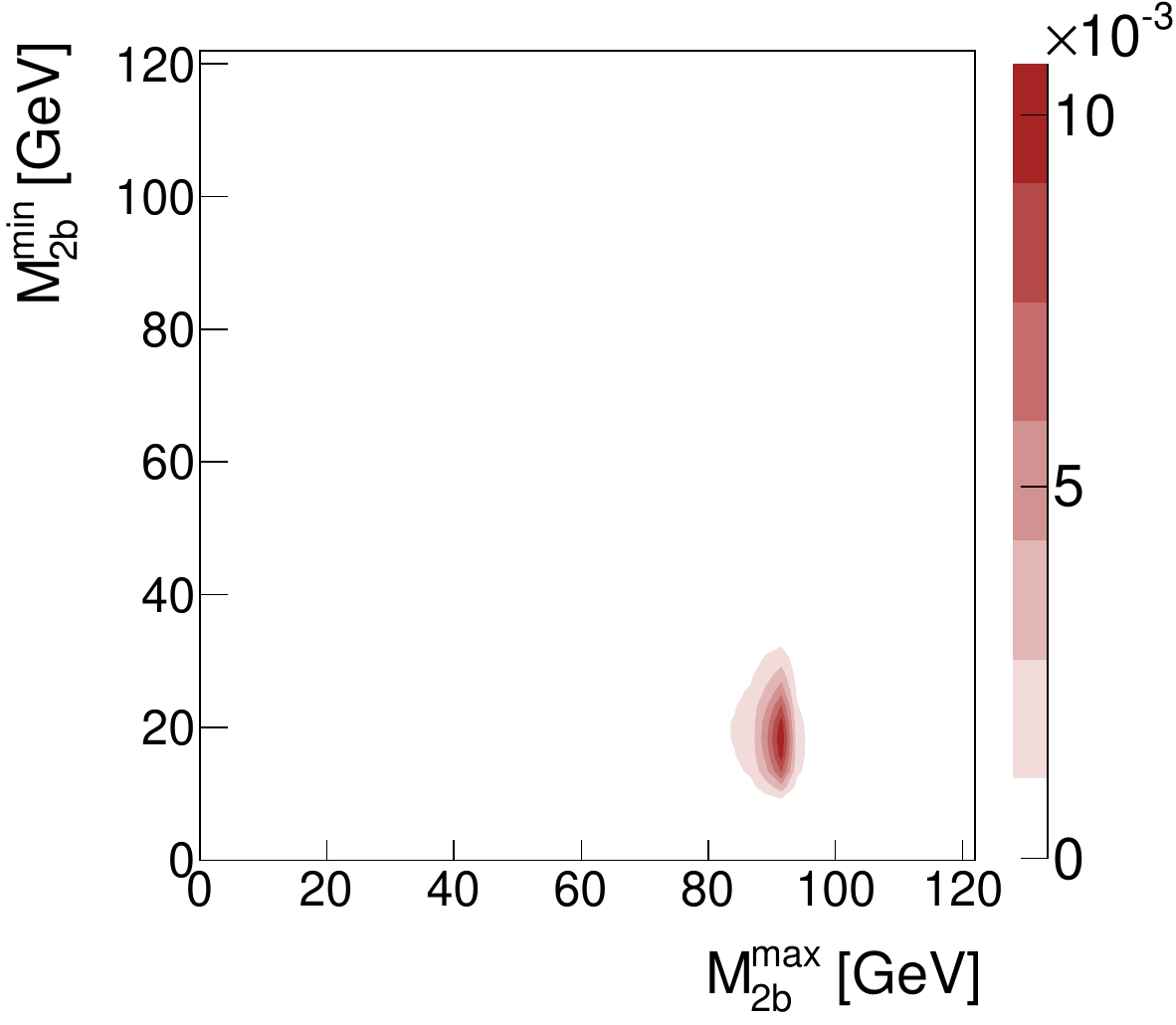}%
    \includegraphics[width=0.33\textwidth]{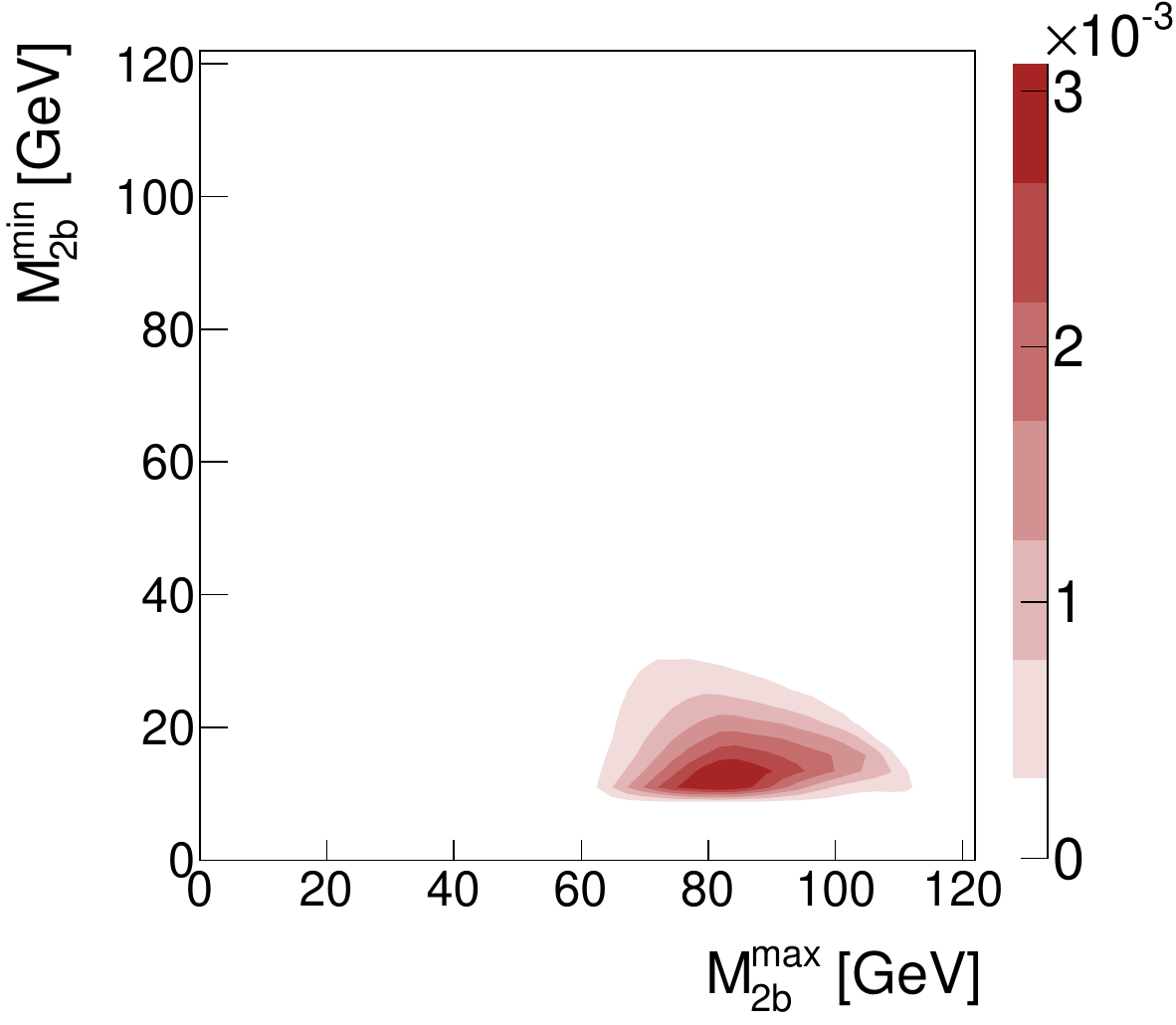}%
    \\
    \includegraphics[width=0.33\textwidth]{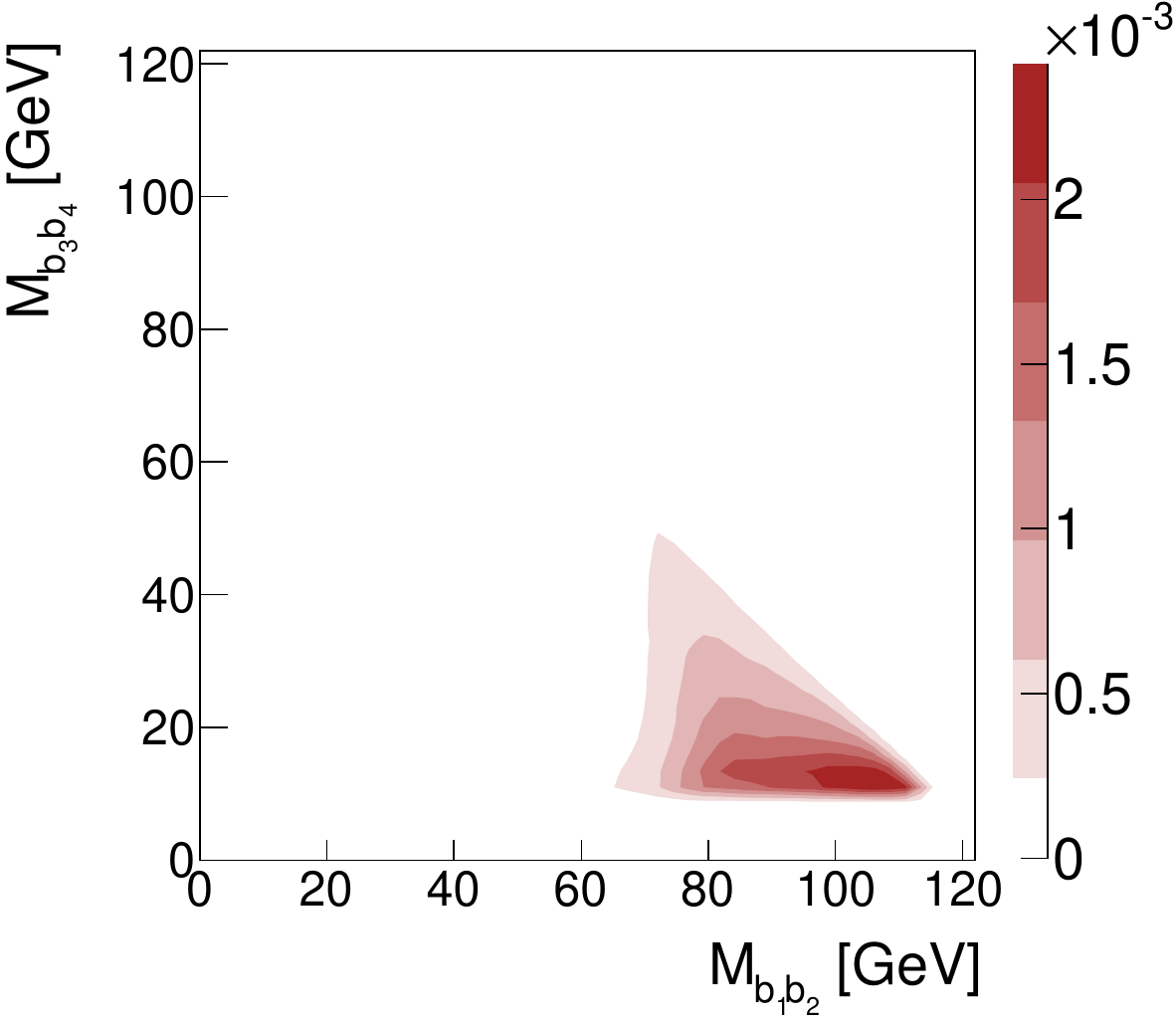}%
    \includegraphics[width=0.33\textwidth]{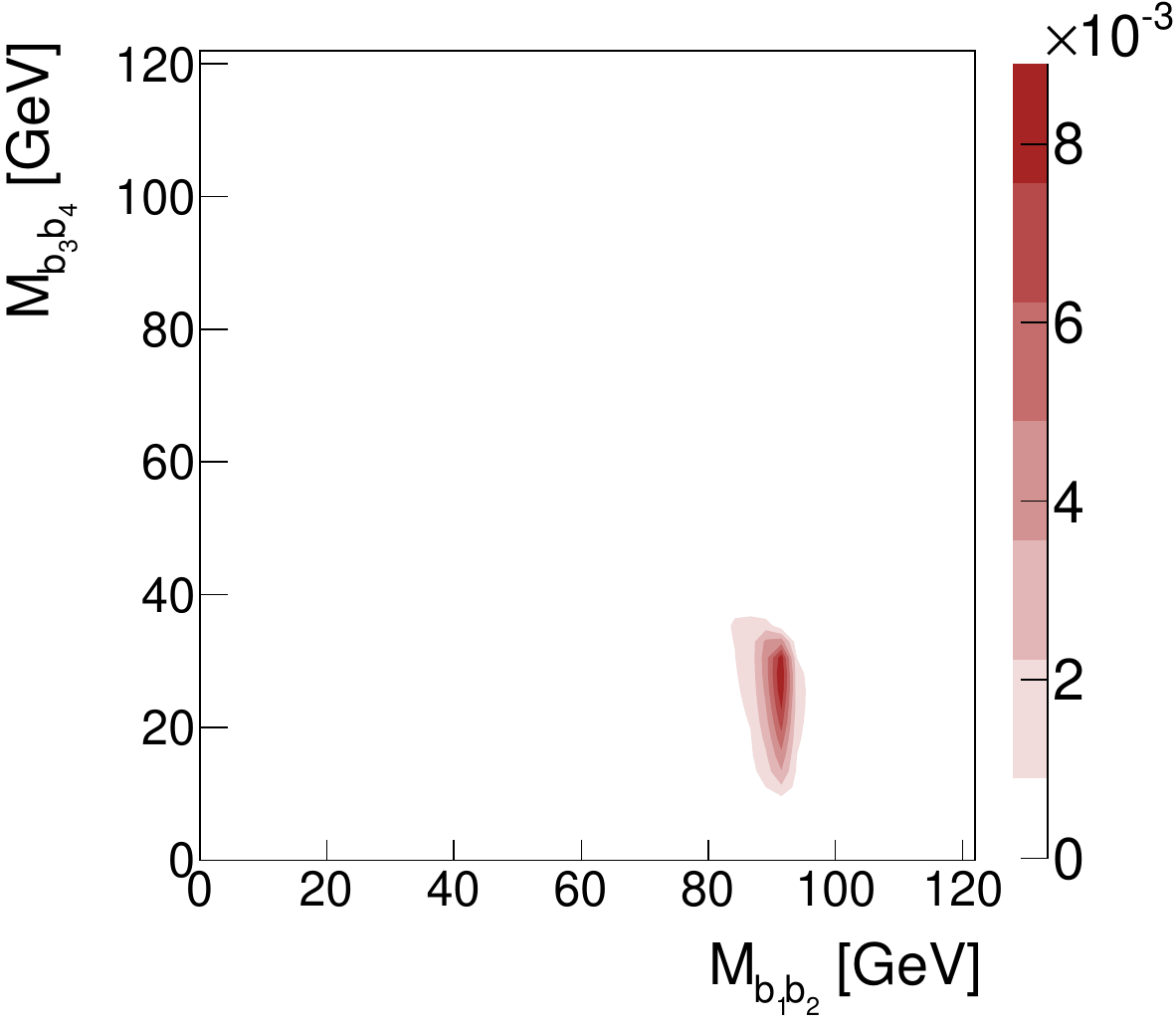}%
    \includegraphics[width=0.33\textwidth]{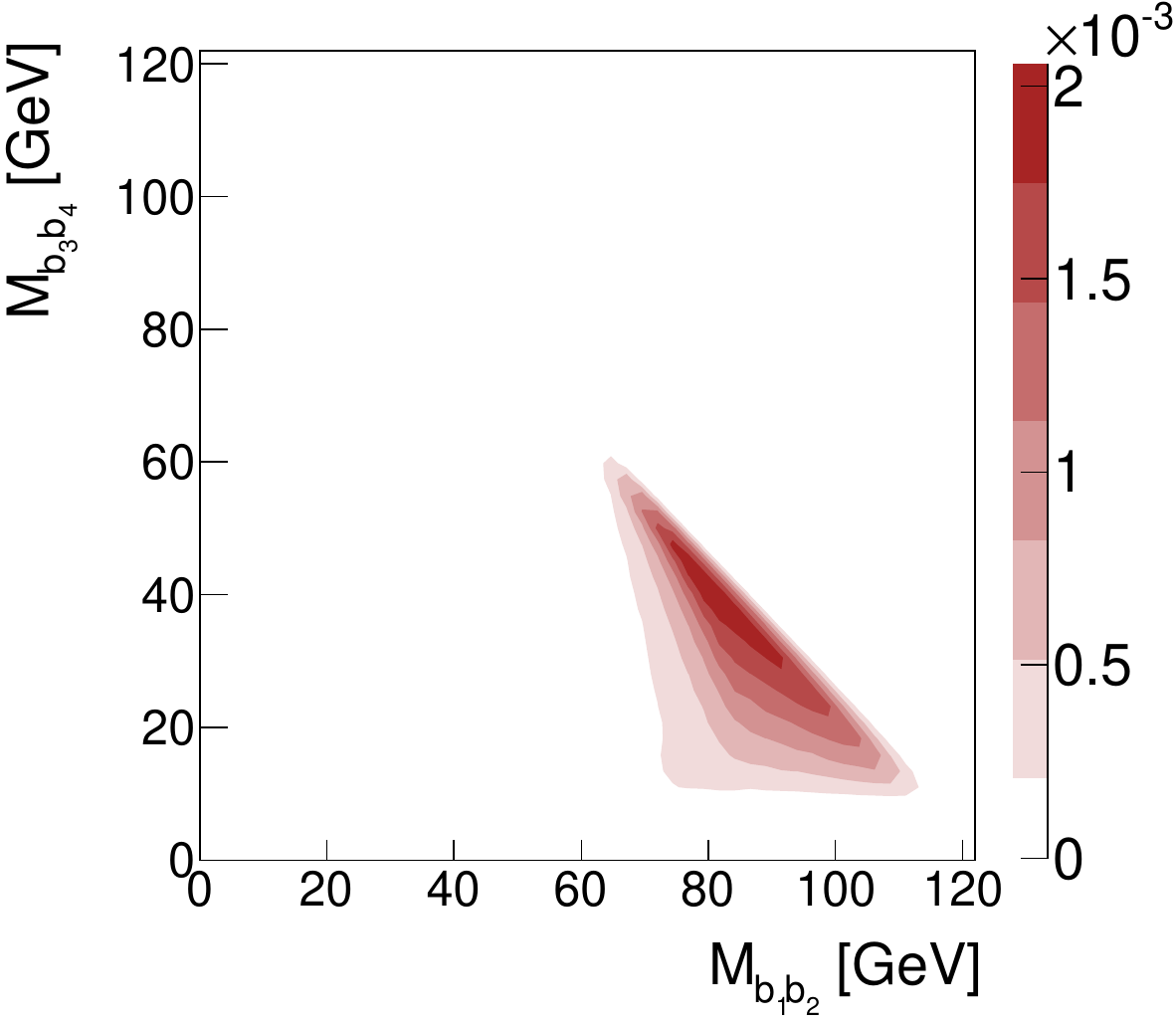}%
    \caption{
Normalized double-differential invariant mass distributions for 
$H \to b\bar{b} \to b\bar{b}g \to b\bar{b}b\bar{b}$ (left), 
$H \to ZZ^* \to b\bar{b}b\bar{b}$ (middle), and 
$H \to gg \to b\bar{b}b\bar{b}$ (right). 
\textbf{Top:} invariant mass of the $b$-quark pair with the maximum value versus that of the pair with the minimum value. 
\textbf{Bottom:} invariant mass of the two highest-energy $b$-quarks $(b_1,b_2)$ versus that of the remaining pair $(b_3,b_4)$. 
} . \label{fig:decay-2d-m2b-minmax}
\end{figure}

Two-dimensional invariant mass distributions provide more discriminating power than one-dimensional ones, both in separating signal from background and in distinguishing different signal topologies. 
Fig.~\ref{fig:decay-2d-m2b-random} illustrates this using Dalitz-type plots for the three subprocesses contributing to $H \to b\bar{b}b\bar{b}$. 
In the upper row, the invariant mass of two randomly selected $b$-quarks is plotted against that of the complementary pair, reflecting the experimental situation where the $b$ and $\bar{b}$ cannot be differentiated. 
In the lower row, the invariant masses of the two $b\bar{b}$ pairs are shown against each other, representing the idealised case where charge identification is possible. 
This comparison demonstrates that, in the idealised case, the separation of the different subprocesses contributing to $H \to b\bar{b}b\bar{b}$    is much more pronounced than 
in the realistic case where no distinction is made between $b$- and $\bar b$-jets.

In order to disentangle the different subprocesses contributing to the $H \to b\bar{b}b\bar{b}$ signal and to separate signal from background in a realistic analysis, let us  examine further two-dimensional invariant mass distributions. 
The top row of Fig.~\ref{fig:decay-2d-m2b-minmax} shows the invariant mass of the $b$-quark pair with the maximum value plotted against that of the pair with the minimum value. 
This representation extends the corresponding one-dimensional distribution discussed above by exploiting correlations between the two variables. 

The bottom row of Fig.~\ref{fig:decay-2d-m2b-minmax} presents the invariant mass of the two highest-energy $b$-quarks $(b_1,b_2)$ versus that of the remaining pair $(b_3,b_4)$. 
These two-dimensional distributions provide a more powerful handle to distinguish the three signal subprocesses:
\begin{enumerate}
    \item For $H\to b\bar{b}\to b\bar{b}g\to b\bar{b}b\bar{b}$, the distribution is stretched horizontally, with $M_{b_1b_2}$ concentrated in the 80--115~GeV range while $M_{b_3b_4}$ peaks at 10--20~GeV. 
    \item For $H\to ZZ^*\to b\bar{b}b\bar{b}$, the distribution is vertical, with $M_{b_1b_2}$ clustering around the $Z$ resonance and $M_{b_3b_4}$ in the 10--35~GeV range. 
    \item For $H\to gg\to b\bar{b}b\bar{b}$, the distribution forms a narrow diagonal strip extending from $(M_{b_1b_2},M_{b_3b_4})=(115,\,10)\,\text{GeV}$ to $(70,\,60)\,\text{GeV}$. 
\end{enumerate}
Altogether, the two-dimensional approach reveals distinctive horizontal, vertical, and diagonal patterns for the three subprocesses. These features, largely washed out in one-dimensional projections, provide clear discrimination and underscore the utility of the two-dimensional method in disentangling different Higgs decay topologies.

\begin{figure}[htb]
 \includegraphics[trim={0 1cm 0 1cm}, clip,width=\textwidth]{figs/legend1.pdf}\\ 
    \begin{subfigure}[a]{0.49\textwidth}
        \includegraphics[trim={0 0cm 0 1cm}, clip,width=\textwidth]{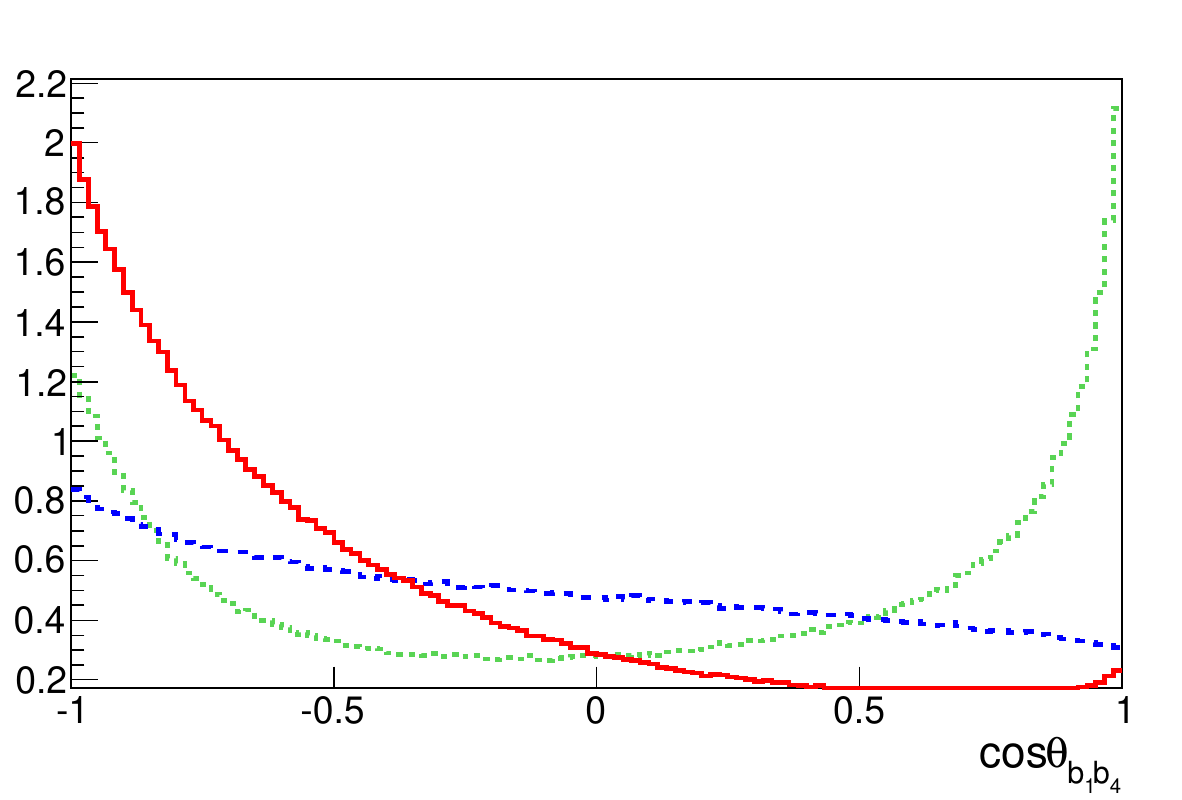}
        \caption{}
    \end{subfigure}
    \begin{subfigure}[a]{0.49\textwidth}
        \includegraphics[trim={0 0cm 0 1cm}, clip,width=\textwidth]{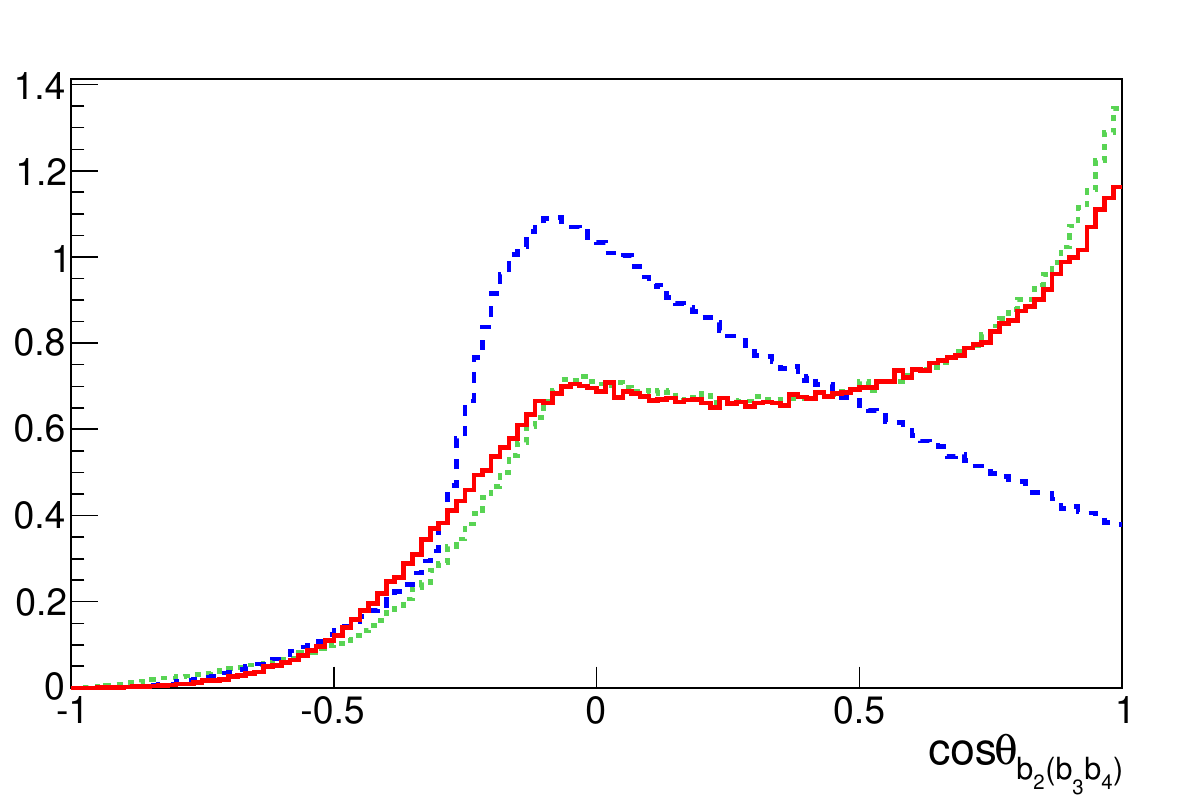}
        \caption{}
    \end{subfigure}
    \\
    \begin{subfigure}[c]{0.49\textwidth}
        \includegraphics[width=\textwidth]{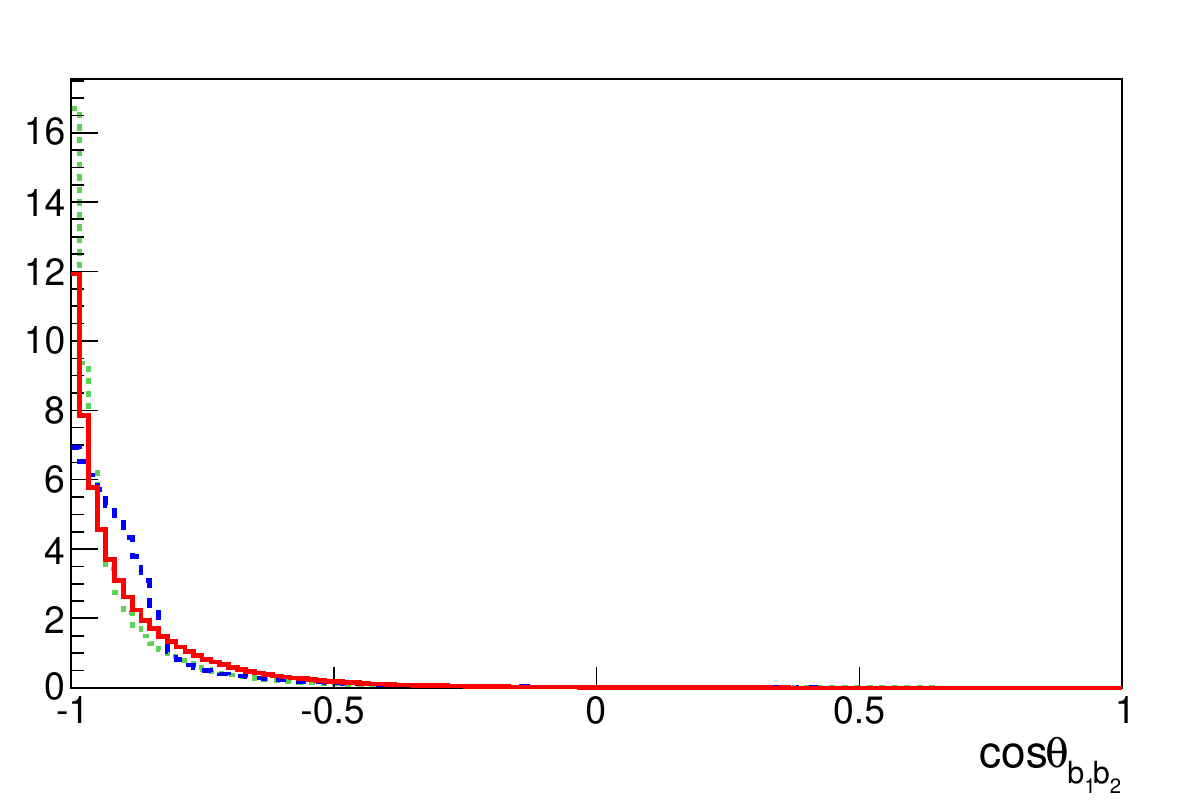}
        \caption{}
    \end{subfigure}
    \begin{subfigure}[d]{0.49\textwidth}
        \includegraphics[width=\textwidth]{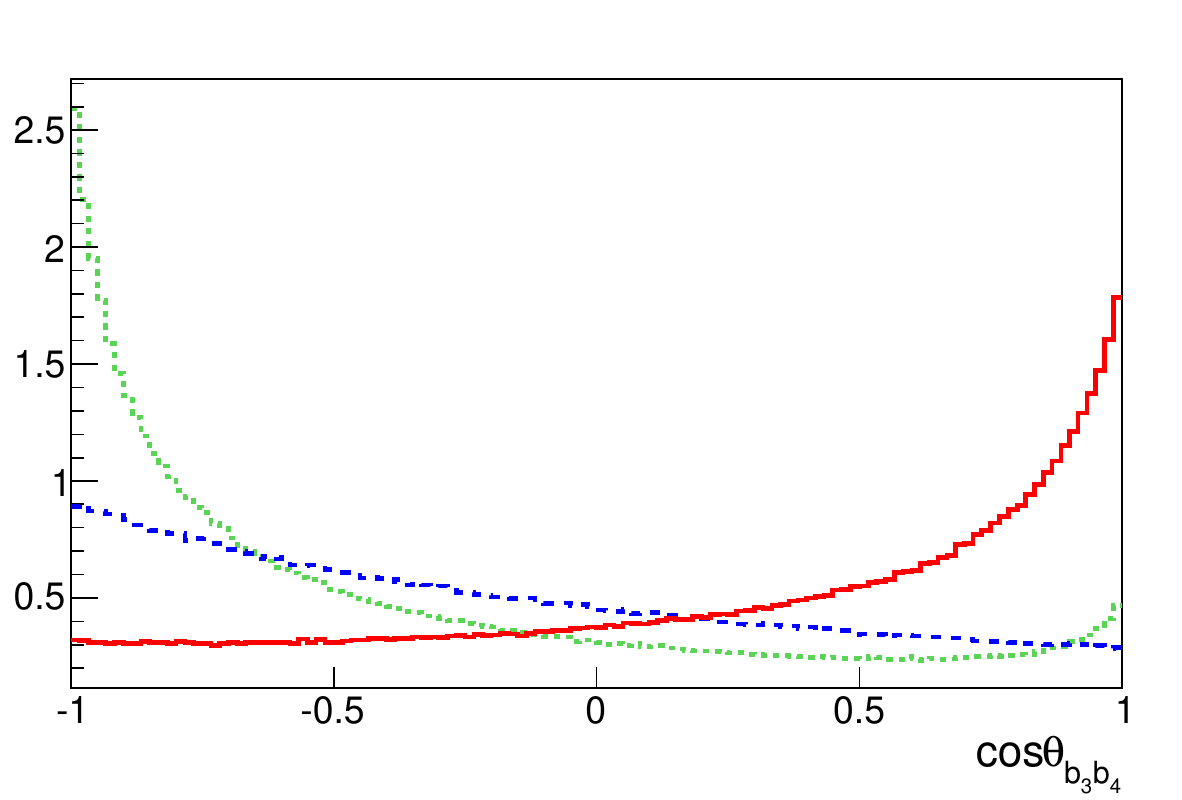}
        \caption{}
    \end{subfigure}
    \caption{Normalized angular distributions of $\cos\theta$ for Higgs decay subprocesses 
    $H\to b\bar{b}\to b\bar{b}g\to b\bar{b}b\bar{b}$ (red solid line), 
    $H\to ZZ^*\to b\bar{b}b\bar{b}$ (blue dashed line), and 
    $H\to gg\to b\bar{b}b\bar{b}$ (green dotted line). 
    {\bf (a)} Angle between the highest- and lowest-energy $b$-quarks, $b_1$ and $b_4$. 
    {\bf (b)} Angle between $b_2$ and the cluster ($b_3,b_4$). 
    {\bf (c)} Angle between the two highest-energy $b$-quarks, $b_1$ and $b_2$. 
    {\bf (d)} Angle between the two lowest-energy $b$-quarks, $b_3$ and $b_4$. 
    }
    \label{fig:hdecay-angular}
\end{figure}

Angular distributions provide additional discrimination power between different Higgs decay subprocesses. 
We define $\cos\theta_{ij}$ as the cosine of the angle between the momenta of $b_i$ and $b_j$, and 
$\cos\theta_{i(jk)}$ as the cosine of the angle between $b_i$ and the momentum of the cluster formed by $(b_j,b_k)$. 

The distribution of $\cos\theta_{b_1b_4}$ in Fig.~\ref{fig:hdecay-angular}(a), defined as the angle between the 
highest- and lowest-energy $b$-quarks, already separates $H\to ZZ^*\to b\bar{b}b\bar{b}$ from QCD-induced channels. 
Figure~\ref{fig:hdecay-angular}(b) shows $\cos\theta_{b_2(b_3b_4)}$, which also distinguishes $ZZ^*$ from QCD processes. 
Although the $b\bar{b}g$ and $gg$ shapes look similar, the dynamics differ: in the $b\bar{b}g$ case the radiated gluon is 
nearly collinear with its parent $b$, while in the $gg$ case the $b_2$ quark from the opposite gluon aligns with the 
$(b_3,b_4)$ cluster, both leading to peaks near~1. 
By contrast, $\cos\theta_{b_1b_2}$ in Fig.~\ref{fig:hdecay-angular}(c) is nearly identical across subprocesses and thus 
uninformative. The $\cos\theta_{b_3b_4}$ distribution in Fig.~\ref{fig:hdecay-angular}(d) is most powerful: gluon 
splitting yields peaks near~1 ($b\bar{b}g$) or $-1$ ($gg$), while the $ZZ^*$ contribution is flat due to the small boost 
of the real $Z$ boson and the possibility that $b_3$ and $b_4$ come from either the real or virtual $Z$.

In summary, Figs.~\ref{fig:hdecay-angular}(a,b) highlight differences between $ZZ^*$ and QCD topologies, while 
Fig.~\ref{fig:hdecay-angular}(d) provides the clearest separation of gluon-splitting from $ZZ^*$. 

\section{$pp\to WH\to W\fb$ signal versus background analysis at the LHC}
\label{section:s-vs-b}

In this section we analyse the $pp \to WH \to W\fb$ signal and the corresponding irreducible backgrounds, focusing on the High-Luminosity LHC (HL-LHC) regime. 
All evaluations and simulations presented below are performed at $\sqrt{s}=14$~TeV, the design energy foreseen for the HL-LHC run.\footnote{The HL-LHC design energy is $\sqrt{s}=14$~TeV, slightly above the current Run~3 value of $\sqrt{s}=13.6$~TeV, and is expected to be reached once HL-LHC operations commence.}

In Sec.~\ref{sec:Decay} we identified kinematic observables that discriminate between different $H\to\fb$ decay topologies, particularly the dominant $H\to b\bar b g$ component. We now assess the effectiveness of these variables for selecting the full process $pp\to WH\to W\fb$ against the irreducible backgrounds, employing both a transparent cut--based strategy and an advanced multivariate (BDT) approach.

The motivation for multivariate tools is clear: the LHC sensitivity to $H\to 4b$ in associated production is limited by a small signal--to--background ratio, as quantified below. 

For parton--level matrix element calculations and event generation, we use \textsc{CalcHEP}~3.8.10~\cite{Belyaev:2012qa}, with CTEQ6L PDFs and a dynamical QCD scale $Q=\hat s$.
The full-amplitude results, including interference effects, were cross-checked with \textsc{CompHEP}~4.5.2~\cite{CompHEP:2004qpa}.
The events are then passed to \textsc{PYTHIA}~8.245~\cite{Sjostrand:2014zea} for parton showering and hadronisation, and to \textsc{Delphes}~3.4.2~\cite{deFavereau:2013fsa} for detector simulation.

The Higgs boson mass used throughout the analysis is fixed to $M_H=125$~GeV.

\subsection{Signal versus background at the parton level}

\noindent
\textbf{Signal cross section.} For the $WH$ signal we use the inclusive cross sections and $K$--factors recommended by the LHC Higgs Cross Section Working Group (HXSWG)~\cite{Ferrera:2011bk,Ferrera:2013yqa,CERNYellowReport14TeV}, see also \url{https://twiki.cern.ch/twiki/bin/view/LHCPhysics/CERNYellowReportPageAt14TeV}. Associated $VH$ production is known at NNLO QCD with NLO electroweak corrections~\cite{Denner:2014cla, Harlander:2014wda, Altenkamp:2012sx, Denner:2011id, Brein:2003wg, Ciccolini:2003jy}. At $\sqrt{s}=14$~TeV the HXSWG tables provide the reference normalisation, $\sigma_{pp\to W^{+}H}=0.922~\text{pb}$ and $\sigma_{pp\to W^{-}H}=0.591~\text{pb}$ for $M_H=125$~GeV, corresponding to the total inclusive $WH$ production cross section $\sigma_{pp\to WH}=1.513~\text{pb}$. In our study, we first evaluate $pp\to W^{+}H$ and $pp\to W^{-}H$ at tree level, obtaining $\sigma^{\text{LO}}(W^+H)=0.805~\text{pb}$ and $\sigma^{\text{LO}}(W^-H)=0.497~\text{pb}$. The ratio to the HXSWG predictions defines the normalisation factors that account for higher--order corrections. These factors are then consistently applied to the  $pp\to WH\to Wb\bar{b}b\bar{b}$ ($2\to5$) process used in our simulations, ensuring that our event samples reproduce the state-of-the-art inclusive cross sections. The corresponding higher--order corrections amount to a net $K$--factor of $\mathcal{O}(1.1$--$1.2)$, in line with the HXSWG recommendations.

\noindent
\textbf{Background estimation.}
Before discussing the treatment of higher--order QCD effects for the $pp\to W+4b$ background, it is important to stress that its perturbative behaviour is fundamentally different from that of the signal process $pp\to WH$.
While $WH$ production is a Drell--Yan--type electroweak process whose NLO QCD corrections are known to be modest, with inclusive $K_{\rm NLO}\simeq 1.1$--$1.2$, the QCD production of heavy flavour in association with a $W$ boson constitutes a genuinely multi--scale process starting at ${\cal O}(\alpha_s^n)$ (with $n\ge 2$).
In this case, NLO corrections open qualitatively new production mechanisms driven by $qg$--initiated subprocesses and gluon splitting into heavy flavour, which can substantially reorganise the dominant contributions to the cross section.
As a result, sizeable NLO enhancements in $W+$heavy--flavour production do not contradict the perturbative stability of $WH$, but instead reflect the very different dynamical origin of the two processes.

The dominant irreducible background to the $pp\to WH\to W+4b$ signal arises from QCD production of the $pp\to W+4b$ final state.
A dedicated fixed--order NLO QCD prediction for $pp\to W+4b$ in the fiducial setup considered here is not currently available.
To estimate the potential size of higher--order effects, we therefore use the known perturbative behaviour of the inclusive $pp\to Wb\bar b$ process as guidance, since it captures the same key QCD mechanisms---in particular $qg$--initiated channels and heavy--flavour production via gluon splitting---that are expected to dominate the corrections in $W+4b$ production.

Existing NLO QCD studies of inclusive $Wb\bar b$ production have demonstrated that QCD corrections can be large, driven primarily by the opening of new $qg$--initiated partonic channels at NLO and by enhanced real--emission contributions~\cite{Campbell:2003hd,FebresCordero:2006sj,Anger:2017glm}.
The magnitude of these corrections has been quantified explicitly at NLO in Refs.~\cite{Campbell:2003hd,FebresCordero:2006sj}, where inclusive NLO/LO ratios significantly above unity are observed.
More recently, the fully inclusive NNLO QCD study of Ref.~\cite{Buonocore:2022pqq} has provided a modern benchmark for this behaviour, confirming that inclusive NLO enhancements of order ${\cal O}(3)$ can arise, depending on the observable definition.
\\
However, for the present analysis the relevant phase--space region is the Higgs--mass window in the $b\bar b$ invariant mass, rather than the fully inclusive rate.
Ref.~\cite{Campbell:2003hd} explicitly presents the differential $M_{b\bar b}$ spectrum and shows an enhancement of order two in the vicinity of the Higgs--mass region.
We have performed an independent cross-check of this behaviour using \textsc{MadGraph5\_aMC@NLO}~\cite{Alwall:2014hca}
(version \texttt{MG5\_aMC\_v3\_6\_6}), considering only the pure QCD contribution to $pp\to Wb\bar b$,
with diagrams involving intermediate Higgs- or $Z$-boson exchange removed.

\begin{figure}[htbp]
  \centering
  \includegraphics[width=0.60\textwidth]{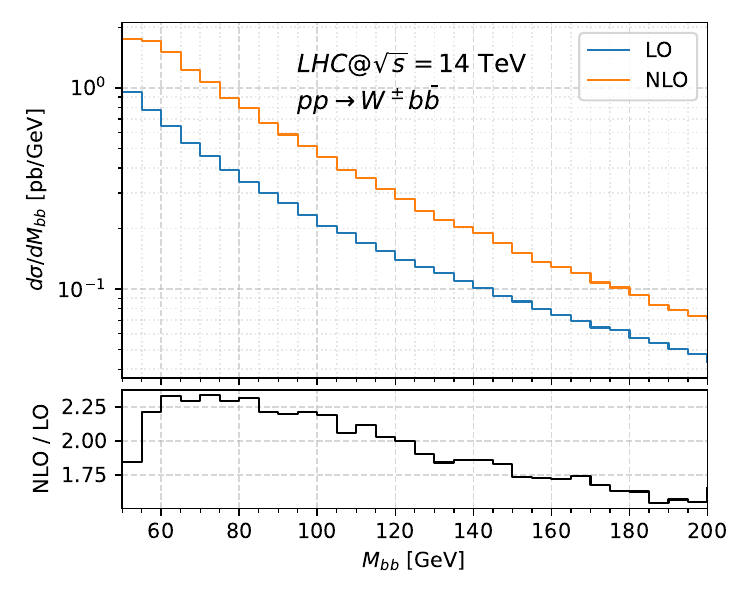}
  \caption{\label{fig:Mbb_NLO_LO}
  Differential $M_{b\bar b}$ distribution at LO and NLO (upper panel) and their ratio (lower panel)
  for the pure QCD contribution to $pp\to Wb\bar b$, computed with
  \textsc{MadGraph5\_aMC@NLO}~\cite{Alwall:2014hca}
  (version \texttt{MG5\_aMC\_v3\_6\_6}).
  The results are shown after applying the common parton--level preselection defined in
  Eq.~\eqref{eq:lhc-cuts}. In the Higgs--mass region, the NLO/LO ratio is close to $2$.}
\end{figure}

The quoted LO and NLO cross sections and the $M_{b\bar b}$ spectra shown in Fig.~\ref{fig:Mbb_NLO_LO}
are obtained after applying the common parton--level preselection defined in Eq.~\eqref{eq:lhc-cuts}.
For this preselection, we obtain $\sigma_{\rm LO}=34.3$~pb and $\sigma_{\rm NLO}=69.4$~pb.
As shown in Fig.~\ref{fig:Mbb_NLO_LO}, the differential NLO/LO ratio in the region $M_{b\bar b}\simeq M_H$
is close to two.
We therefore adopt $K_{\rm NLO}\simeq 2$ as the appropriate normalisation factor for the background estimate
in the Higgs--mass region relevant for $H\to4b$.
Since this factor is inferred from $Wb\bar b$ rather than from a dedicated NLO calculation of $W+4b$, it should be understood as a phenomenological normalisation of the dominant non-resonant background.
Throughout this study, LO matrix elements are used to model the shapes of kinematic distributions.

The common parton--level preselection applied to both signal and background is
\begin{equation}
\label{eq:lhc-cuts}
\begin{aligned}
p_T^b &> 10~\text{GeV}, \qquad &&\text{for each } b \text{ quark},\\
\Delta R_{bb} &\equiv \sqrt{(\Delta\eta_{bb})^2+(\Delta\phi_{bb})^2}>0.5, \qquad &&\text{for every } b\text{--quark pair},\\
50~\text{GeV} &< M_{4b}<200~\text{GeV}. &&
\end{aligned}
\end{equation}

This preselection is deliberately loose: it removes the soft and collinear regions while retaining the full Higgs-mass region and the adjacent continuum needed for the subsequent shape analysis.

\begin{table}[htbp]
  \begin{center}
    \begin{tabular}{|l|c|} \hline\hline
      Subprocess & Cross section (fb) \\ \hline\hline
      \multicolumn{2}{|c|}{Signal} \\ \hline
      \multicolumn{2}{|l|}{$pp \to WH$, $H\to \fb$} \\ \hline
      \quad $H\to b\bar{b}\to b\bar{b}g\to b\bar{b}b\bar{b}$ & $3.11\cdot10^{-1}$ \\
      \quad $H\to gg\to \fb$ & $7.40\cdot10^{-3}$ \\
      \quad $H\to ZZ\to \fb$ & $2.76\cdot10^{-1}$ \\
      \quad Total (including interference) & $0.577$ \\ \hline\hline
      \multicolumn{2}{|c|}{Background} \\ \hline
      $pp \to WHb\bar{b} \to W\fb$ & $0.482$ \\ \hline
      \multicolumn{2}{|l|}{$pp\to W\fb$ (non-resonant, no Higgs)} \\ \hline
      \quad QCD & $76.1$ \\
      \quad EW-QCD & $13.8$ \\
      \quad EW & $1.15\cdot10^{-1}$ \\
      \quad Total (including interference) & $90.4$ \\ \hline
      Total background & $90.9$ \\ \hline\hline
    \end{tabular}
  \end{center}
  \caption{\label{tab:lhc-prod}
Cross sections for signal and background after applying the selection cuts of Eq.~\eqref{eq:lhc-cuts}.
For the signal, selected contributions are shown separately, together with the full result including interference.
All values are obtained at LO parton level and rescaled by the corresponding $K$-factors.
In particular, the dominant non-resonant $W+4b$ background is normalised using $K_{\rm NLO}\simeq 2$,
inferred from the behaviour of $Wb\bar b$ production in the Higgs-mass region, as discussed in the text.}

\end{table}

The summary of signal and background cross sections is given in Table~\ref{tab:lhc-prod}.

After the loose parton-level preselection, the total background is still larger than the signal by a factor of about $90.9/0.577\simeq 1.6\times 10^2$.
This motivates us to look for various kinematical observables which would allow us to separate signal and background.

We first discuss such distributions at parton level and then, in the next subsection, perform the signal-versus-background analysis after fast detector simulation.

\begin{figure}[htb]
\includegraphics[width=0.5\textwidth,center]{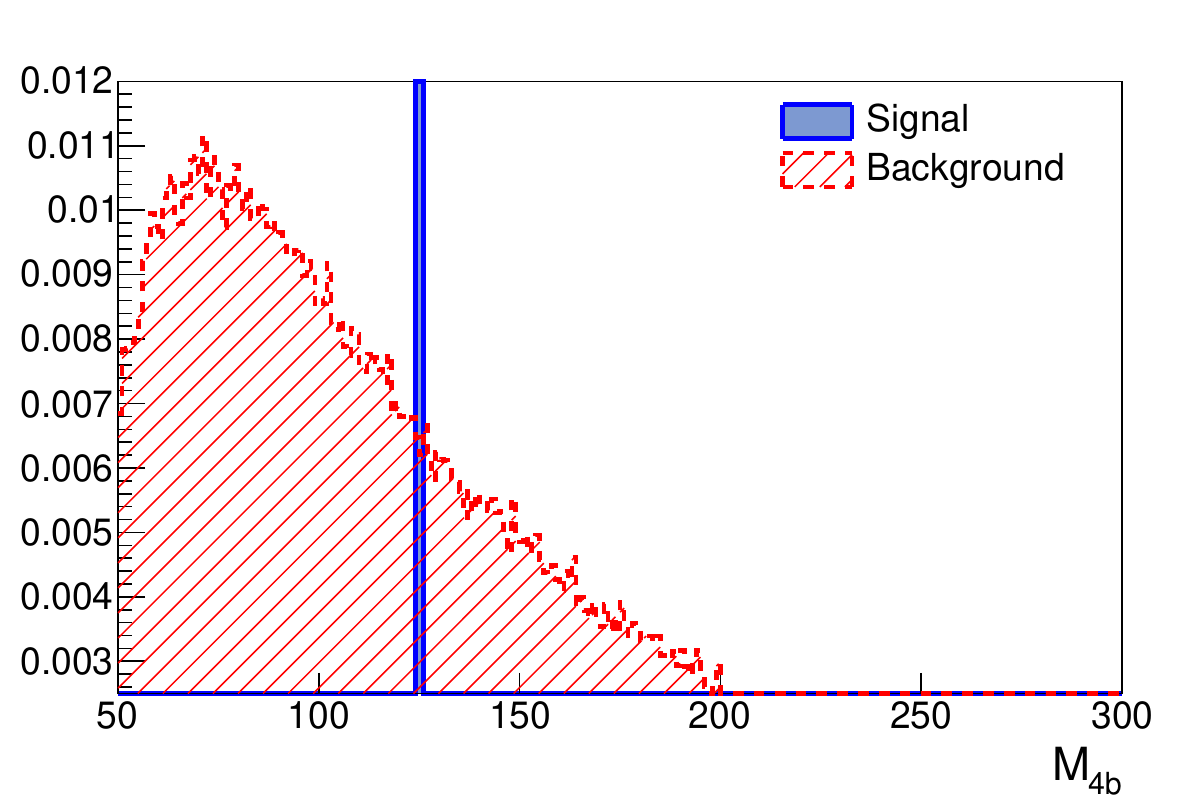}
	\caption{\label{fig:m4b}
        Normalized distribution of the four-$b$ invariant mass for the signal (blue) and background (red) processes for the $pp \to Wb\bar{b}b\bar{b}$ final state at the parton level.
        }
\end{figure}

\paragraph{One-dimensional distributions.}

We start with the central observable of the study, the invariant mass of the four-$b$ system, shown in Fig.~\ref{fig:m4b}.
At parton level, the signal forms a narrow structure around $M_H$, while the non-resonant background is continuous.
A narrow $M_{4b}$ window would therefore provide strong discrimination at this idealised level.
However, as shown below in the detector-level analysis, the power of this variable is substantially reduced by jet reconstruction and energy-smearing effects.
For this reason, we also explore additional observables: the energies of the $b$ quarks in the reconstructed Higgs rest frame, transverse momenta in the lab frame, invariant masses of two- and three-$b$ subsystems, and angular correlations in the Higgs rest frame.

{
For the variables defined in the reconstructed Higgs rest frame, the selected $b$ quarks are ordered by decreasing energy,
$E_1>E_2>E_3>E_4$.
This convention is independent of the lab-frame transverse-momentum ordering used for $p_T$ variables.
Thus, $M_{12}$ and $M_{34}$ denote the invariant masses of the two most energetic and two least energetic $b$ quarks in the reconstructed Higgs rest frame, respectively.}
\begin{figure}[htbp]
  \begin{subfigure}[a]{0.48\textwidth}
    \includegraphics[width=\textwidth]{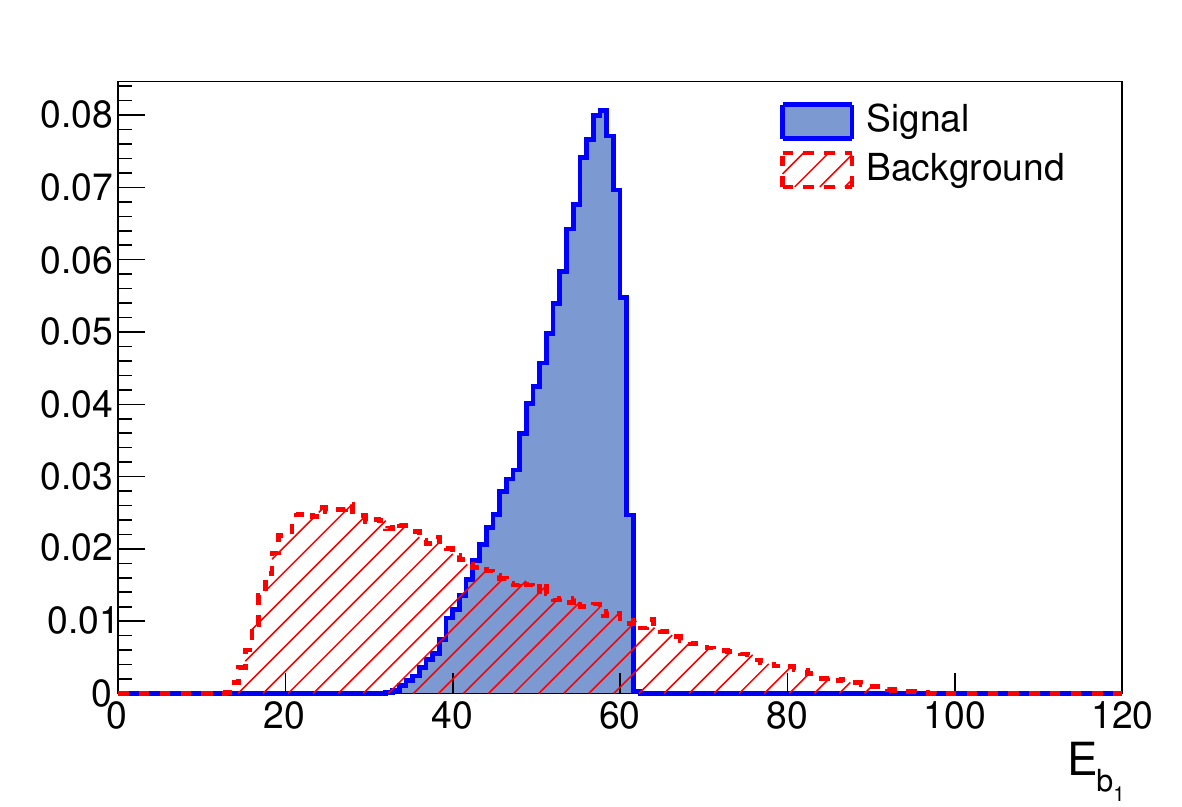}
    \caption{}
  \end{subfigure}
  \begin{subfigure}[a]{0.48\textwidth}
    \includegraphics[width=\textwidth]{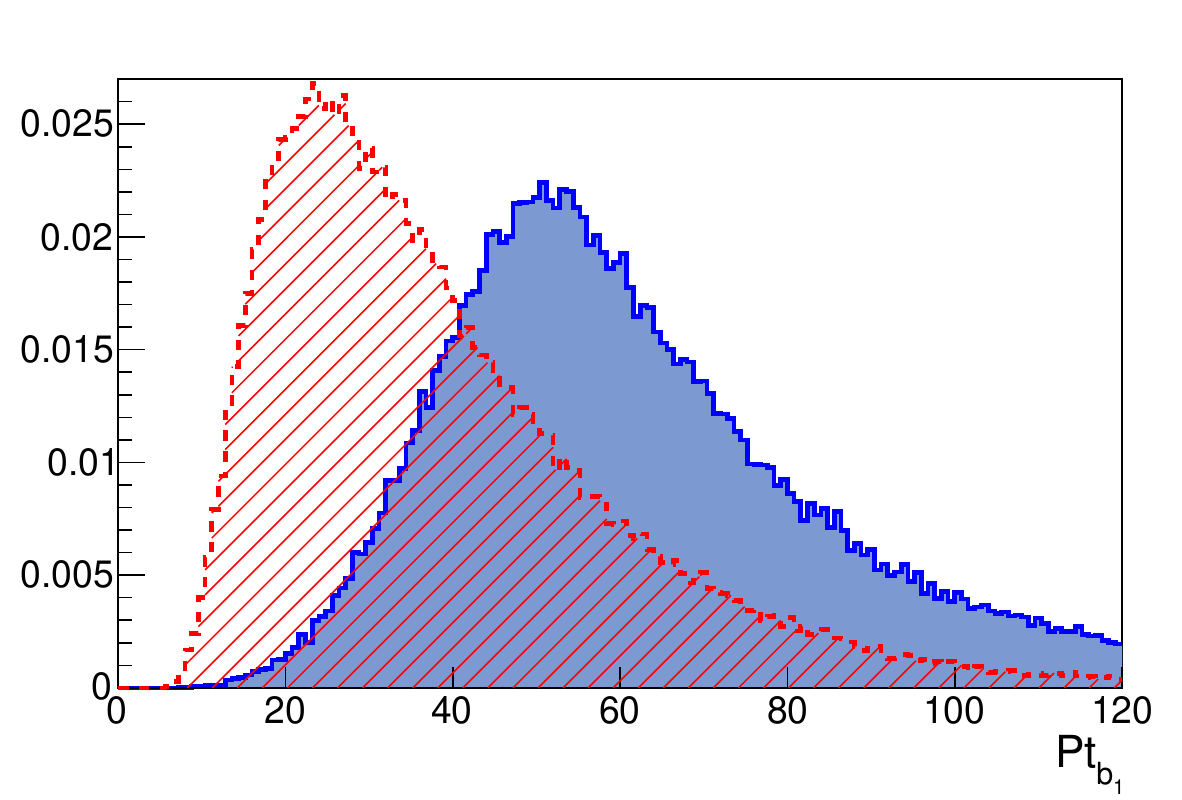}
    \caption{}
  \end{subfigure}
\\  \begin{subfigure}[c]{0.48\textwidth}
    \includegraphics[width=\textwidth]{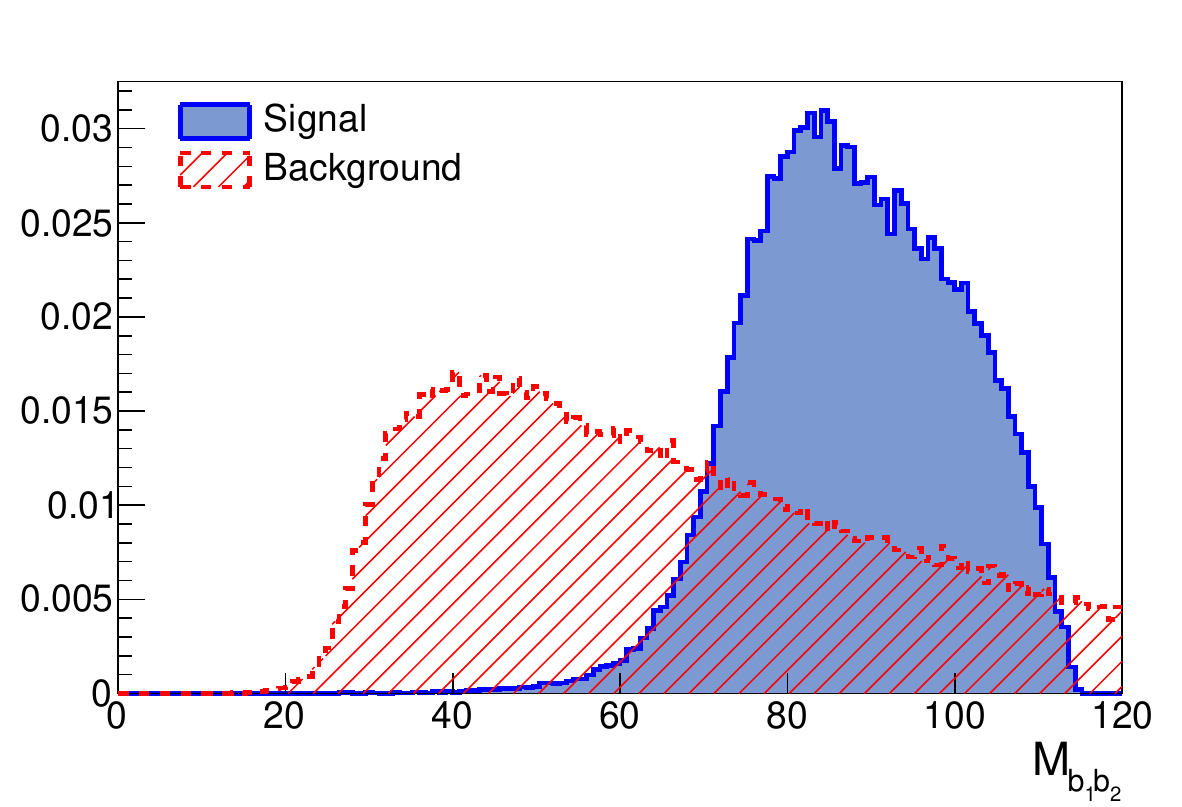}
    \caption{}
  \end{subfigure}
  \begin{subfigure}[d]{0.48\textwidth}
    \includegraphics[width=\textwidth]{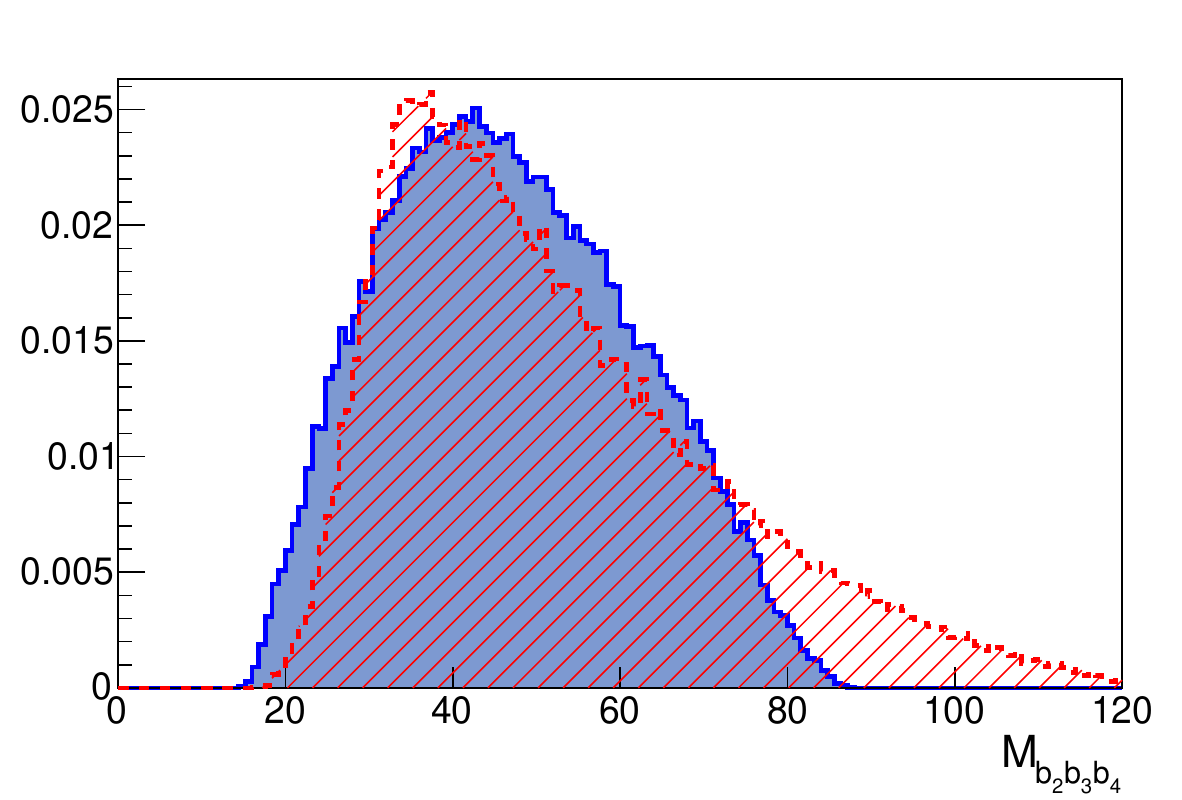}
    \caption{}
  \end{subfigure}
  \\
  \begin{subfigure}[c]{0.48\textwidth}
    \includegraphics[width=\textwidth]{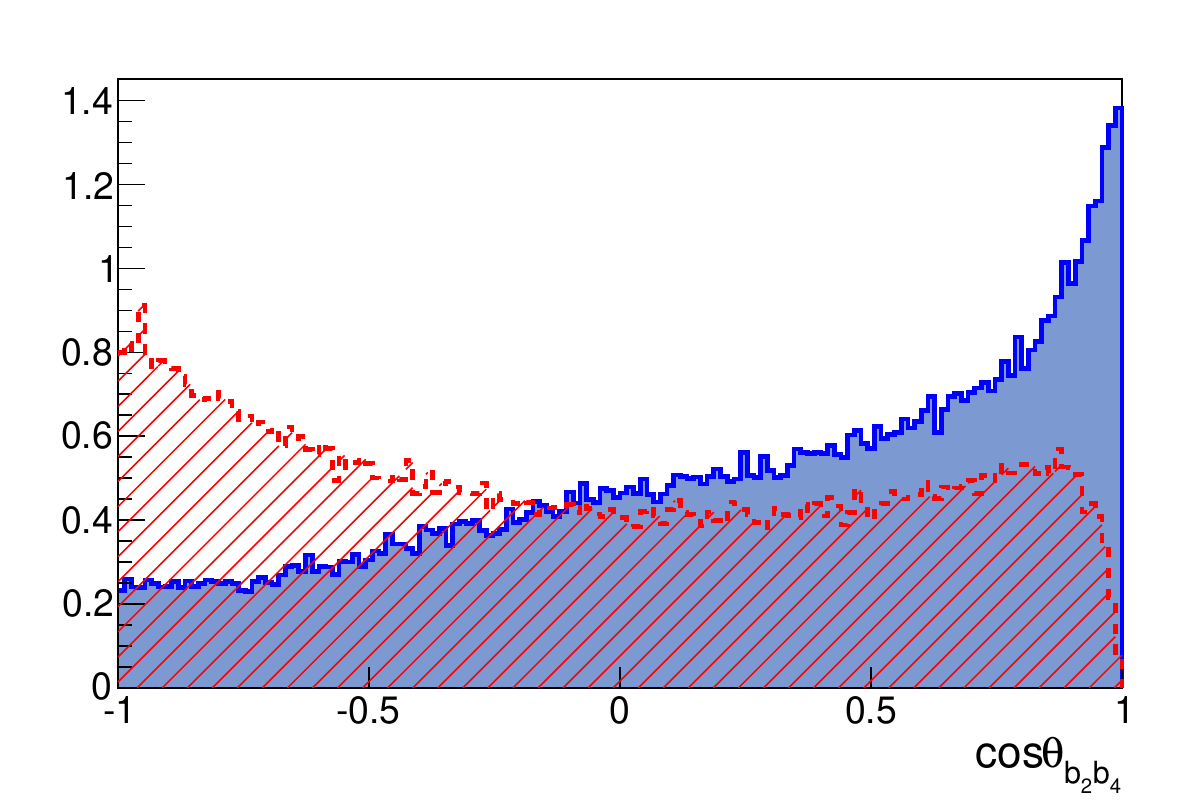}
    \caption{}
  \end{subfigure}
  \begin{subfigure}[d]{0.48\textwidth}
    \includegraphics[width=\textwidth]{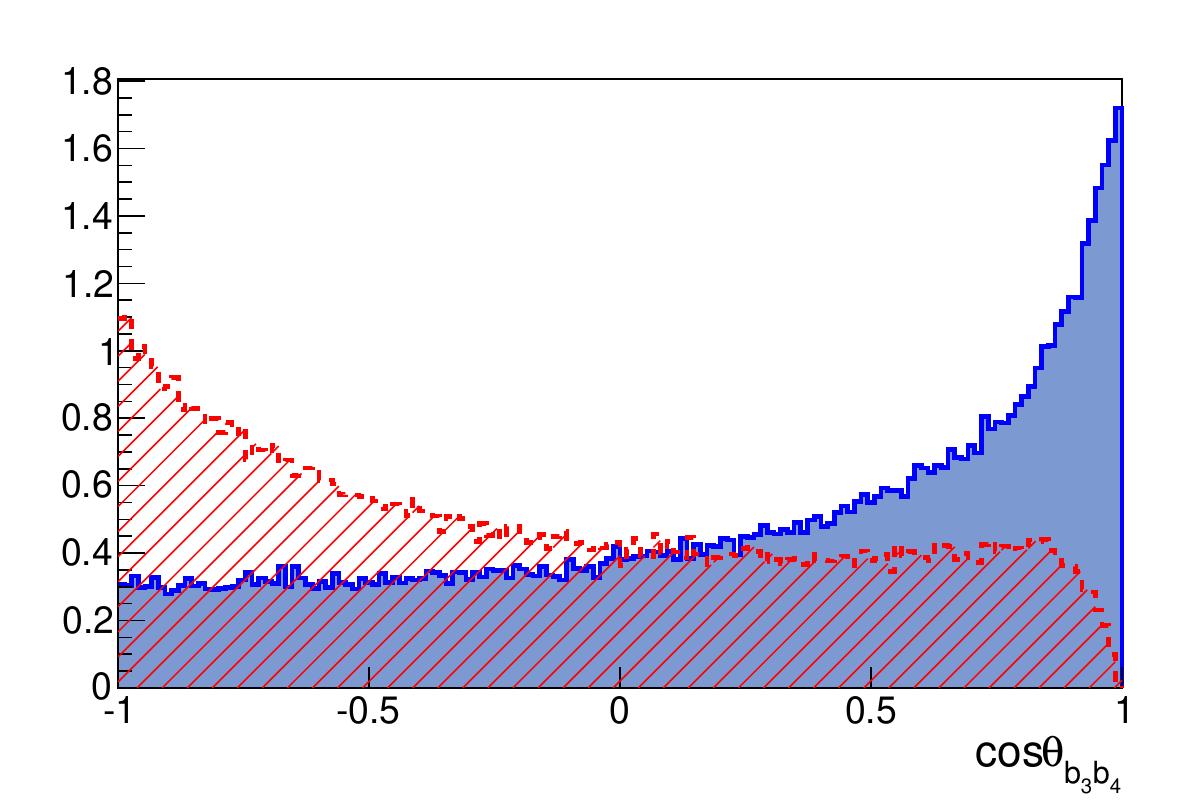}
    \caption{}
  \end{subfigure}
  \caption{\label{fig:BDT-vars-part}
  Normalized parton-level distributions for the $pp\to W b\bar b b\bar b$ signal (blue) and background (red):
  (a) $E_1$, the energy of the most energetic $b$ quark in the reconstructed Higgs rest frame;
  (b) $p_{T,1}$, the transverse momentum of the leading $b$ quark in the lab frame;
   (c)  $M_{12}$, the invariant mass of the two most energetic $b$ quarks in the reconstructed Higgs rest frame;
   (d)$M_{234}$, the invariant mass of the three least energetic $b$ quarks in the reconstructed Higgs rest frame;
   (e) $\cos\theta_{24}$;
  (f) $\cos\theta_{34}$.}
\end{figure}

In Fig.~\ref{fig:BDT-vars-part}(a), we show the distribution of $E_1$, the energy of the most energetic $b$ quark in the reconstructed Higgs rest frame.
The signal peaks near $M_H/2$, as expected from the dominant $H\to b\bar b g$ topology, where two energetic $b$ quarks retain a large fraction of the Higgs energy.
In contrast, the background is shifted towards lower energies and falls at small energies as a consequence of the $p_T^b>10$~GeV preselection.
Fig.~\ref{fig:BDT-vars-part}(b) presents the transverse momentum of the leading $b$ quark in the lab frame:
the signal is harder, while the background is broader and shifted to lower $p_T$.
These two panels already show that simple energy and $p_T$ variables carry useful discriminating information.

Fig.~\ref{fig:BDT-vars-part}(c) shows $M_{12}$ in the reconstructed Higgs rest frame.
For the signal, this variable is concentrated at large invariant masses, reflecting the two hard $b$ quarks produced in the primary Higgs decay chain.
The background instead populates lower masses more strongly.

Panel (d), $M_{234}$, is less sharply separating than $M_{12}$ but still shows a systematic difference between signal and background, with signal events extending to larger invariant masses.
Panels (e) and (f) display angular correlations in the reconstructed Higgs rest frame:
for $\cos\theta_{24}$, signal events are enhanced near $+1$, consistent with collinear configurations generated by gluon splitting, while the background is more broadly distributed.
For $\cos\theta_{34}$, the signal again shows an enhancement near $+1$, whereas the background has a stronger component towards negative values, reflecting more back-to-back configurations typical of QCD heavy-flavour production.
These angular observables provide discriminating power complementary to the invariant-mass variables.

\paragraph{Two-dimensional distributions.}
The one-dimensional distributions discussed above demonstrate that several variables are individually useful, but they do not show whether the discriminating information is correlated.
This is important because the dominant $H\to b\bar b g\to b\bar b b\bar b$ topology is not characterised by a single variable alone: it typically contains a hard two-$b$ system together with a softer, often more collinear pair from gluon splitting.
The background, by contrast, is generated by continuum QCD heavy-flavour production and can populate similar one-dimensional ranges while having different correlations between the same variables.
For this reason, we examine two-dimensional distributions before moving to the multivariate analysis.

\begin{figure}[htb]
  \includegraphics[width=0.48\textwidth]{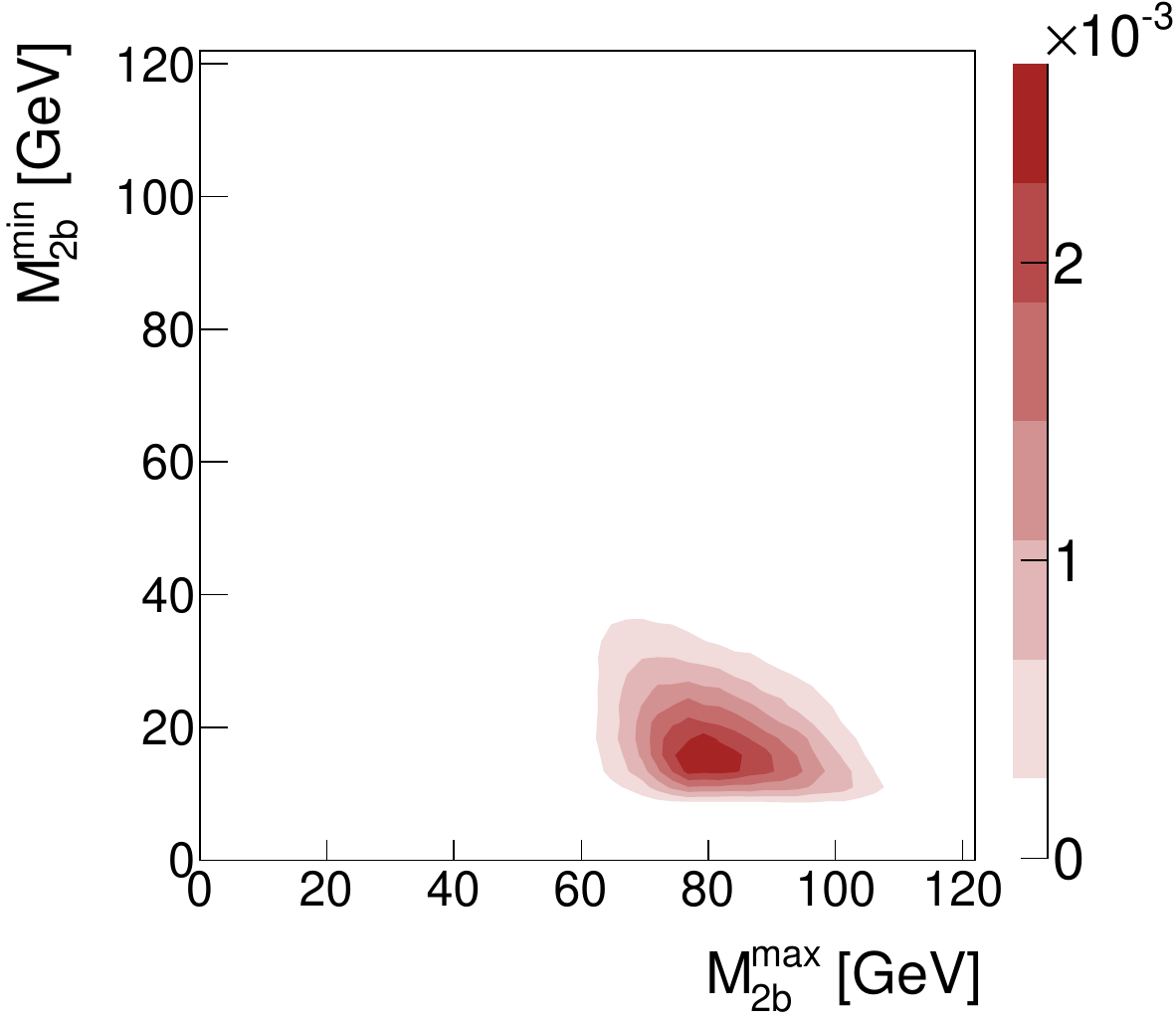}%
  \includegraphics[width=0.48\textwidth]{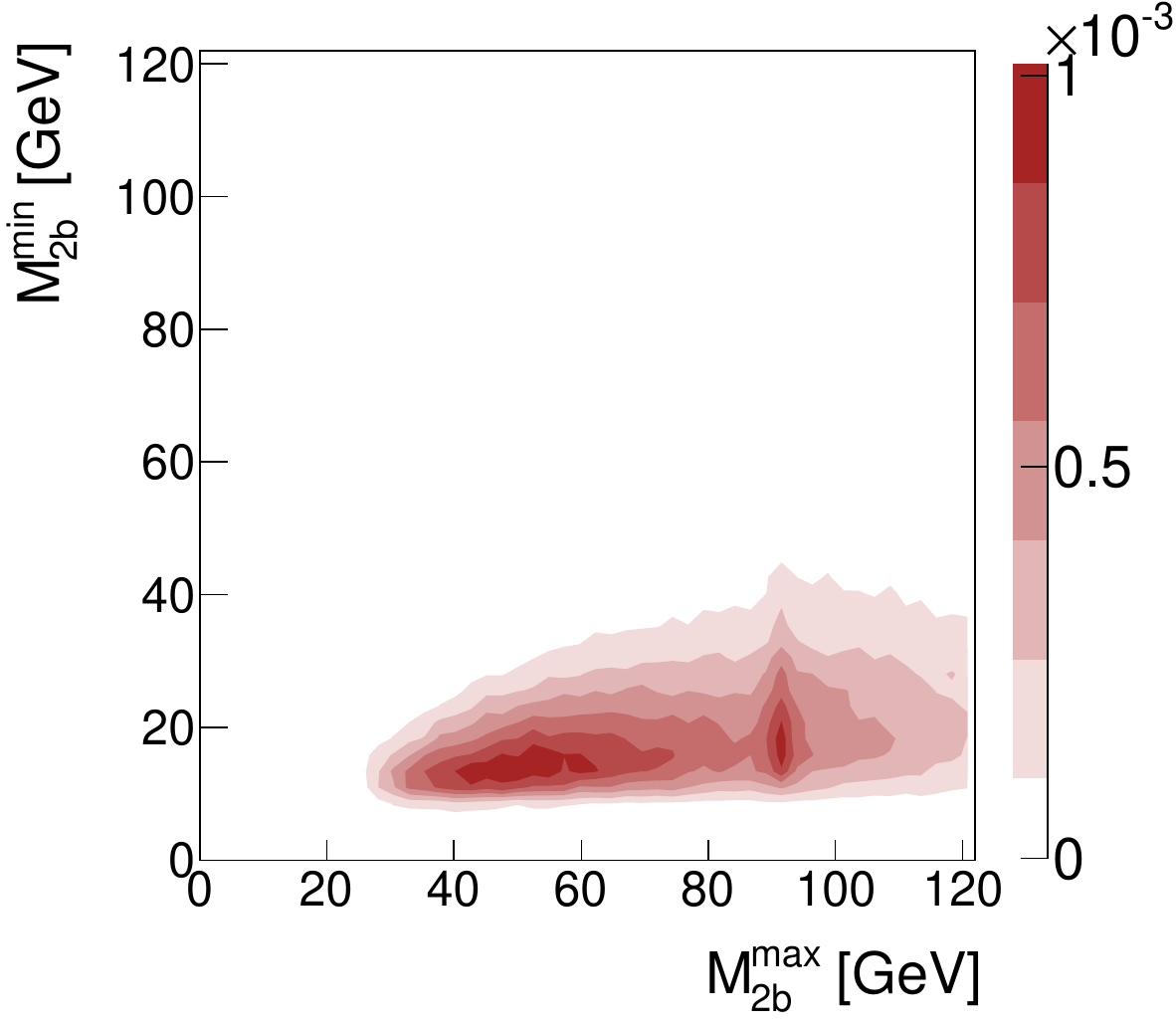}%
  \\
  \includegraphics[width=0.48\textwidth]{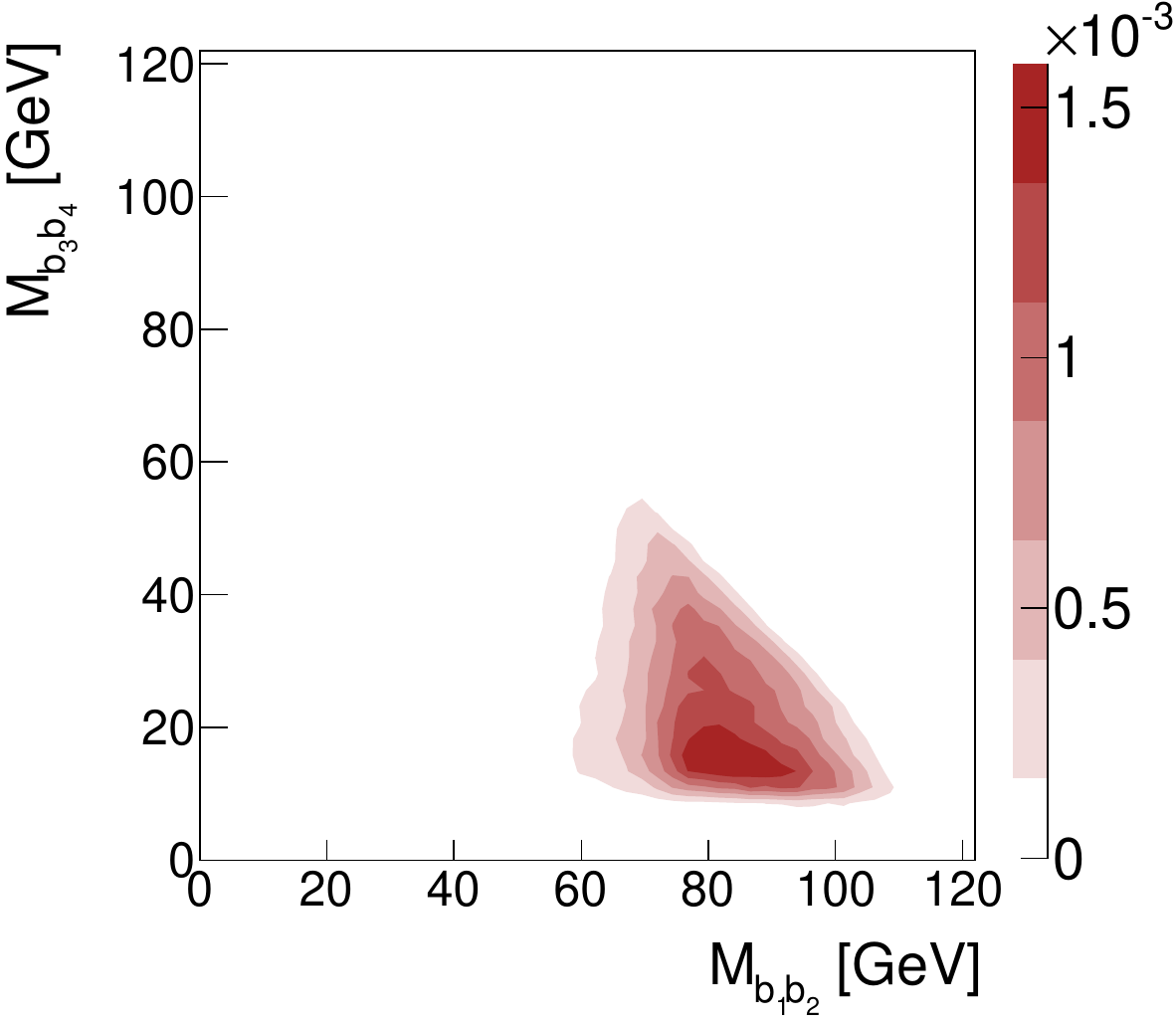}%
  \includegraphics[width=0.48\textwidth]{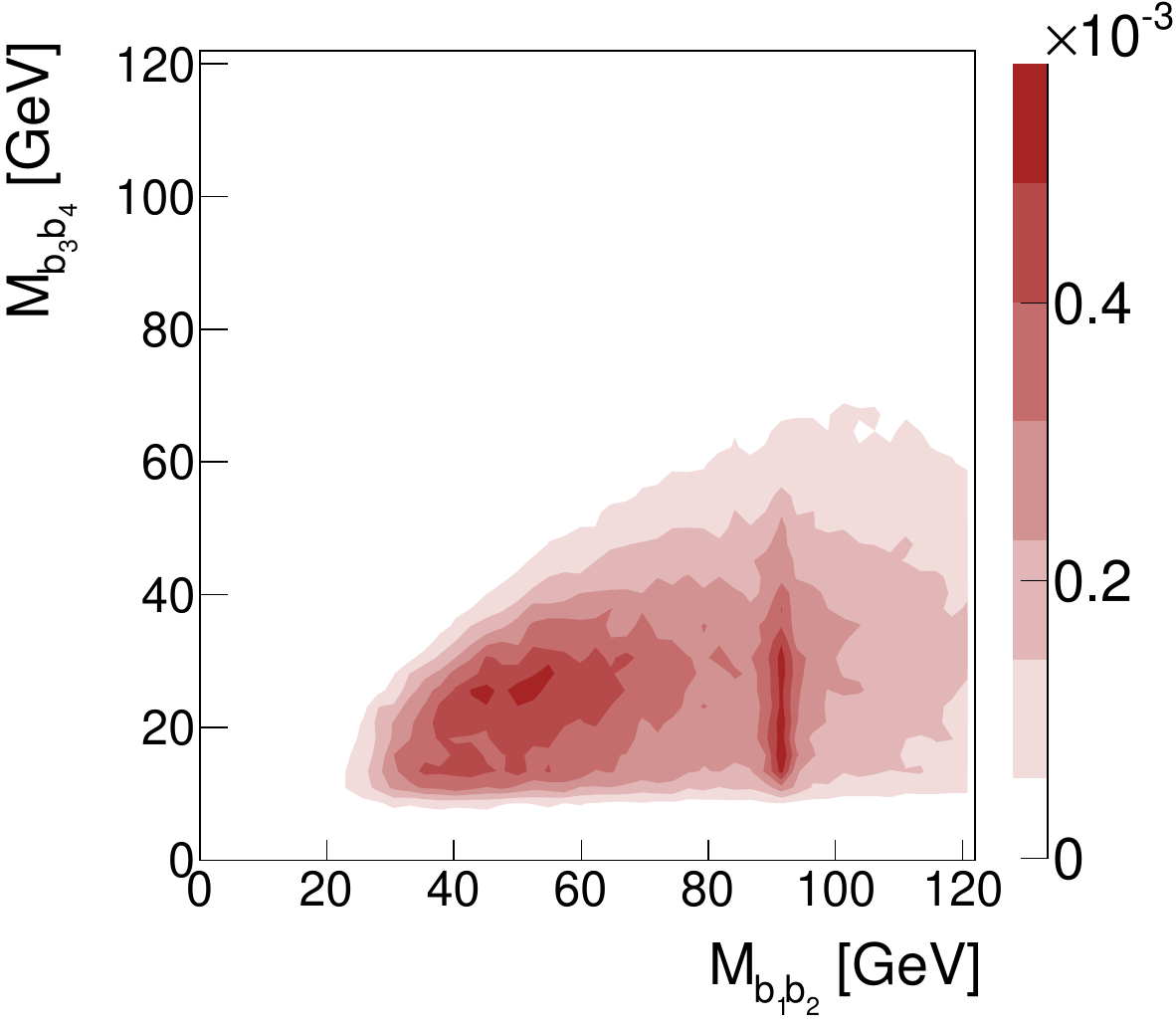}%
  \caption{\label{fig:dalitz_pre}
  Normalized double-differential distributions in the reconstructed Higgs rest frame:
  signal (left) and background (right).
  The signal-plus-background panels are not shown, since the background dominates the mixture after preselection and therefore obscures the signal-specific correlation patterns.
  {\bf Top:} $M_{bb}^{\max}$ versus $M_{bb}^{\min}$, where the maximum and minimum are taken over all possible two-$b$ combinations.
  {\bf Bottom:} $M_{12}$ versus $M_{34}$, where the $b$ quarks are ordered by their energies in the reconstructed Higgs rest frame.}
\end{figure}

The top row of Fig.~\ref{fig:dalitz_pre} shows the correlation between the largest and smallest two-$b$ invariant masses in each event.
For the signal, events are not distributed uniformly over the kinematic plane.
Instead, they occupy a characteristic region with relatively large $M_{bb}^{\max}$ and small-to-moderate $M_{bb}^{\min}$.
This reflects the asymmetric structure of the dominant $H\to b\bar b g$ contribution: one pair tends to reconstruct a hard subsystem associated with the primary $b\bar b$ configuration, while another pair can be formed from softer quarks associated with gluon splitting.
The background has a different pattern: it extends more broadly towards low values of both invariant masses and also shows a narrow structure associated with resonant $Z\to b\bar b$ production.
Therefore the two-dimensional plane separates not only by the absolute values of the invariant masses, but also by the way the event populates correlated hard-soft pair configurations.

The bottom row of Fig.~\ref{fig:dalitz_pre} uses the energy-ordered variables $M_{12}$ and $M_{34}$.
This representation is more directly connected to the decay topology.
For signal events, $M_{12}$, built from the two most energetic $b$ quarks in the reconstructed Higgs rest frame, is typically large, while $M_{34}$, built from the two least energetic $b$ quarks, is smaller.
Thus the signal forms a correlated band rather than an unstructured cloud.
This behaviour is expected when the two hardest $b$ quarks carry most of the Higgs energy and the softer pair often originates from the splitting of a radiated gluon.
For the background, the population is shifted towards lower values of both $M_{12}$ and $M_{34}$ and is more diffuse.
The comparison demonstrates that the variables are not independent: the discriminating power comes from their correlation.
A rectangular cut on either $M_{12}$ or $M_{34}$ alone would lose information that is retained in the two-dimensional distribution.

The reason for not showing the signal-plus-background panels is also visible from Table~\ref{tab:lhc-prod}.
Since the background after preselection is larger than the signal by roughly two orders of magnitude, the combined distribution is visually dominated by the background.
It therefore adds little information for understanding the signal topology, while reducing the space available for the direct signal-background comparison.

\begin{figure}[htb]
  \includegraphics[width=0.49\textwidth]{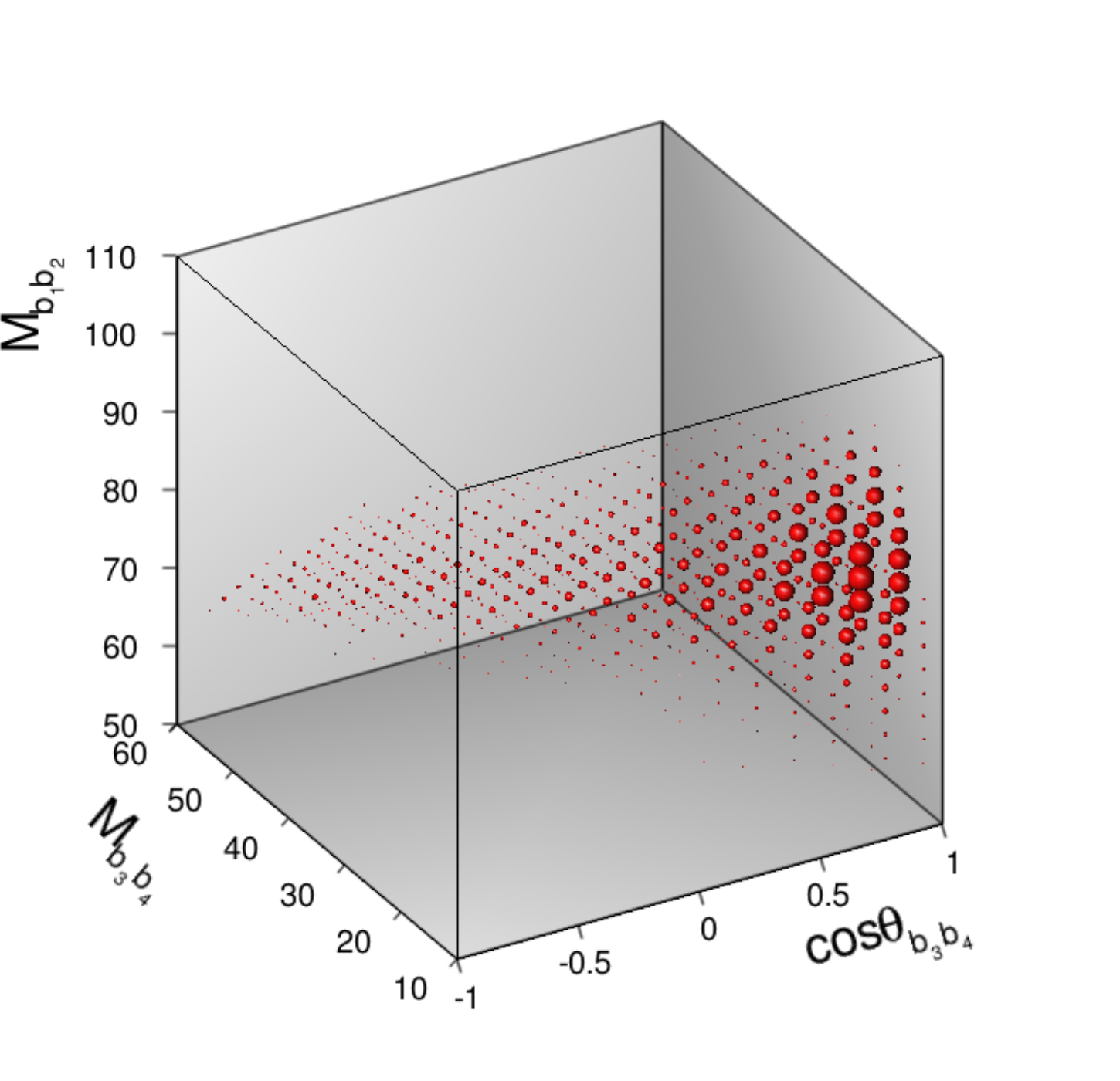}%
  \includegraphics[width=0.49\textwidth]{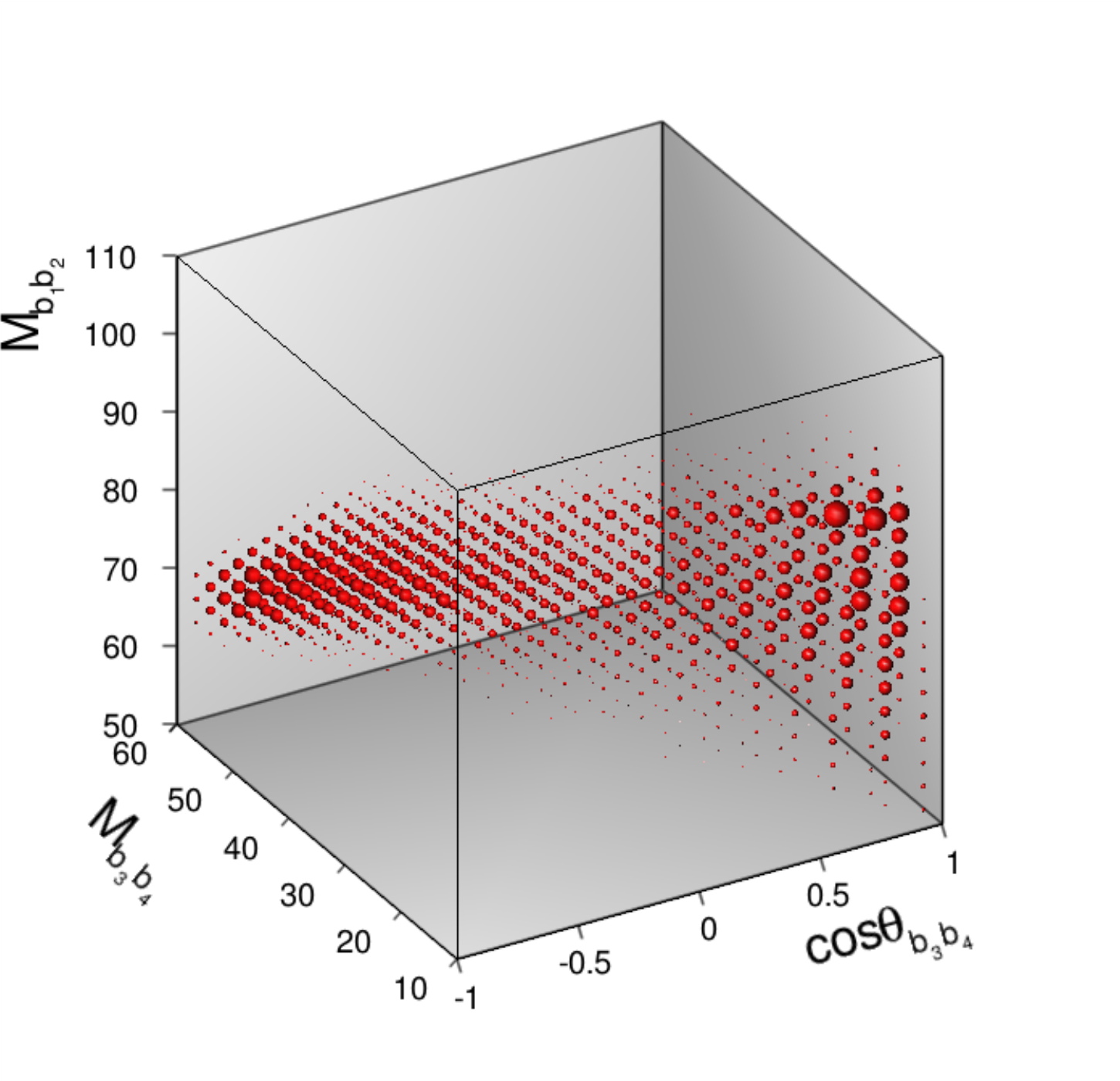}%
  \\
  \includegraphics[width=0.49\textwidth]{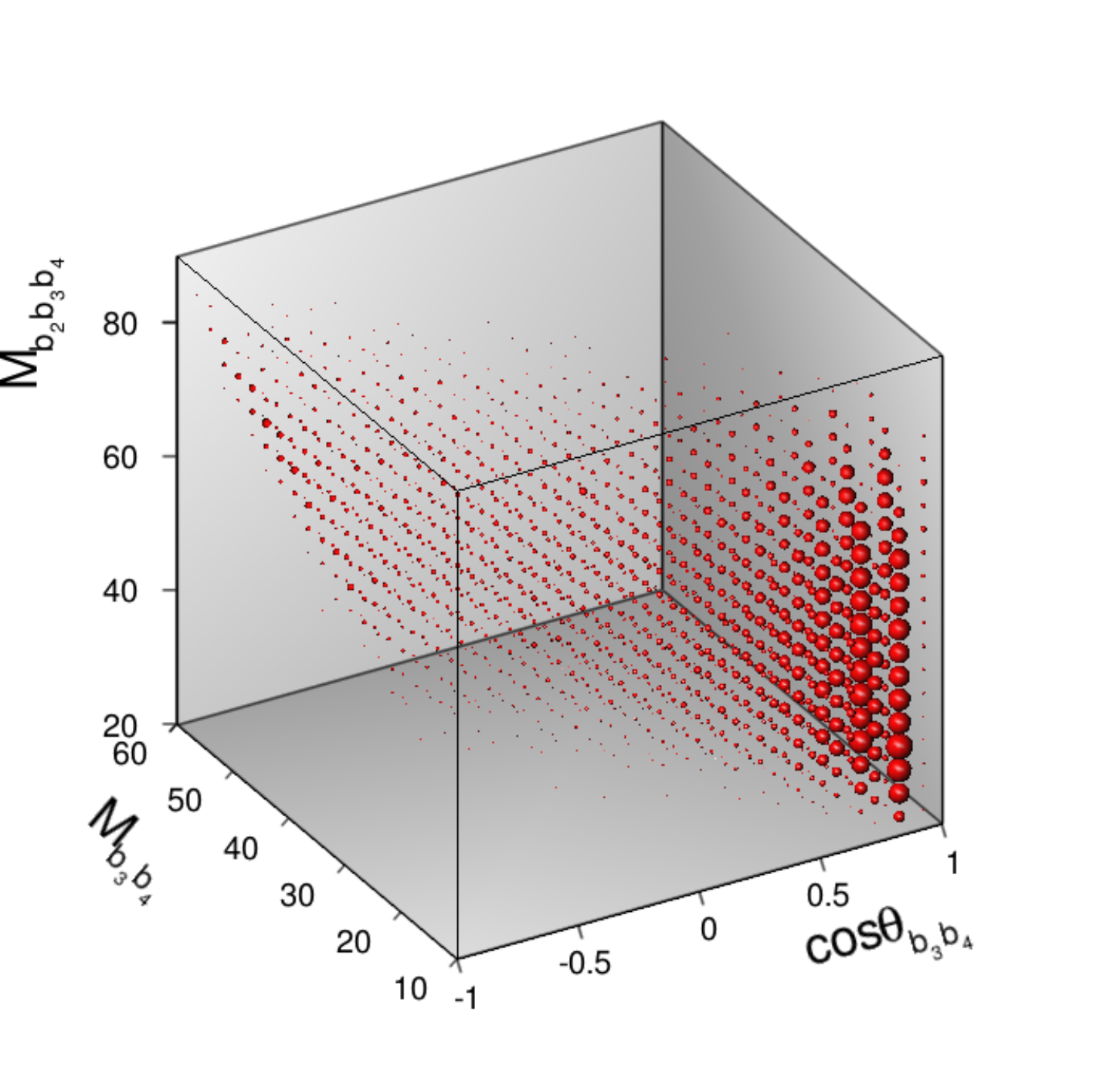}%
  \includegraphics[width=0.49\textwidth]{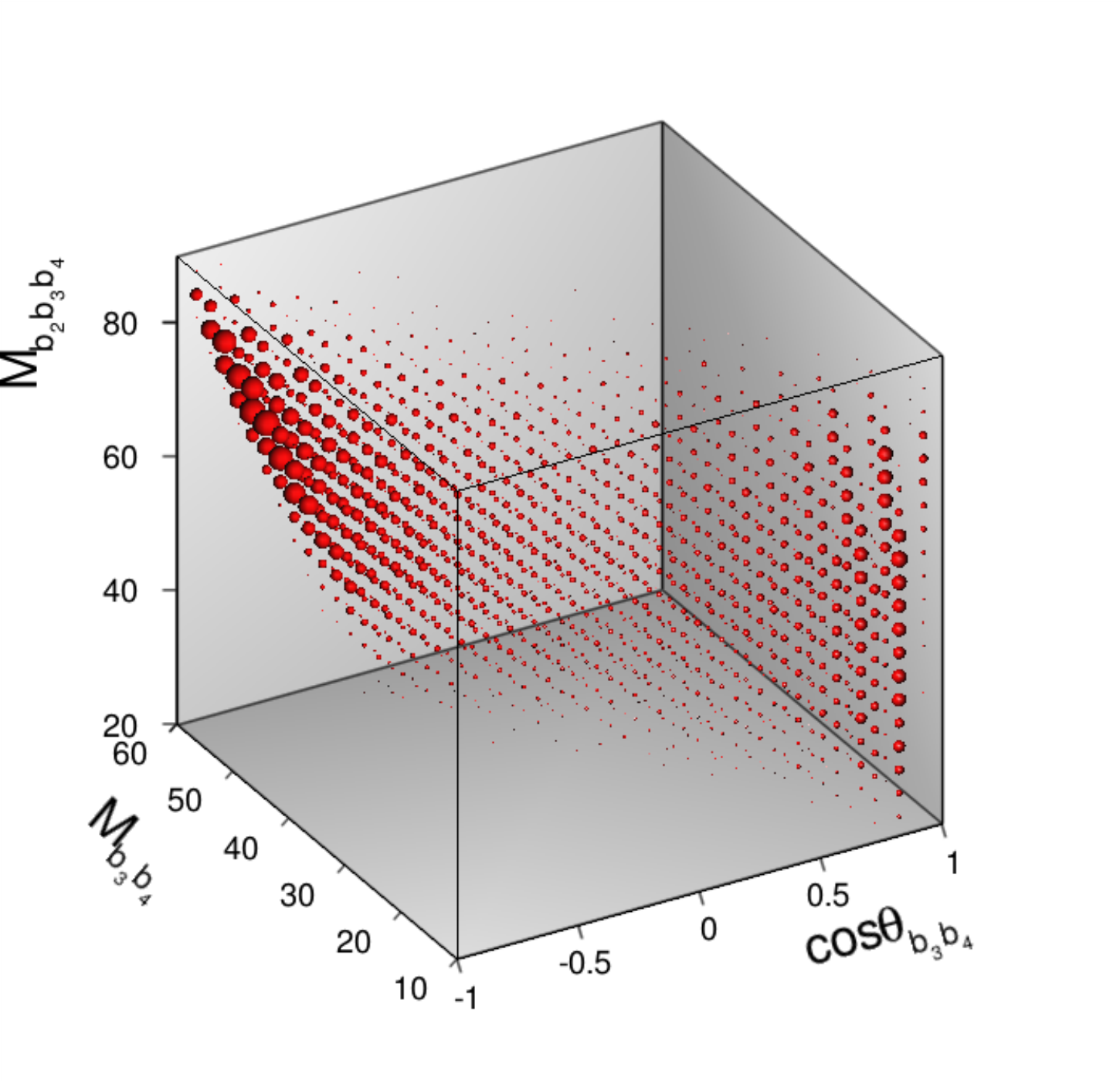}%
  \caption{\label{fig:dens3D}
  Probability-density maps:
  signal (left), background (right).
  {\bf Top:} $(\cos\theta_{34},M_{12},M_{34})$.
  {\bf Bottom:} $(\cos\theta_{34},M_{234},M_{34})$.}
\end{figure}

\paragraph{Three-dimensional distributions.}
The two-dimensional distributions already show that correlations are essential.
However, the signal topology also contains angular information, especially from the tendency of the softer $b$ quarks to arise from gluon splitting.
Therefore, in Fig.~\ref{fig:dens3D} we combine invariant-mass information with the angular variable $\cos\theta_{34}$.
The purpose of these three-dimensional density maps is not merely to visualise another set of variables, but to test whether the signal occupies a compact region in the joint space of masses and angles, while the background remains broadly distributed.

In the top row of Fig.~\ref{fig:dens3D}, the signal is concentrated in a region with large $M_{12}$, relatively small $M_{34}$, and enhanced $\cos\theta_{34}$.
This is precisely the pattern expected from the dominant $H\to b\bar b g$ topology: the two most energetic $b$ quarks form a hard subsystem, while the softer pair tends to be more collinear when it originates from gluon splitting.
The background does not exhibit the same compact correlation.
It is spread over a wider range of $M_{12}$ and $M_{34}$ and includes more events with negative or moderate values of $\cos\theta_{34}$.
Thus, the angular variable does not simply duplicate the information contained in the invariant masses; it separates configurations that can overlap in the two-dimensional mass plane.

The bottom row, based on $(\cos\theta_{34},M_{234},M_{34})$, provides a complementary correlation.
The variable $M_{234}$ probes the invariant mass of the subsystem made of the three less energetic $b$ quarks.
For signal events, this variable remains correlated with the soft-pair invariant mass and with the collinearity of the two least energetic $b$ quarks.
This produces a more localised high-density region.
The background is again broader, because continuum $W+4b$ production does not impose the same parent-resonance and gluon-splitting structure.
These three-dimensional distributions therefore make explicit why a multivariate classifier is appropriate: the separation is not controlled by a single observable, but by correlated patterns among energies, invariant masses and angular variables.

\subsection{Signal versus Background analysis at fast detector simulation level using multivariate analysis.}


To develop a realistic strategy for identifying signal events under LHC experimental conditions,
we use the complete simulation chain described at the beginning of Sec.~\ref{section:s-vs-b}.
Signal and background events, including the relevant interference effects at parton level,
are showered and hadronised using \textsc{PYTHIA}, and then passed through detector simulation,
jet reconstruction and $b$-tagging using \textsc{Delphes}.

We used the CMS detector card corresponding to the HL-LHC detector configuration. Jets were reconstructed using the anti-$k_T$ algorithm with radius parameter $R=0.5$, and the built-in Delphes $b$-tagging parametrisation was used to model realistic $b$-jet identification efficiencies and mistag rates.

At the event-generation stage, only minimal generator-level requirements were applied where needed to regulate soft and collinear regions of the LO matrix elements. The physics selection relevant for the detector-level analysis was imposed after showering, hadronisation and detector simulation, using reconstructed jets and Delphes $b$-tagging. In particular, the analysis requires at least four reconstructed $b$-tagged jets. This detector-level requirement introduces an effective acceptance selection beyond the loose parton-level preselection discussed in Sec.~\ref{section:s-vs-b}.

For the detector-level analysis we study distributions analogous to the parton-level observables shown in Fig.~\ref{fig:BDT-vars-part}, but now using reconstructed $b$-tagged jets rather than parton-level $b$ quarks. This comparison is important because variables that are powerful at parton level may be degraded by jet reconstruction, finite energy resolution, combinatorial effects and $b$-tagging. We therefore first examine how the most relevant one-dimensional observables behave after the full simulation chain before using them in the BDTG analysis.

We focus below on the full signal and full background samples, which are the physically relevant inputs for the detector-level BDTG analysis. All distributions and performance estimates shown in this subsection include the complete signal and background contributions, with interference effects included.

\begin{figure}[!htb]
	\includegraphics[width=0.5\textwidth,center]{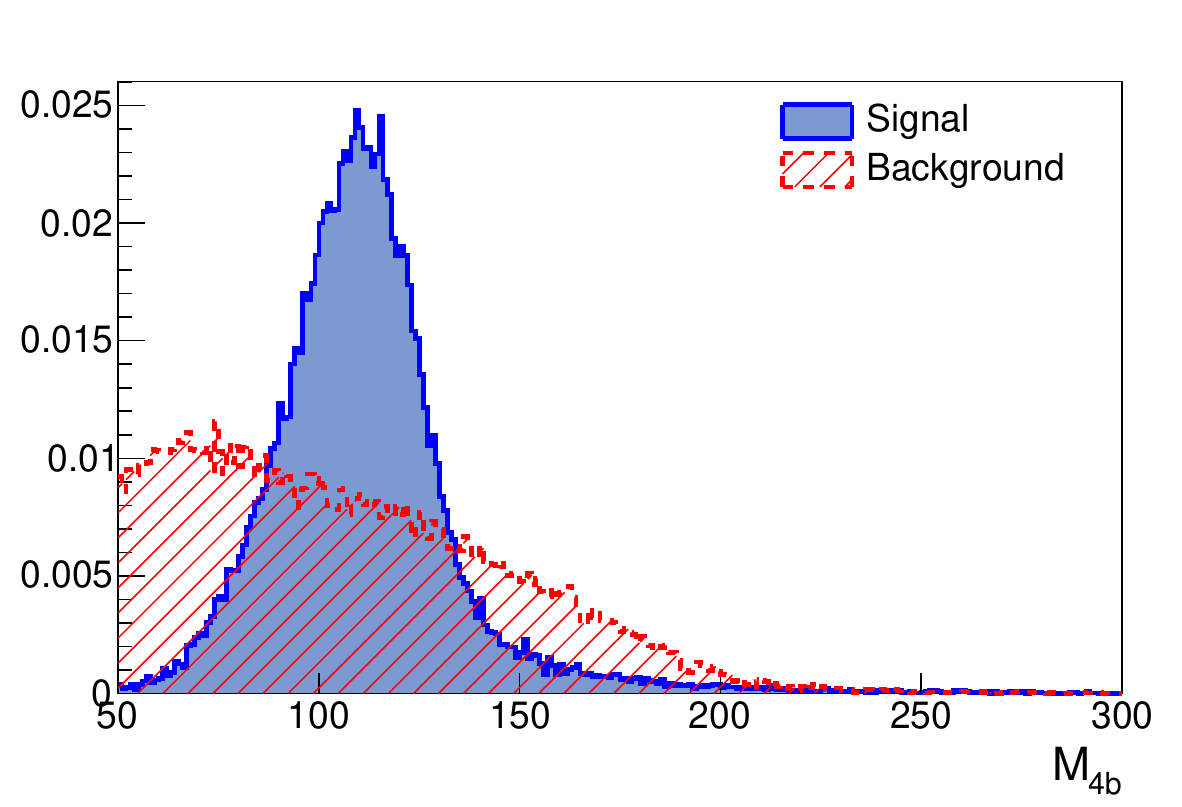}
	\caption{\label{fig:varlhcfull4b} Normalized distribution of the four-$b$-jet invariant mass for the full signal (blue) and full background (red) processes for the $pp \to Wb\bar{b}b\bar{b}$ final state at the LHC after the full simulation chain, including Delphes detector simulation and built-in $b$-tagging.}
\end{figure}

Figure~\ref{fig:varlhcfull4b} shows the detector-level distribution of the reconstructed four-$b$-jet invariant mass for the full signal and full background samples. Compared with the parton-level distribution in Fig.~\ref{fig:m4b}, the signal peak is substantially broadened. This is expected: the reconstructed $M_{4b}$ distribution is affected by jet-energy resolution, showering, hadronisation and the fact that reconstructed jets do not exactly reproduce the original parton-level four-momenta. As a result, the very narrow invariant-mass cut that would appear attractive at parton level would be inefficient at detector level: it would remove a sizeable fraction of the signal while the large continuum background would still remain. This is the first indication that a simple cut-based analysis is not optimal and that additional kinematic information must be exploited.

\begin{figure}[htbp]
	\begin{subfigure}[a]{0.46\textwidth}
		\includegraphics[width=\textwidth]{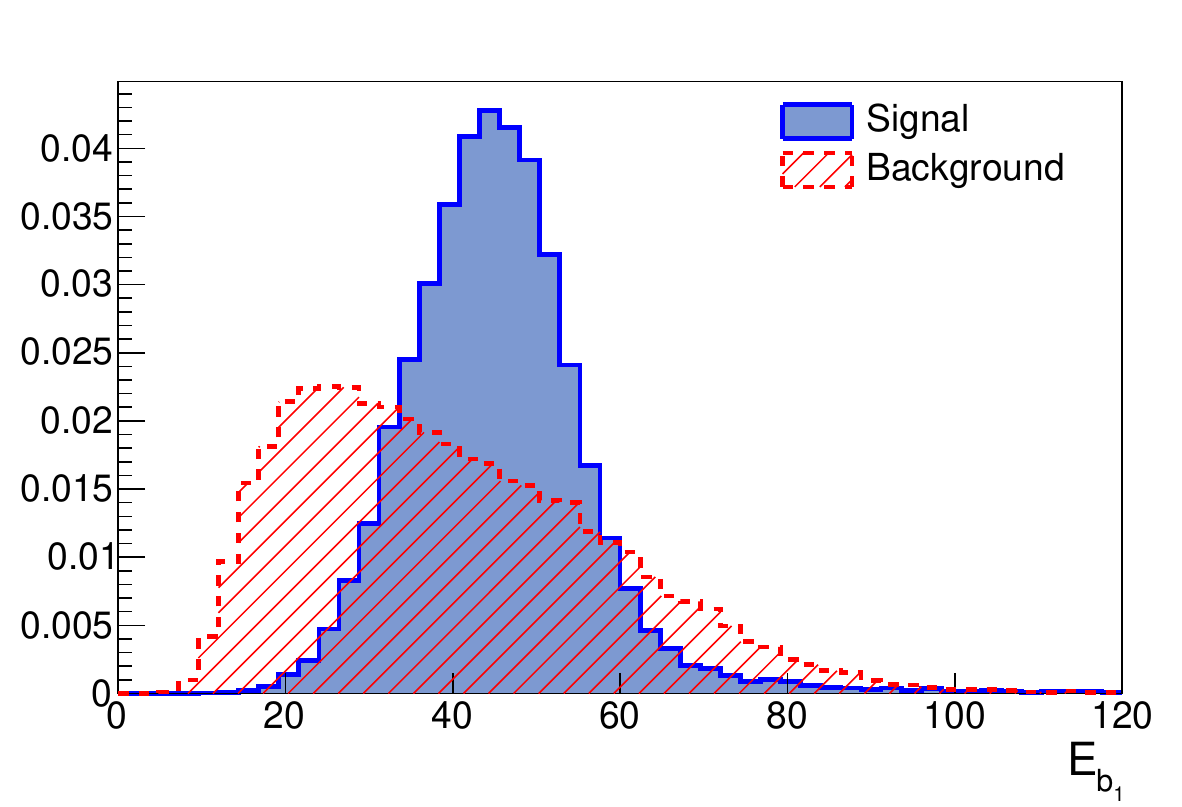}
		\caption{}
	\end{subfigure}
	\begin{subfigure}[a]{0.46\textwidth}
		\includegraphics[width=\textwidth]{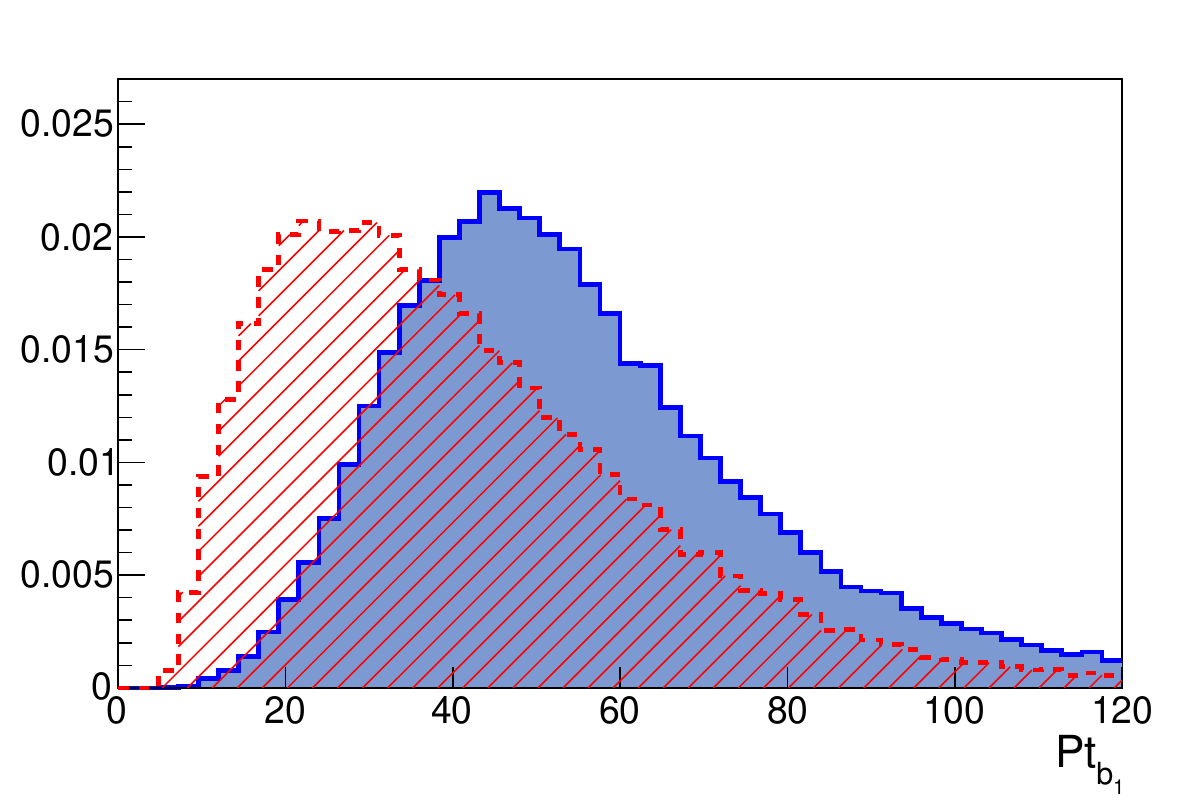}
		\caption{}
	\end{subfigure}
	\\
	\begin{subfigure}[c]{0.46\textwidth}
		\includegraphics[width=\textwidth]{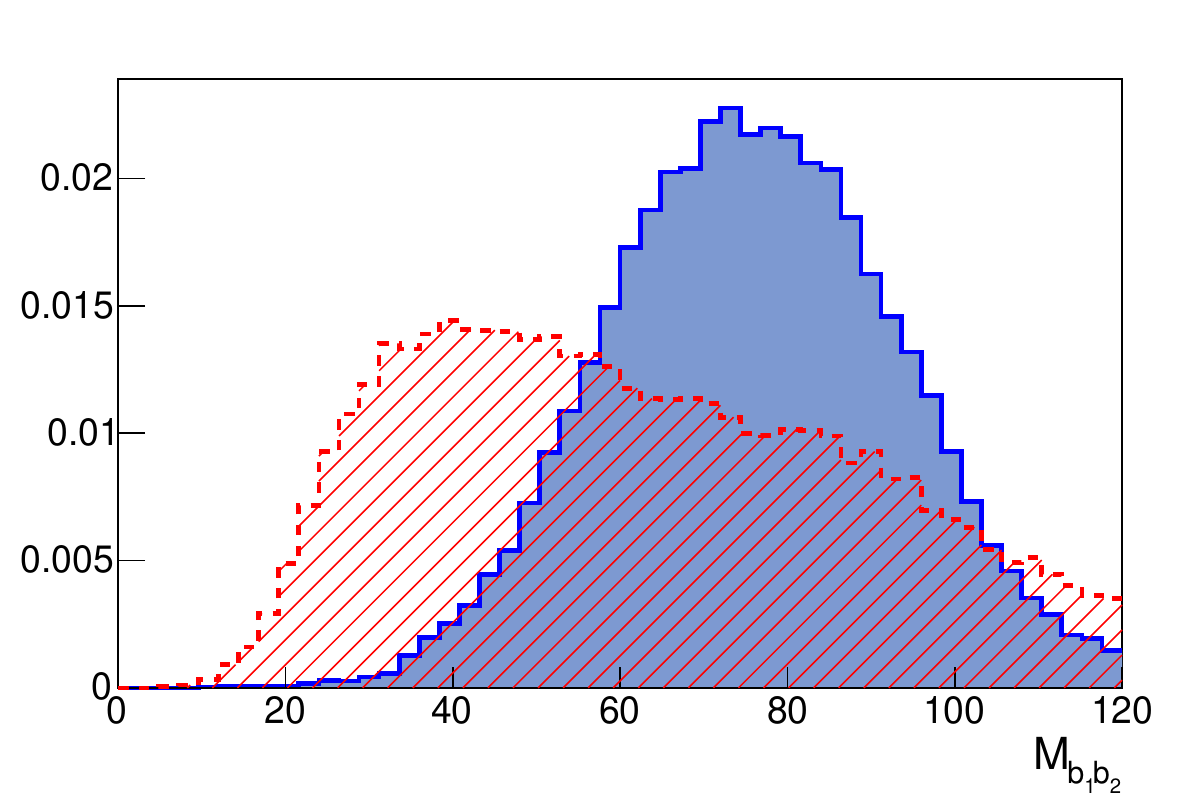}
		\caption{}
	\end{subfigure}
	\begin{subfigure}[d]{0.46\textwidth}
		\includegraphics[width=\textwidth]{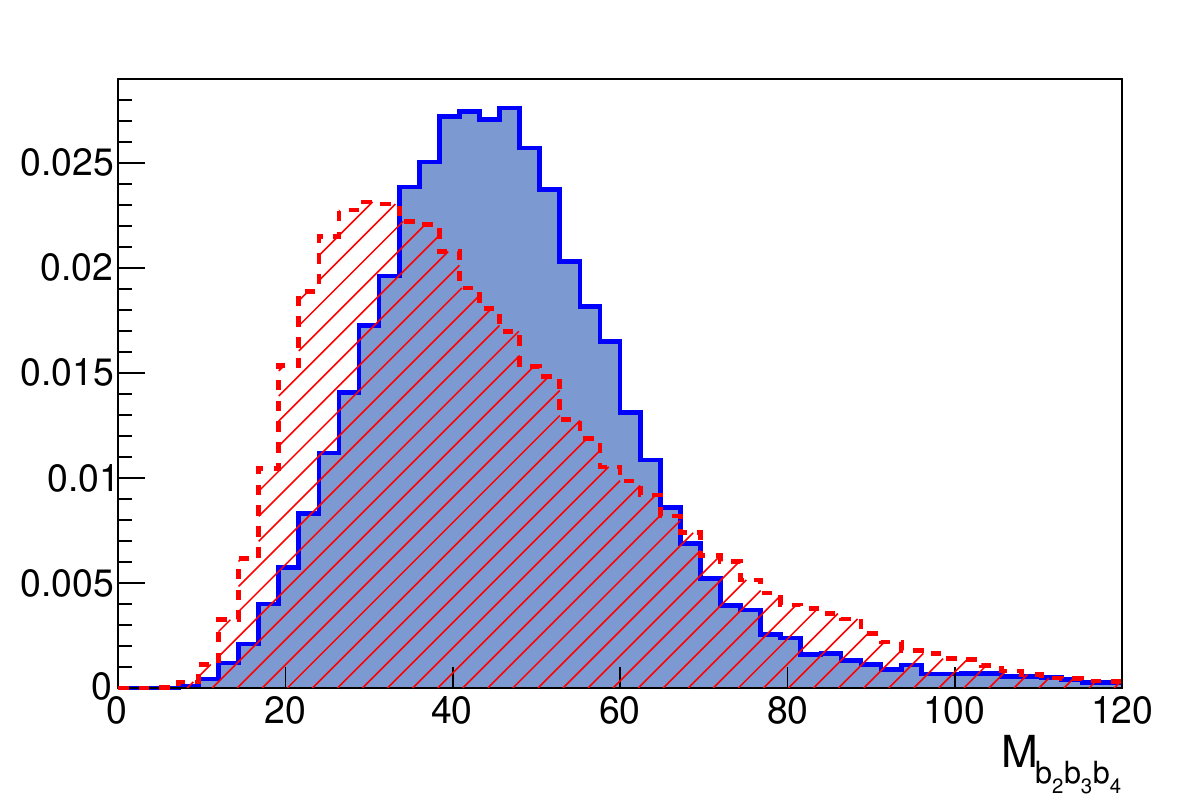}
		\caption{}
	\end{subfigure}
	\\
	\begin{subfigure}[c]{0.46\textwidth}
		\includegraphics[width=\textwidth]{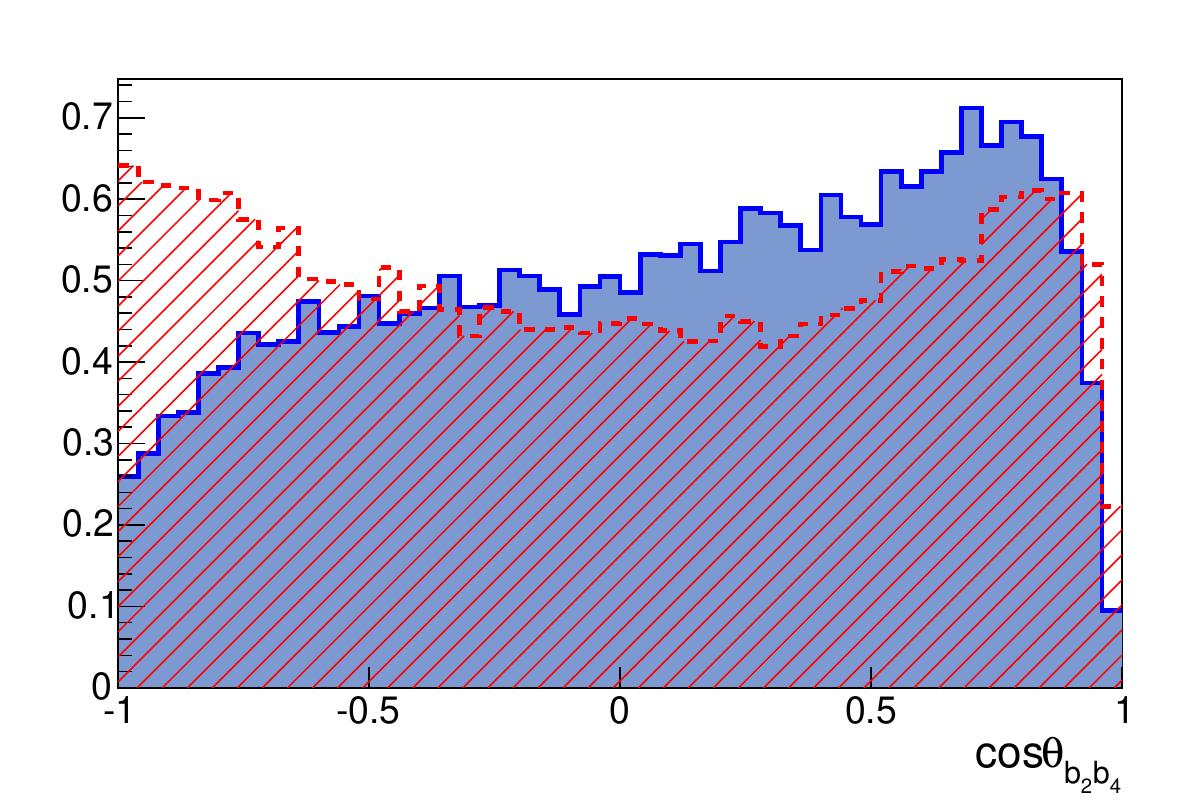}
		\caption{}
	\end{subfigure}
	\begin{subfigure}[d]{0.46\textwidth}
		\includegraphics[width=\textwidth]{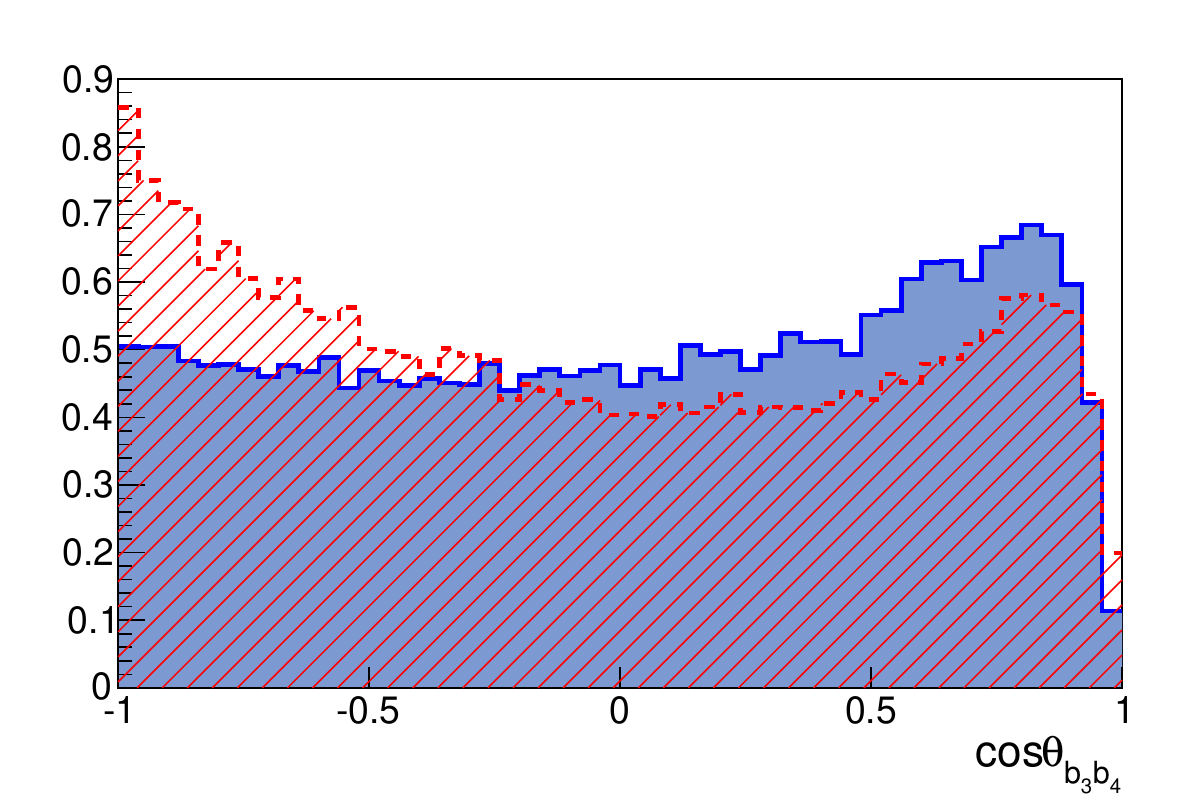}
		\caption{}
	\end{subfigure}
	\caption{\label{fig:varlhcfull} Normalized detector-level distributions of the selected kinematical variables for the full signal (blue) and full background (red) processes for the $pp \to Wb\bar{b}b\bar{b}$ final state at the LHC after the full simulation chain, including Delphes detector simulation and built-in $b$-tagging: {\bf (a)} $E_1$, the energy of the most energetic $b$ jet in the reconstructed Higgs rest frame; {\bf (b)} $p_{T,1}$, the transverse momentum of the leading $b$ jet in the lab frame; {\bf (c)} $M_{12}$; {\bf (d)} $M_{234}$; {\bf (e)} $\cos\theta_{24}$; {\bf (f)} $\cos\theta_{34}$.}
\end{figure}

Comparing the detector-level distributions in Fig.~\ref{fig:varlhcfull} with the corresponding parton-level distributions in Fig.~\ref{fig:BDT-vars-part} reveals which observables are robust against detector and reconstruction effects. The energy distribution of the leading $b$ object in the reconstructed four-$b$ rest frame, Fig.~\ref{fig:varlhcfull}(a), is visibly broadened compared with Fig.~\ref{fig:BDT-vars-part}(a). At parton level the signal has a sharper peak associated with the hard $b$ quark from the dominant $H\to b\bar b g$ topology, while after reconstruction this peak is smeared by jet-energy resolution and showering effects. Nevertheless, the signal remains harder than the background, so this variable retains discriminating power.

A similar but milder effect is seen for the transverse momentum of the leading reconstructed $b$ jet, Fig.~\ref{fig:varlhcfull}(b), compared with Fig.~\ref{fig:BDT-vars-part}(b). The signal distribution remains shifted towards larger $p_T$, whereas the background still populates the softer region more strongly. This confirms that the lab-frame hardness of the leading $b$ jet survives the detector simulation and remains useful for signal-background separation.

The invariant-mass variables also remain informative. The distribution of $M_{12}$ in Fig.~\ref{fig:varlhcfull}(c), compared with Fig.~\ref{fig:BDT-vars-part}(c), is broader at detector level, but the signal still favours larger values than the background. The variable $M_{234}$, shown in Fig.~\ref{fig:varlhcfull}(d), gives a somewhat weaker one-dimensional separation, but it carries complementary information because it probes the subsystem formed by the three less energetic reconstructed $b$ jets.

The angular observables are affected most strongly. In Figs.~\ref{fig:varlhcfull}(e) and \ref{fig:varlhcfull}(f), the sharp parton-level angular structures seen in Figs.~\ref{fig:BDT-vars-part}(e) and \ref{fig:BDT-vars-part}(f) are smoothed by reconstruction and combinatorial effects. This is especially visible for the signal, where the addition of the subleading $H\to ZZ^\ast\to 4b$ and $H\to gg\to 4b$ components also dilutes the angular pattern of the dominant $b\bar b g$ topology. Even so, the signal and background shapes remain different, particularly in the high-$\cos\theta$ region. Therefore the angular variables are still useful, but mainly in combination with invariant masses and energy variables rather than as standalone cuts.

Overall, the detector-level comparison shows that many of the parton-level observables retain useful discrimination after the full simulation. However, none of them alone provides sufficient separation. This motivates the use of a multivariate classifier that can exploit correlations among the reconstructed energies, transverse momenta, invariant masses and angular variables.


\begin{figure}[htb]
\includegraphics[width=0.45\textwidth]{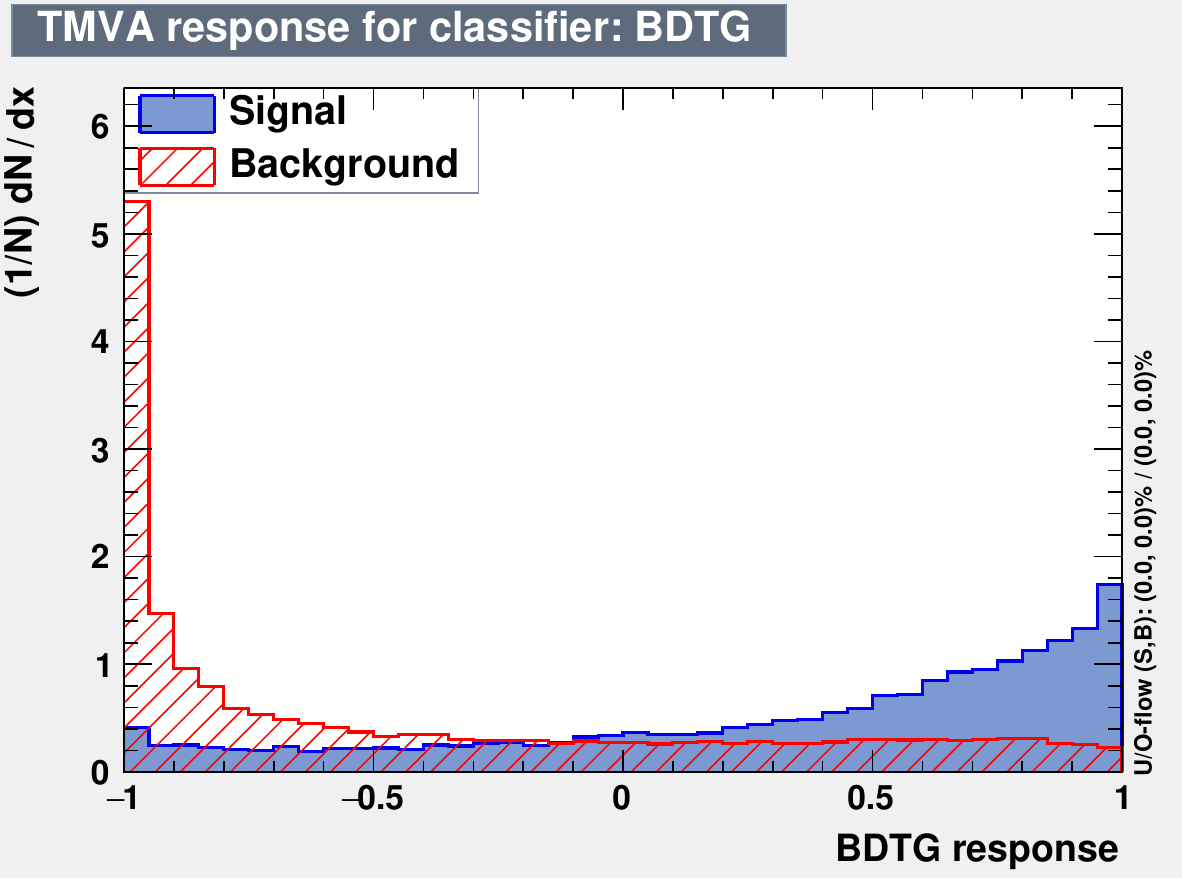}\ \
\includegraphics[width=0.45\textwidth]{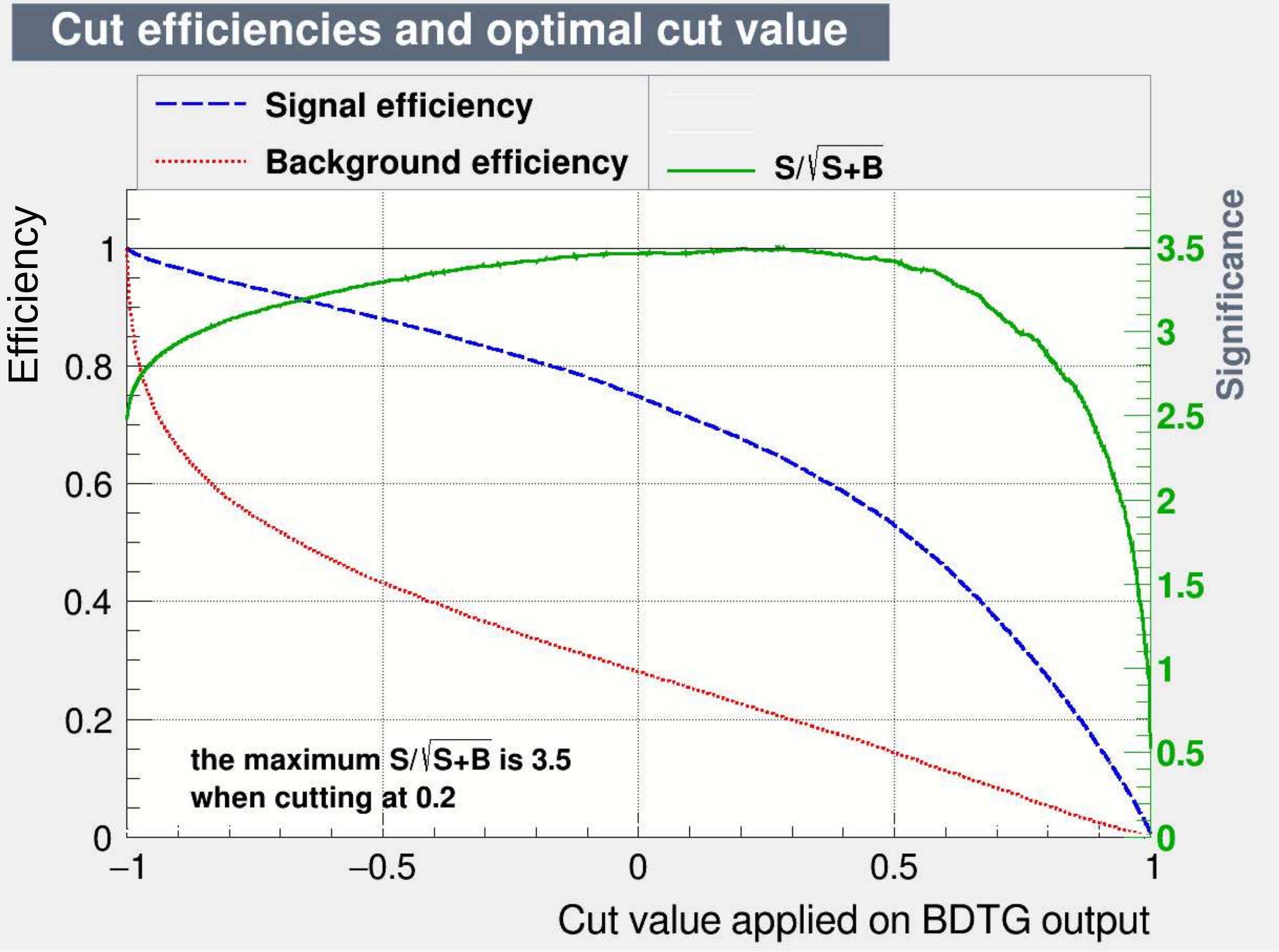}%
\caption{\label{fig:classifierlhc}
		 Left: BDTG classifier response for the $pp \to W H \to W b\bar{b}b\bar{b}$ signal and the corresponding background at the HL-LHC. Right: signal and background efficiencies, together with the statistical significance $S/\sqrt{S+B}$, as functions of the lower cut on the BDTG response, for an integrated luminosity of $3000~\text{fb}^{-1}$.}
\end{figure}


The Monte Carlo samples described above, together with the reconstructed kinematic variables discussed in the previous paragraphs, were used to construct a multidimensional classifier separating the signal from the background. We used TMVA~\cite{TMVA:2007ngy}, implemented in the ROOT framework version 6.24/06~\cite{Brun:1997pa}. Among the available TMVA methods, we chose the Gradient Boosted Decision Trees algorithm, denoted BDTG, because it efficiently exploits nonlinear correlations among weakly separating variables and is well suited for problems with large backgrounds.

Events are required to contain at least four $b$-tagged jets.
If more than four $b$-tagged jets are present, the four jets with the largest transverse momenta are selected.
We label these jets as $b_1,b_2,b_3,b_4$, ordered by decreasing transverse momentum in the lab frame, with $b_1$ denoting the hardest selected jet.

For the BDTG training we use the following $20$ input variables:
\begin{enumerate}
  \item the transverse momenta of the four selected $b$-tagged jets,
  $p^T_{b_1}$, $p^T_{b_2}$, $p^T_{b_3}$, and $p^T_{b_4}$;

  \item the energies of the four selected $b$-tagged jets in the reconstructed Higgs rest frame,
  defined as the centre-of-mass frame of the selected four-$b$ system.
  These energies are ordered as $E_1 > E_2 > E_3 > E_4$, where the ordering refers to the reconstructed Higgs rest frame:
  $E_1$, $E_2$, $E_3$, and $E_4$;

  \item the minimum and maximum invariant masses among all possible two-$b$ combinations,
  $M^{\min}_{b_i b_j}$ and $M^{\max}_{b_i b_j}$;

  \item the invariant mass of the two most energetic $b$ jets in the reconstructed Higgs rest frame,
  $M_{12}$;

  \item the invariant mass of the two least energetic $b$ jets in the reconstructed Higgs rest frame,
  $M_{34}$;

  \item the invariant mass of the three least energetic $b$ jets in the reconstructed Higgs rest frame,
  $M_{234}$;

  \item the invariant mass of the selected four-$b$ system,
  $M_{4b}$;

  \item the cosines of the angles between all pairs of the four selected $b$ jets in the reconstructed Higgs rest frame,
  $\cos\theta_{ij}$, giving six variables.
\end{enumerate}
Altogether, the BDTG is trained using $4+4+2+1+1+1+1+6=20$ input variables.

The left panel of Fig.~\ref{fig:classifierlhc} shows the BDTG response for the full signal and full background samples. The two distributions are clearly separated: signal events preferentially populate the high-BDTG region, while the background is concentrated at lower BDTG values. This confirms that the set of variables listed above contains enough correlated information to probe the rare $H\to 4b$ decay mode in associated $WH$ production.

The right panel of Fig.~\ref{fig:classifierlhc} shows the signal efficiency, background efficiency and statistical significance as functions of the lower cut on the BDTG response. We define the statistical significance as
\begin{equation}
    \label{eq:lhc-significance}
    \alpha = \frac{S}{\sqrt{S+B}} \, ,
\end{equation}
where $S$ and $B$ are the expected numbers of signal and background events after the cut. This Gaussian approximation is justified here because the event samples after the relevant cuts still contain large statistics: even after the tighter high-purity cut discussed below, the expected yields are $S=198$ and $B=3786$ at $3000~\text{fb}^{-1}$.

Besides the statistical significance, the signal-to-background ratio, $S/B$, is an important quantity because it controls the robustness of the analysis against background normalisation and shape uncertainties.
A cut that maximises $S/\sqrt{S+B}$ is not necessarily the best working point once such uncertainties are considered.
For this reason, we quote both the statistically optimal working point and a tighter, higher-purity working point.

The expected event yields at $3000~\text{fb}^{-1}$ are summarised in Table~\ref{tab:lhc-bdt-yields}. The first row gives the number of events after the detector simulation and the requirement of at least four reconstructed $b$-tagged jets, before applying any BDTG cut. The second row corresponds to the BDTG cut that maximises the statistical significance, ${\rm BDTG}>0.2041$. This gives $S=406$, $B=13076$, and $\alpha=3.49$, with $S/B\simeq 3.1\%$. The third row shows a tighter, higher-purity working point, ${\rm BDTG}>0.76$, for which the significance remains close to $3\sigma$, while the purity improves to $S/B\simeq 5.2\%$.

\begin{table}[htbp]
\centering
\begin{tabular}{|l|c|c|c|c|c|c|c|}
\hline\hline
{ Working point} &
{ BDTG cut} &
{ $S$} &
{ $B$} &
{ $S/B$} &
{ $S/\sqrt{S+B}$} &
{ Signal eff.} &
{ Background eff.} \\
\hline\hline
{ Before BDTG cut} &
{ --} &
{ 600} &
{ 58254} &
{ $1.03\%$} &
{ $2.47$} &
{ 1.000} &
{ 1.000} \\
\hline
{ Maximum statistical significance} &
{ $0.2041$} &
{ 406} &
{ 13076} &
{ $3.10\%$} &
{ $3.49$} &
{ 0.6759} &
{ 0.2245} \\
\hline
{ Higher-purity working point} &
{ $0.76$} &
{ 198} &
{ 3786} &
{ $5.23\%$} &
{ $3.14$} &
{ 0.330} &
{ 0.0650} \\
\hline\hline
\end{tabular}
\caption{\label{tab:lhc-bdt-yields}
Expected numbers of signal and background events for the LHC analysis at $3000~\text{fb}^{-1}$ after Delphes detector simulation and the requirement of at least four reconstructed $b$-tagged jets. The table shows the yields before the BDTG cut, after the cut that maximises the statistical significance, and after a tighter high-purity cut. The signal and background efficiencies are given relative to the event numbers before the BDTG cut.}
\end{table}

The maximum of the statistical significance occurs near ${\rm BDTG}=0.2041$, where $\alpha=3.49$ for $3000~\text{fb}^{-1}$.
This working point is optimal in the purely statistical sense, but the resulting purity is only about $3\%$.
A tighter cut, ${\rm BDTG}>0.76$, reduces the signal efficiency from $0.676$ to $0.330$, but suppresses the background more strongly, reducing the background efficiency from $0.225$ to $0.065$.
As a result, the signal-to-background ratio improves from about $3.1\%$ to about $5.2\%$, while the statistical significance remains close to $3\sigma$.
This higher-purity point is therefore a useful alternative working point, especially for an analysis where background modelling uncertainties are important.

The HL-LHC programme may ultimately deliver integrated luminosities beyond the nominal $3000~\text{fb}^{-1}$ target, with scenarios reaching about $4.5~\text{ab}^{-1}$ per experiment discussed in Ref.~\cite{Zerlauth:2025HLLHC}. If both ATLAS and CMS accumulated and combined approximately $4.5~\text{ab}^{-1}$ each, the effective luminosity would be about $9~\text{ab}^{-1}$. Scaling the high-purity working point from $3000~\text{fb}^{-1}$ to $9~\text{ab}^{-1}$ gives
\begin{equation}
    \alpha(9~\text{ab}^{-1}) \simeq 3\,\sqrt{\frac{9}{3}} \simeq 5.2 \, .
\end{equation}
This suggests that, even using the more conservative high-purity BDTG working point with $S/B\simeq 5\%$, a combined HL-LHC analysis could potentially reach the level required for discovery, of the $H\to 4b$ decay mode. This extrapolation is necessarily indicative, since a full experimental combination would require a detailed treatment of correlated and uncorrelated systematic uncertainties.

Although the LHC analysis therefore has realistic discovery potential, the remaining background is still large. Even at the high-purity working point, thousands of background events survive the BDTG cut. This limits the ability of the LHC to perform precision measurements of the $H\to4b$ branching ratio or to disentangle accurately the individual components of the decay, such as the dominant $b\bar b g$ contribution and the $ZZ^\ast$ contribution. For such precision studies, a cleaner experimental environment is needed. As shown in the next section, a future linear collider such as the ILC, where the dominant QCD background is strongly reduced, can provide a substantially better setting for measuring the properties of the $H\to4b$ decay.

\section{$4b$ signature from the Higgs boson at the ILC}
\label{section:ilc}

\subsection{Analysis setup}
\label{sec:ilc-setup}

We now turn to the $H\to b\bar b b\bar b$ signature at a future $e^+e^-$ collider.
Compared with the LHC, the ILC offers much cleaner experimental conditions: the initial state is known, the QCD multijet background is strongly reduced, and the process $e^+e^-\to ZH$ allows one to study Higgs decays in a controlled environment.
This makes the ILC well suited for measuring the $H\to4b$ branching ratio and for studying the internal structure of this rare decay.

The analysis is performed for the ILC running at $\sqrt{s}=250$~GeV, following the collider setup of the ILC Technical Design Report~\cite{Behnke:2013xla}.
At this energy, the Higgs-strahlung process $e^+e^-\to ZH$ is close to its maximal production rate.
We use the following reference parameters: instantaneous luminosity $10^{34}~{\rm cm}^{-2}{\rm s}^{-1}$, corresponding to an annual integrated luminosity of about $100~{\rm fb}^{-1}$, beam population $2\times10^{10}$, bunch length $0.3$~mm, and transverse beam-size scale $\sigma_x+\sigma_y\simeq700$~nm.

For inclusive Higgs-strahlung production at $\sqrt{s}=250$~GeV, the cross section is
$\sigma(e^+e^-\to ZH)=253.1$~fb when ISR and beamstrahlung effects are not included.
The maximum of the corresponding no-ISR curve is $\sigma(e^+e^-\to ZH)=254.8$~fb at $\sqrt{s}=244.7$~GeV.
Including ISR and beamstrahlung effects reduces the cross section to
\begin{equation}
\sigma(e^+e^-\to ZH)_{\rm ISR+BS}=222.7~{\rm fb}.
\end{equation}
This value is used as the normalisation for the ILC signal event sample.

The signal cross section is controlled by the branching ratio of the rare Higgs decay $H\to4b$.
Using ${\rm Br}(H\to4b)=1.66\times10^{-3}$, we obtain
\begin{equation}
\sigma(e^+e^-\to ZH)\,{\rm Br}(H\to4b)
=
222.7~{\rm fb}\times1.66\times10^{-3}
\simeq0.370~{\rm fb},
\end{equation}

once ISR and beamstrahlung effects are included.
Thus, the apparent difference between the no-ISR estimate based on $253.1$~fb and the value used for the full process is not an additional physical suppression of the decay, but is mainly due to the use of different production normalisations.

\begin{table}[htbp]
\centering
\begin{tabular}{|l|c|}
\hline\hline
Quantity & Value \\
\hline\hline
$\sqrt{s}$ & $250$~GeV \\
$\sigma(e^+e^-\to ZH)$, no ISR/beamstrahlung & $253.1$~fb \\
$\sigma(e^+e^-\to ZH)$, ISR+beamstrahlung & $222.7$~fb \\
${\rm Br}(H\to4b)$ & $1.66\times10^{-3}$ \\
$\sigma(e^+e^-\to ZH\to Zb\bar b b\bar b)$ & $0.370$~fb \\
\hline\hline
\end{tabular}
\caption{\label{tab:ilc-signal}
Signal normalisation used for the ILC analysis. The $e^+e^-\to ZH\to Zb\bar b b\bar b$ cross section is consistent with the ISR+beamstrahlung Higgs-strahlung rate multiplied by the calculated $H\to4b$ branching ratio.}
\end{table}

The dominant background is the continuum process
\begin{equation}
e^+e^-\to Zb\bar b b\bar b,
\end{equation}
including electroweak and mixed electroweak-QCD contributions.
For the broad reconstructed four-$b$ mass window used in the analysis,
\begin{equation}
\Delta M_{4b}\equiv |M_H-M_{4b}|\leq75~{\rm GeV},
\end{equation}
the background cross section is
\begin{equation}
\sigma_B(e^+e^-\to Zb\bar b b\bar b)=2.912\times10^{-1}~{\rm fb}.
\end{equation}
For comparison, in a very narrow mass window, $\Delta M_{4b}\leq5$~GeV, the purely electroweak contribution is only $5.54\times10^{-4}$~fb.
However, such a narrow cut is not used as the main analysis strategy, because detector smearing and jet reconstruction effects broaden the reconstructed Higgs mass distribution.
The realistic analysis therefore relies on a multivariate treatment of the full kinematic information rather than on a narrow $M_{4b}$ cut alone.

The detector-level simulation follows the same general workflow as in the LHC analysis: parton-level events are showered and hadronised, and then passed through fast detector simulation, jet reconstruction and $b$-tagging with \textsc{Delphes}. For the ILC study we use a dedicated ILC detector card, so that the detector response, jet reconstruction and flavour-tagging parametrisations correspond to an $e^+e^-$ collider setup.
The jet radius used in the jet reconstruction affects the event acceptance.
To quantify this dependence, we evaluated the fraction of simulated events passing the Delphes-level reconstruction for several jet-radius choices.
The corresponding efficiencies are shown in Table~\ref{tab:ilc-delphes-eff}.

\begin{table}[htbp]
\centering
\begin{tabular}{|c|c|c|c|c|}
\hline\hline
Jet radius &
Signal efficiency &
Background efficiency &
$S$ at $300~{\rm fb}^{-1}$ &
$B$ at $300~{\rm fb}^{-1}$ \\
\hline\hline
$R=0.4$ &
$42263/79191=0.534$ &
$38182/91055=0.419$ &
$59.3$ &
$36.6$ \\
\hline
$R=0.5$ &
$24608/79191=0.311$ &
$14731/91055=0.162$ &
$34.5$ &
$14.1$ \\
\hline
$R=1.0$ &
$10920/79191=0.138$ &
$4979/91055=0.0547$ &
$15.3$ &
$4.8$ \\
\hline\hline
\end{tabular}
\caption{\label{tab:ilc-delphes-eff}
Delphes-level reconstruction efficiencies for the ILC signal and background samples for different jet-radius choices. The event yields are obtained using $\sigma_S=0.370$~fb and $\sigma_B=0.291$~fb and are quoted for $300~{\rm fb}^{-1}$.}
\end{table}

Together with the reconstruction efficiencies, Table~\ref{tab:ilc-delphes-eff} gives the expected signal and background event numbers for $300~{\rm fb}^{-1}$, corresponding approximately to three years of ILC operation at the nominal annual integrated luminosity of $100~{\rm fb}^{-1}$.
Table~\ref{tab:ilc-delphes-eff} shows that the acceptance depends noticeably on the jet radius.
This is expected for a four-$b$ final state, because the softer $b$ jets can be close in angle, especially when they originate from gluon splitting.
A smaller jet radius preserves more resolved $b$-jet candidates, while a larger radius increases the probability that nearby hadronic activity is merged into a single reconstructed jet.
This effect is much less problematic at the ILC than at the LHC, but it still has to be accounted for when quoting detector-level efficiencies.

In the numerical analysis below we use $R=0.5$ as the reference jet-radius choice. This value provides a reasonable compromise between resolving nearby $b$ jets in the four-$b$ final state and avoiding an overly aggressive separation of hadronic activity into very narrow jets.


Before performing the signal-background separation, we first examine the main kinematic observables at parton level.
We use the same notation as in Sec.~\ref{section:s-vs-b}: variables labelled by $1,2,3,4$ in the reconstructed Higgs rest frame follow the energy ordering $E_1>E_2>E_3>E_4$, while lab-frame transverse-momentum variables use the corresponding $p_T$ ordering.

\begin{figure}[htbp]
\begin{subfigure}[a]{0.49\textwidth}
\includegraphics[width=\textwidth]{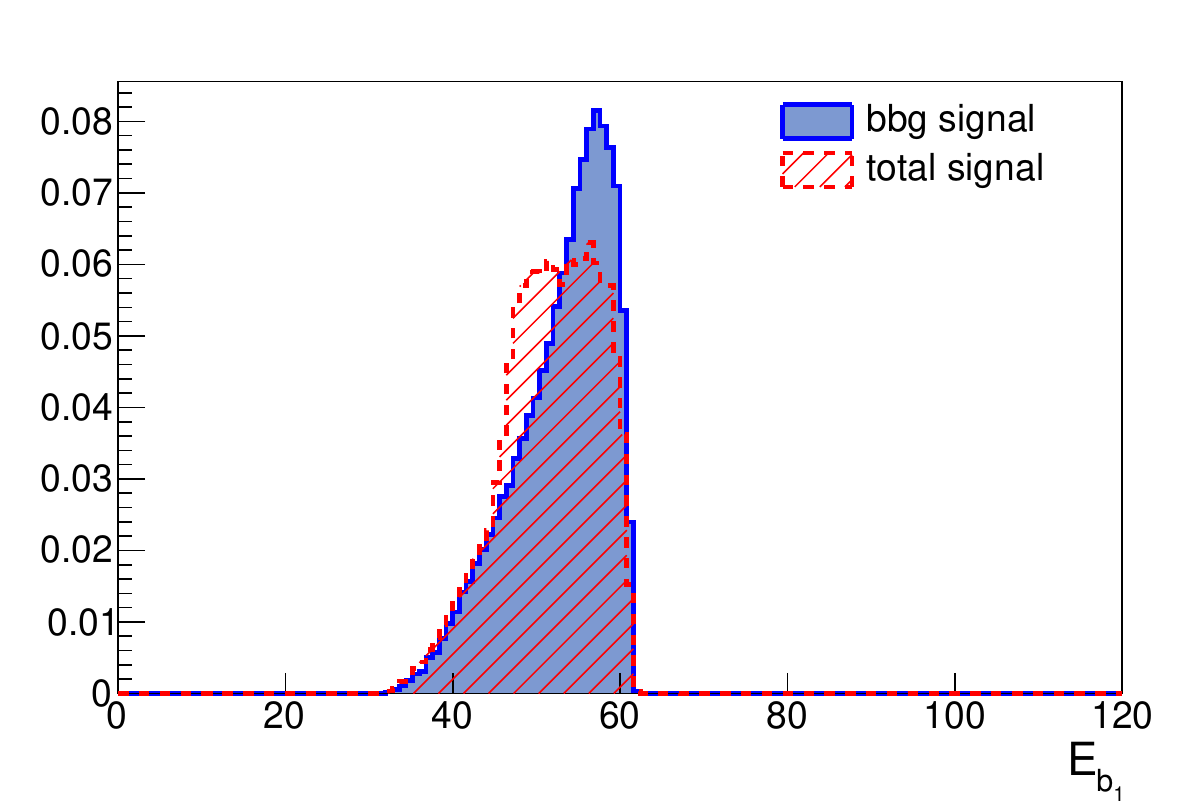}
\caption{}
\end{subfigure}
\begin{subfigure}[a]{0.49\textwidth}
\includegraphics[width=\textwidth]{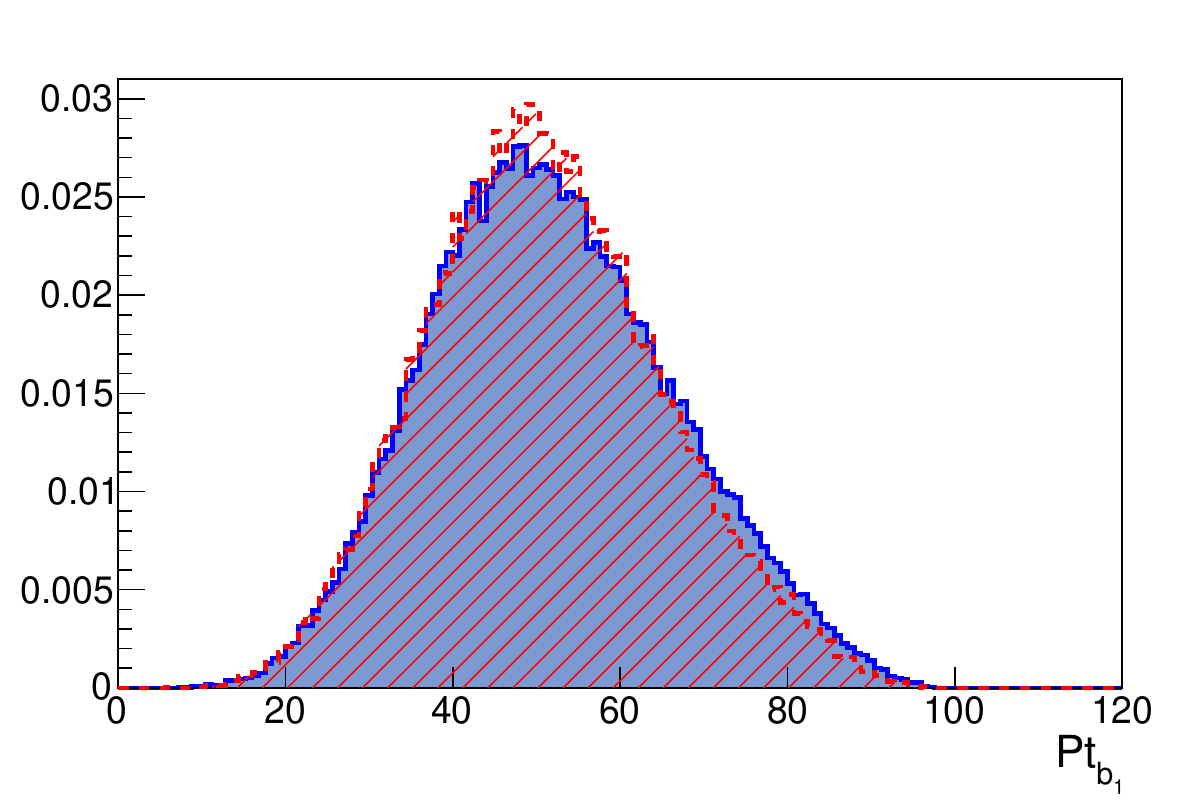}
\caption{}
\end{subfigure}
\\
\begin{subfigure}[c]{0.49\textwidth}
\includegraphics[width=\textwidth]{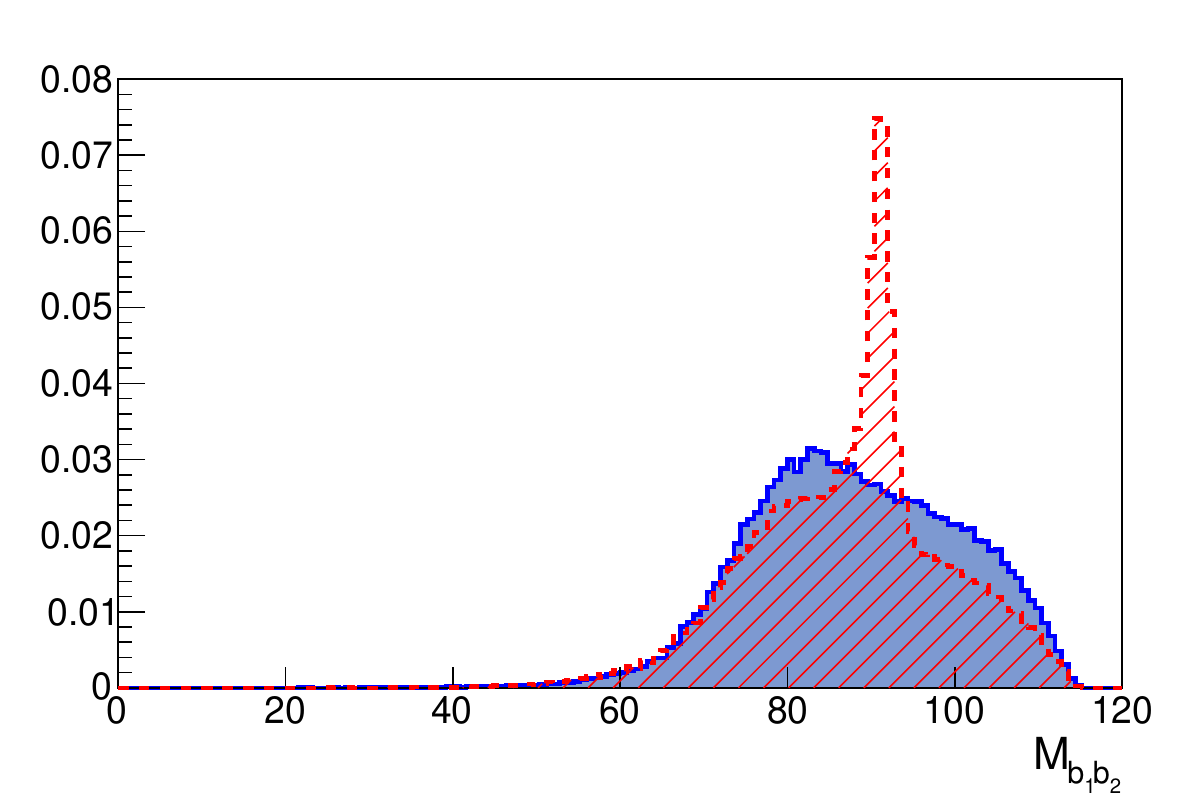}
\caption{}
\end{subfigure}
\begin{subfigure}[d]{0.49\textwidth}
\includegraphics[width=\textwidth]{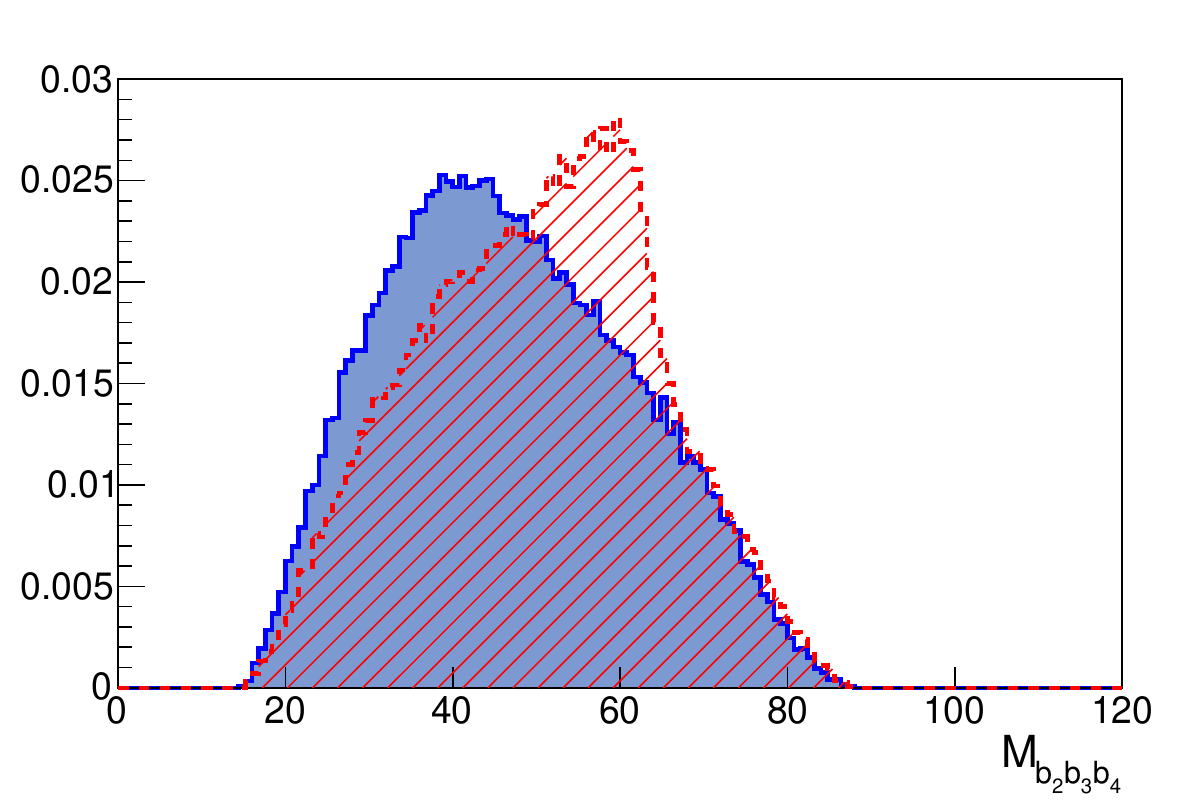}
\caption{}
\end{subfigure}
\\
\begin{subfigure}[c]{0.49\textwidth}
\includegraphics[width=\textwidth]{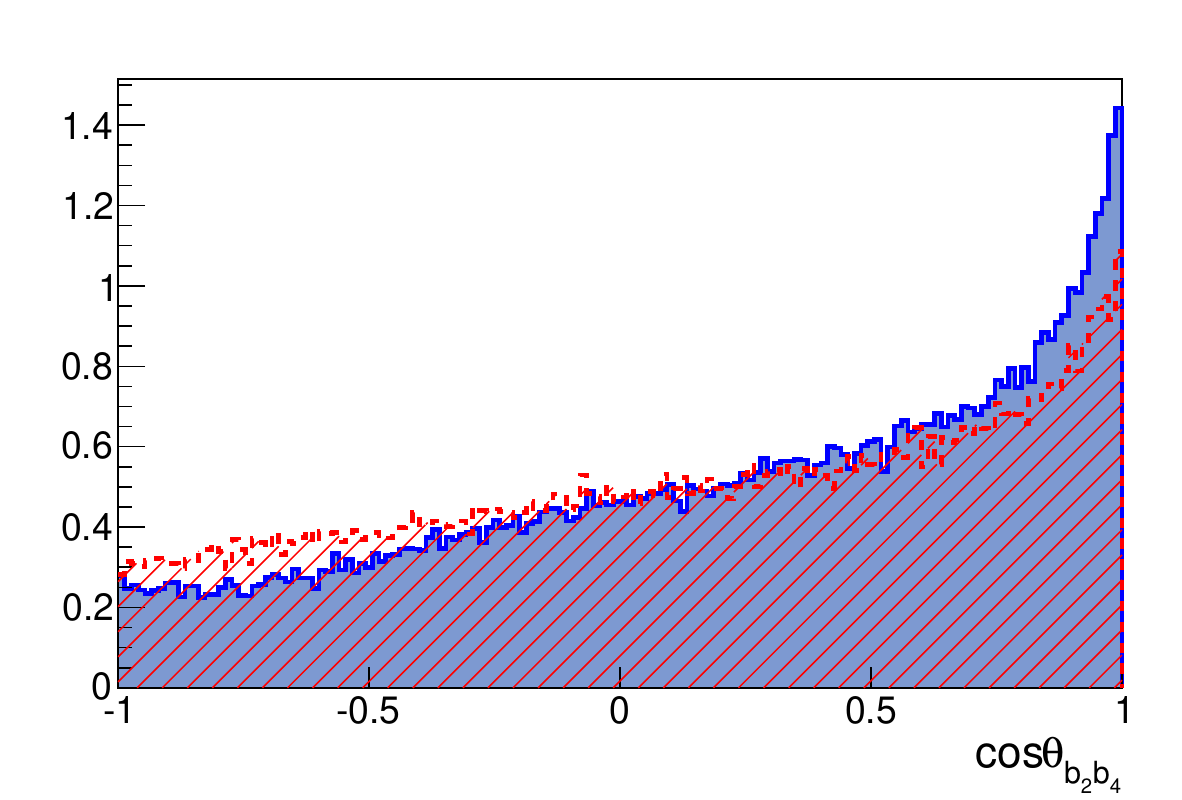}
\caption{}
\end{subfigure}
\begin{subfigure}[d]{0.49\textwidth}
\includegraphics[width=\textwidth]{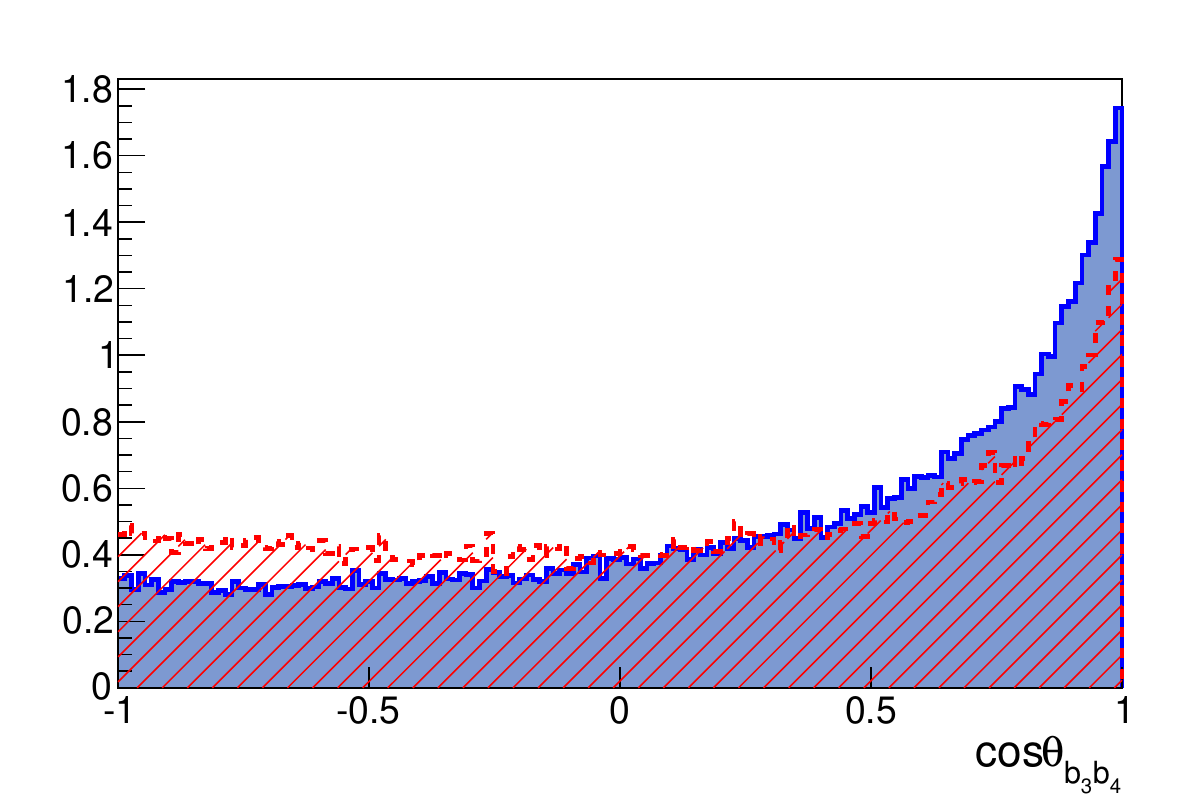}
\caption{}
\end{subfigure}
\caption{\label{fig:ILC_vars1}
Normalized parton-level distributions for the dominant $H\to b\bar b g\to b\bar b b\bar b$ signal component (blue solid) and the full $H\to4b$ signal (red dashed) at the ILC:
{\bf (a)} $E_1$;
{\bf (b)} $p_{T,1}$;
{\bf (c)} $M_{12}$;
{\bf (d)} $M_{234}$;
{\bf (e)} $\cos\theta_{24}$;
{\bf (f)} $\cos\theta_{34}$.}
\end{figure}

Figure~\ref{fig:ILC_vars1} shows that, at parton level, the dominant $b\bar b g$ topology captures most of the characteristic structure of the full $H\to4b$ signal.
The largest differences appear in the invariant-mass variables, especially $M_{12}$ and $M_{234}$, where the $ZZ^\ast$ contribution in the full signal introduces a visible enhancement near the $Z$-boson mass.
The angular variables are less sensitive to the subleading components, although the full signal is slightly smoother than the pure $b\bar b g$ contribution.
This comparison is useful because it shows which features originate from the dominant QCD-radiation topology and which are modified by the full gauge-invariant signal.

\begin{figure}[htbp]
\includegraphics[width=0.5\textwidth,center]{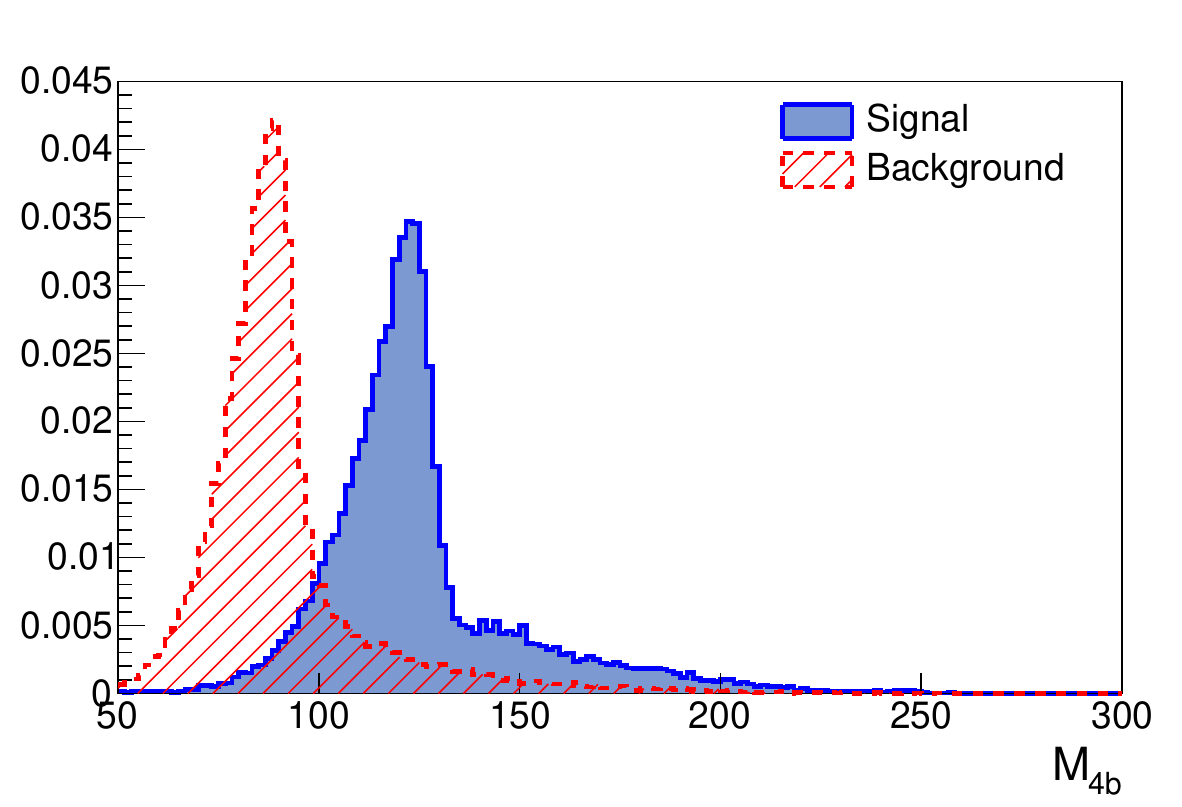}
\caption{\label{fig:ILC_m4b_full}
Normalized distribution of the reconstructed four-$b$-jet invariant mass after Delphes detector simulation and built-in $b$-tagging for the full signal (blue solid) and full background (red dashed) at the ILC.}
\end{figure}

At detector level, we first compare the full signal and full background using the reconstructed four-$b$ invariant mass.
This variable is directly tied to the Higgs resonance and gives the simplest measure of how much separation between signal and background survives detector simulation and $b$-tagging.
The result is shown in Fig.~\ref{fig:ILC_m4b_full}.

The signal distribution in Fig.~\ref{fig:ILC_m4b_full} remains localised around $M_H$, while the background is shifted towards lower invariant masses and decreases in the Higgs-mass region.
Compared with the LHC case, the separation in $M_{4b}$ is much less degraded by detector effects.
This reflects the cleaner ILC environment and the better reconstruction of the four-$b$ system.
Thus, already the reconstructed $M_{4b}$ variable provides useful signal-background discrimination, before the additional kinematic variables are included.

\begin{figure}[htbp]
\begin{subfigure}[a]{0.49\textwidth}
\includegraphics[width=\textwidth]{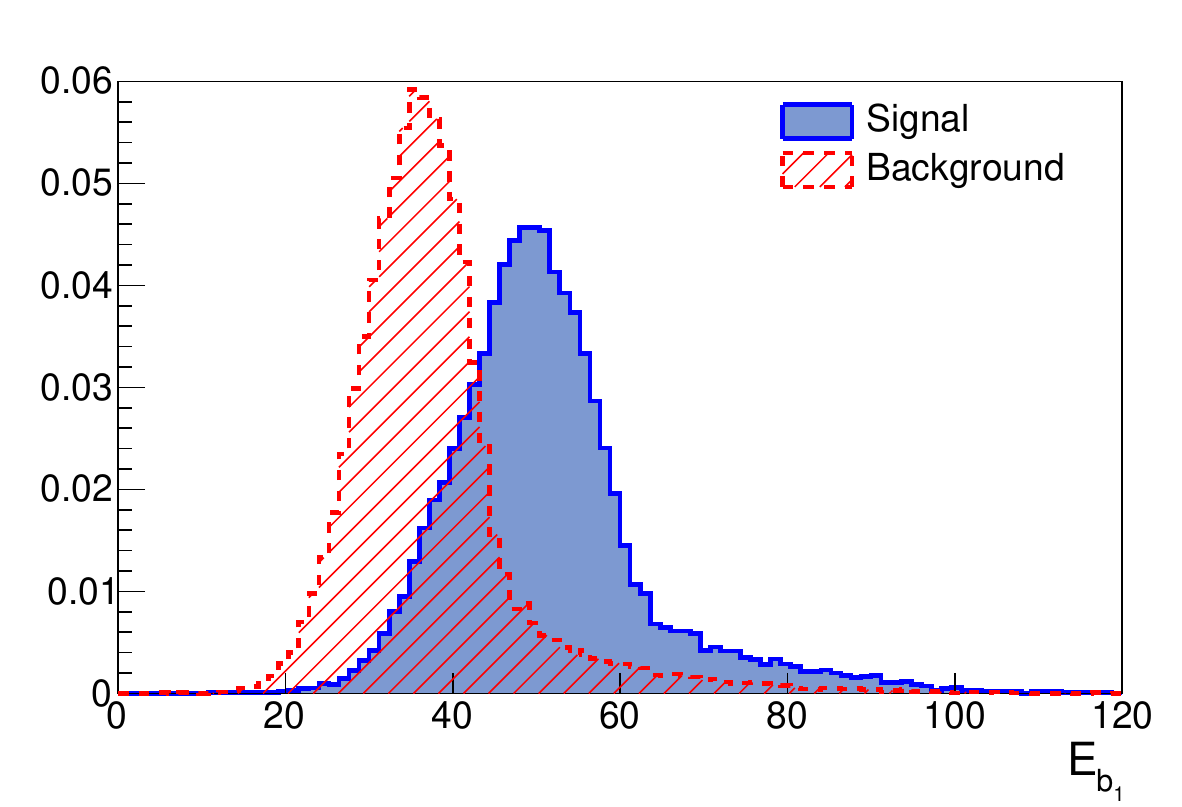}
\caption{}
\end{subfigure}
\begin{subfigure}[a]{0.49\textwidth}
\includegraphics[width=\textwidth]{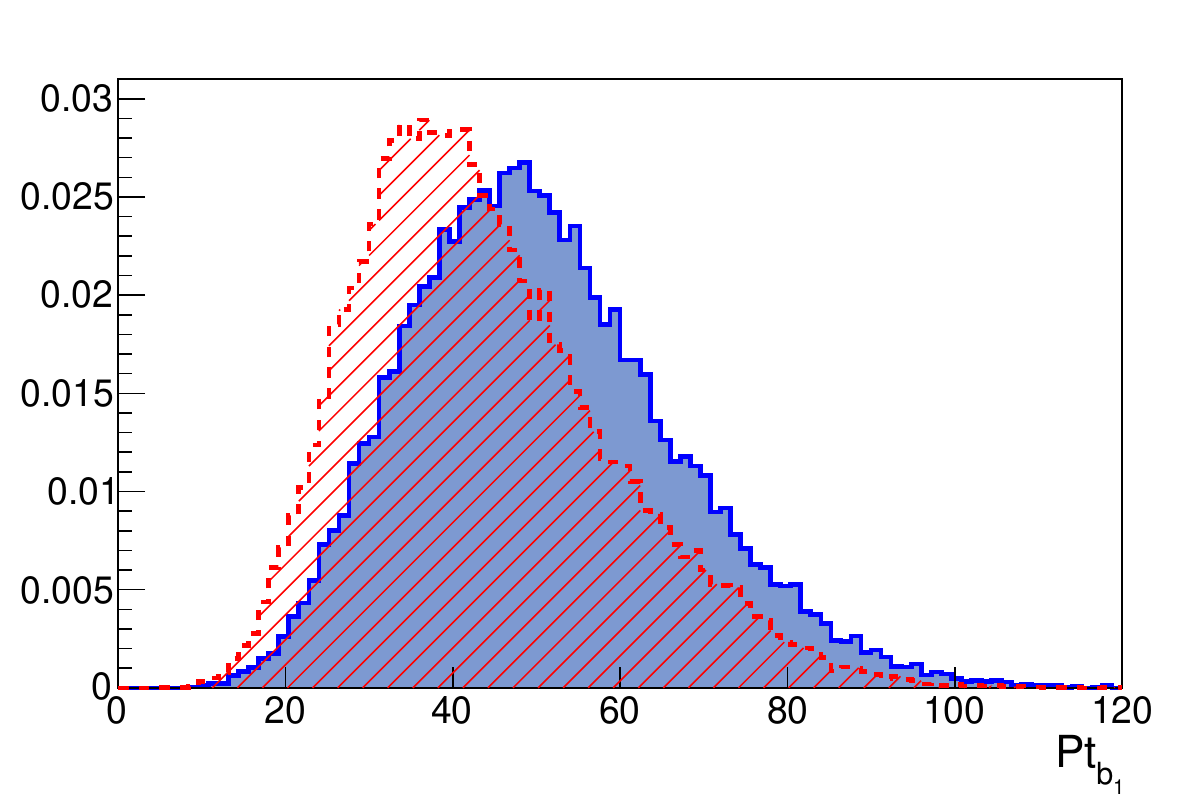}
\caption{}
\end{subfigure}
\\
\begin{subfigure}[c]{0.49\textwidth}
\includegraphics[width=\textwidth]{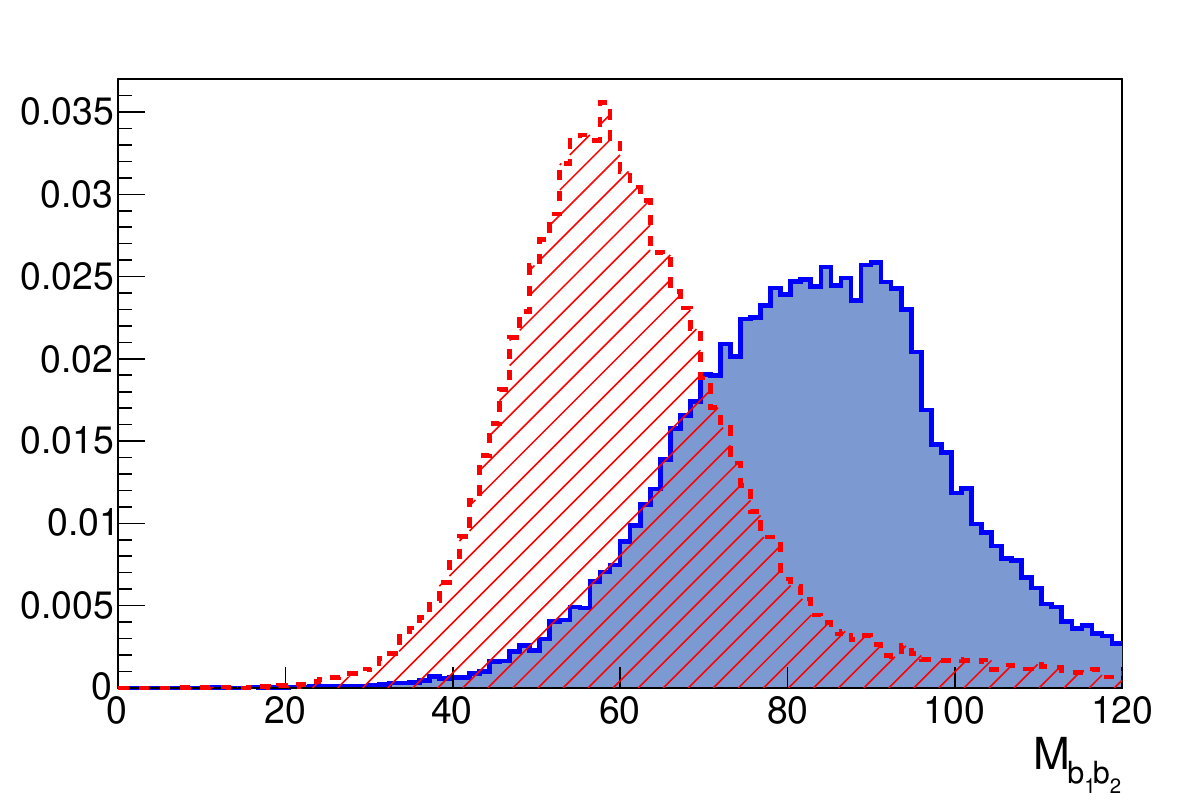}
\caption{}
\end{subfigure}
\begin{subfigure}[d]{0.49\textwidth}
\includegraphics[width=\textwidth]{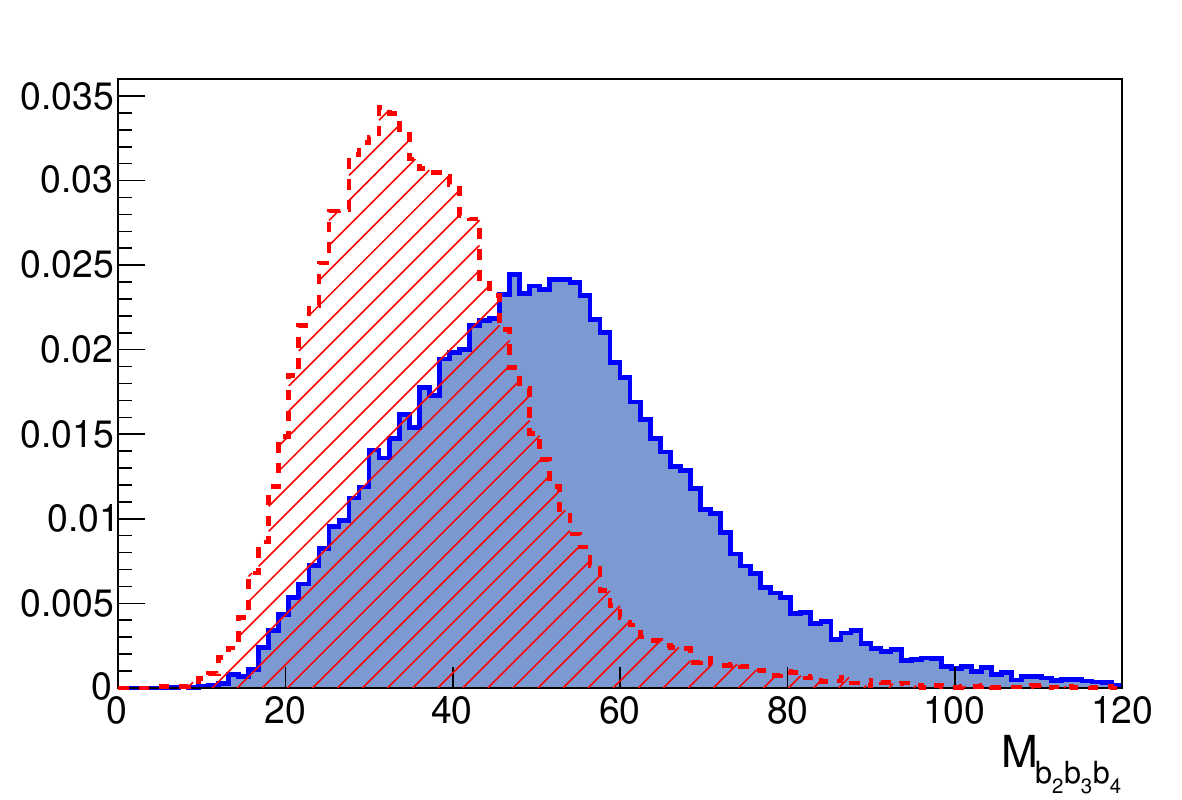}
\caption{}
\end{subfigure}
\\
\begin{subfigure}[c]{0.49\textwidth}
\includegraphics[width=\textwidth]{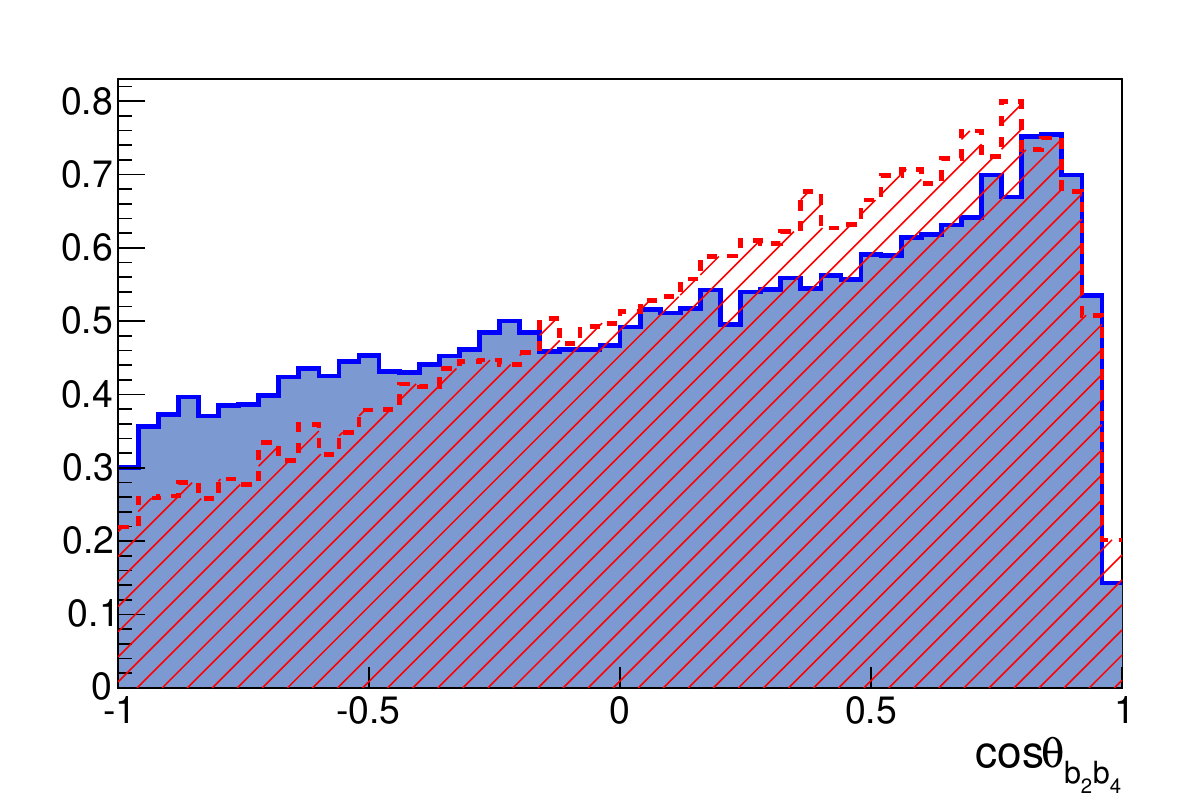}
\caption{}
\end{subfigure}
\begin{subfigure}[d]{0.49\textwidth}
\includegraphics[width=\textwidth]{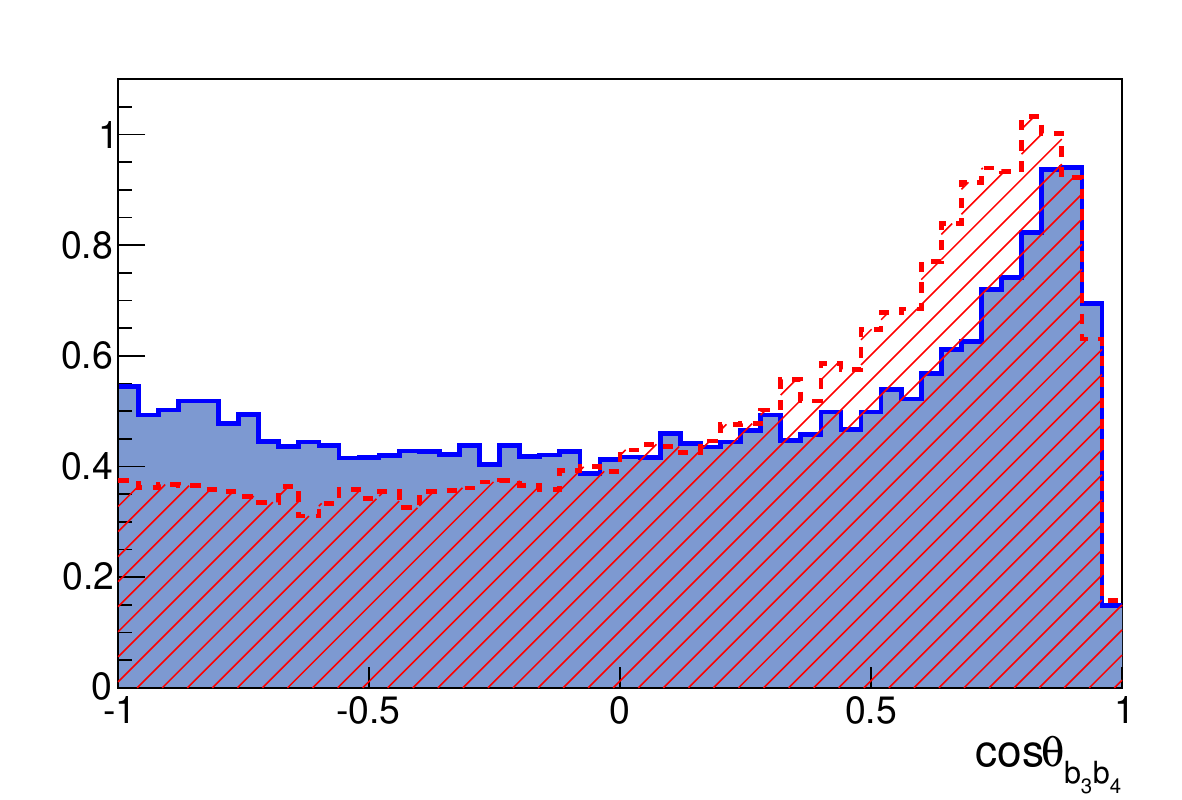}
\caption{}
\end{subfigure}
\caption{\label{fig:ILC_vars3}
Normalized distributions after Delphes detector simulation and built-in $b$-tagging for the full signal (blue solid) and full background (red dashed) processes at the ILC:
{\bf (a)} $E_1$;
{\bf (b)} $p_{T,1}$;
{\bf (c)} $M_{12}$;
{\bf (d)} $M_{234}$;
{\bf (e)} $\cos\theta_{24}$;
{\bf (f)} $\cos\theta_{34}$.}
\end{figure}

The remaining detector-level observables are shown in Fig.~\ref{fig:ILC_vars3}.
The separation is visibly stronger than in the LHC case.
The leading energy $E_1$ and the invariant masses $M_{12}$ and $M_{234}$ show clear displacement between signal and background.
The background is shifted towards lower values in these variables, while the signal retains the kinematic imprint of a parent Higgs resonance.
The angular variables provide additional information, but the dominant improvement compared with the LHC comes from the cleaner reconstruction of the Higgs mass and the much smaller continuum background.
These distributions confirm that the ILC analysis benefits both from the sharp reconstructed $M_{4b}$ peak and from the additional correlations among the four-$b$ kinematic observables.


\begin{figure}[htb]
\includegraphics[width=0.48\textwidth]{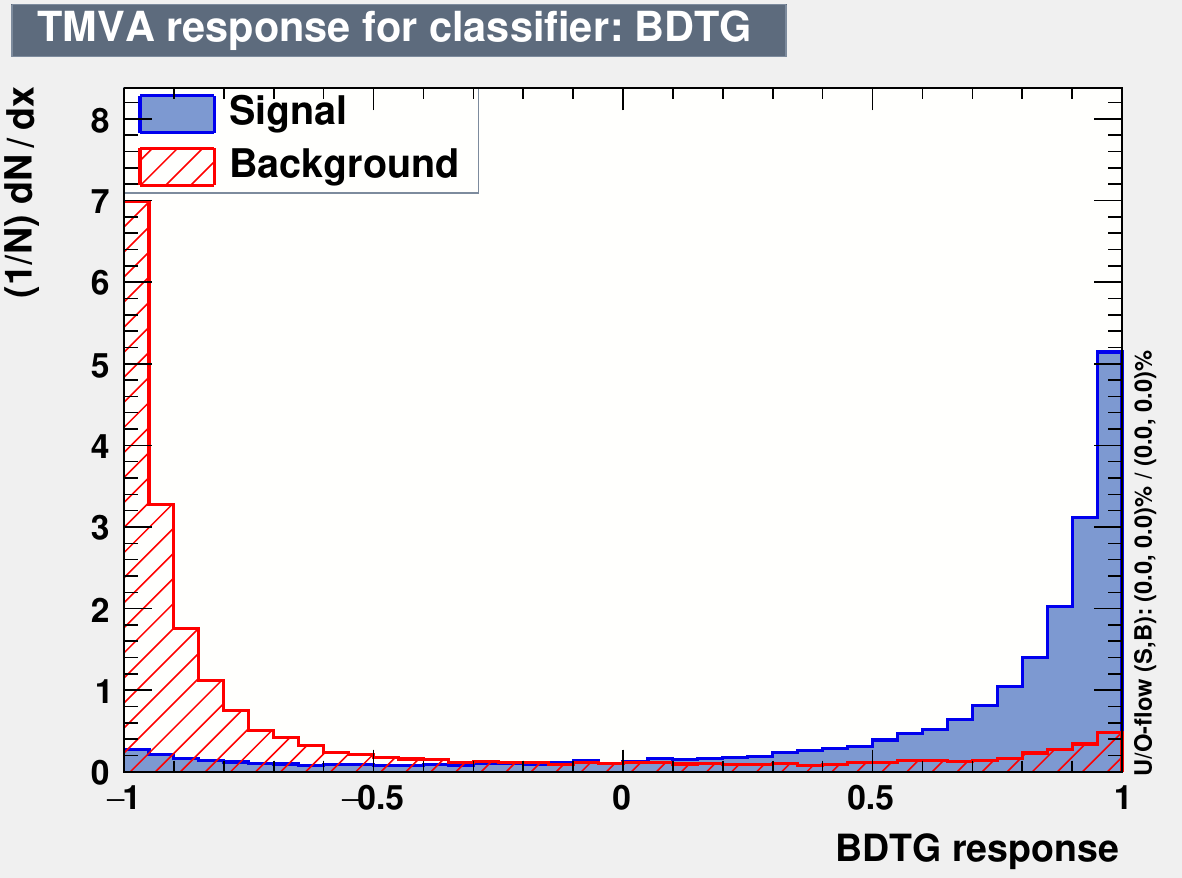}\ \ 
\includegraphics[width=0.48\textwidth]{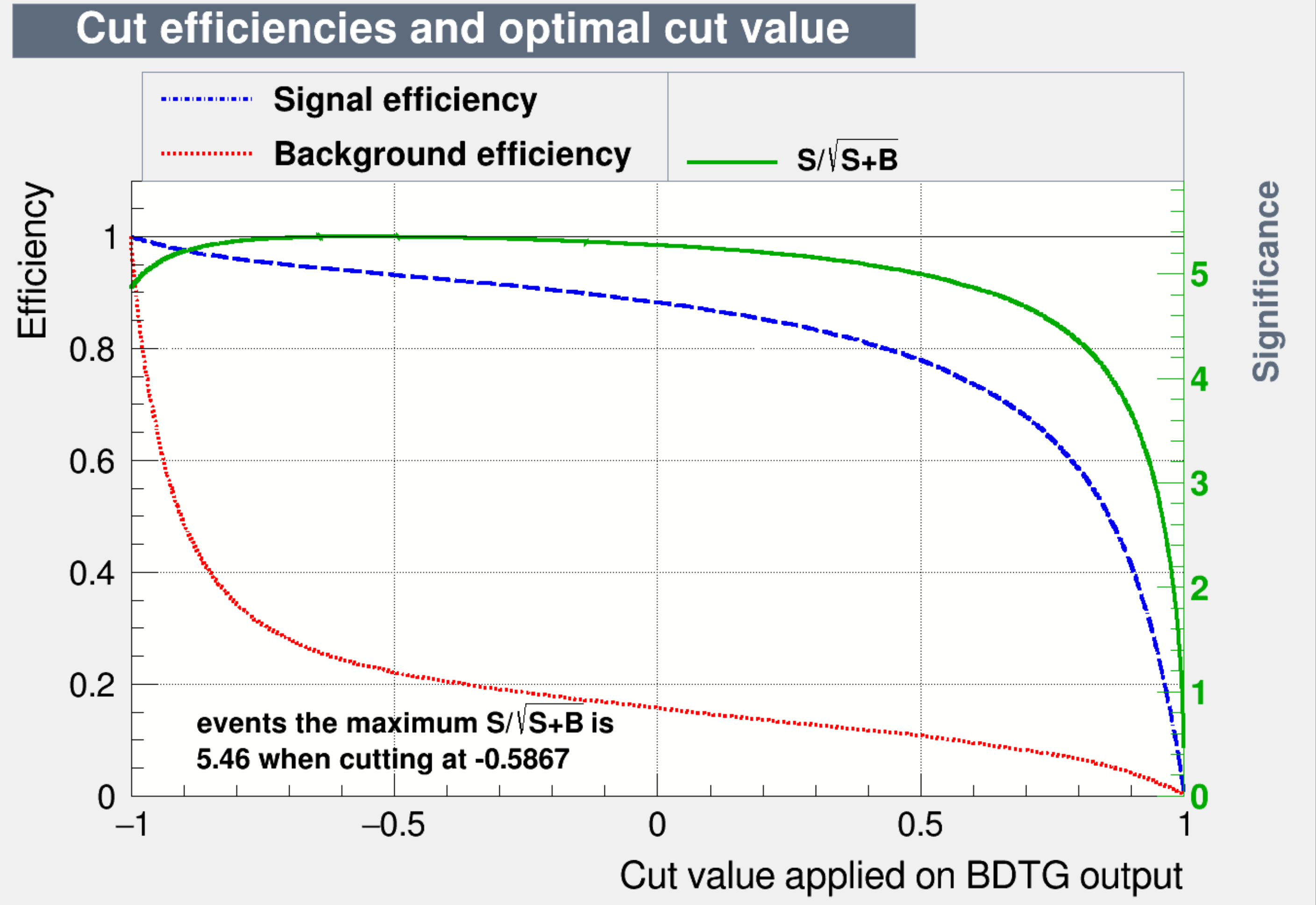}%
\caption{\label{fig:classifierilc}
Left: BDTG classifier response for the $e^+e^- \to ZH \to Zb\bar{b}b\bar{b}$ signal and the corresponding background at the ILC.
Right: signal and background efficiencies, together with the statistical significance $S/\sqrt{S+B}$, as functions of the lower cut on the BDTG response, for an integrated luminosity of $300~{\rm fb}^{-1}$.}
\end{figure}

As for the LHC analysis, we use the TMVA BDTG classifier to exploit correlations among the reconstructed observables.
The same set of variables is used: four transverse momenta, four Higgs-rest-frame energies,  the minimum and maximum two-$b$ invariant masses, $M_{12}$, $M_{34}$, $M_{234}$, $M_{4b}$, and the six angular variables $\cos\theta_{ij}$.
The left panel of Fig.~\ref{fig:classifierilc} shows that the signal and background BDTG responses are strongly separated.
The signal accumulates at large BDTG output values, while the background is concentrated at negative values.
This behaviour is much cleaner than in the LHC case and reflects the reduced continuum background and better kinematic reconstruction at the ILC.

The right panel of Fig.~\ref{fig:classifierilc} shows the signal efficiency, background efficiency and statistical significance as functions of the lower BDTG cut.
For the ILC setup used in this analysis, the event yields at $300~{\rm fb}^{-1}$ are summarised in Table~\ref{tab:ilc-bdt-yields}.
Before the BDTG cut, the expected event numbers are $S=35$ and $B=14$.
The optimal BDTG cut is
\begin{equation}
{\rm BDTG}>-0.5867,
\end{equation}
which gives $S=32.86$ and $B=3.35$ after the cut.
The corresponding statistical significance is
\begin{equation}
\alpha=\frac{S}{\sqrt{S+B}}=5.46,
\end{equation}
already for $300~{\rm fb}^{-1}$.
The signal-to-background ratio improves from $S/B\simeq2.5$ before the BDTG cut to $S/B\simeq9.8$ after the optimal cut.
This is the main qualitative difference with the LHC: at the ILC the BDTG cut does not merely make the signal observable, but produces a high-purity sample suitable for a branching-ratio measurement.

\begin{table}[htbp]
\centering
\begin{tabular}{|l|c|c|c|c|c|c|c|}
\hline\hline
Working point &
BDTG cut &
$S$ &
$B$ &
$S/B$ &
$S/\sqrt{S+B}$ &
Signal eff. &
Background eff. \\
\hline\hline
Before BDTG cut &
-- &
$35.00$ &
$14.0$ &
$2.5$ &
$5.00$ &
$1.00$ &
$1.00$ \\
\hline
Optimal BDTG cut &
$-0.5867$ &
$32.86$ &
$3.35$ &
$9.8$ &
$5.46$ &
$0.94$ &
$0.24$ \\
\hline\hline
\end{tabular}
\caption{\label{tab:ilc-bdt-yields}
Expected signal and background event yields for the ILC analysis at $300~{\rm fb}^{-1}$. The signal yield has been rescaled to the updated signal cross section $\sigma(e^+e^-\to ZH\to Zb\bar b b\bar b)=0.370$~fb.}
\end{table}

We now estimate the expected precision on the branching ratio.
For a counting measurement after the BDTG cut, the number of observed events is
\begin{equation}
N_{\rm obs}=S+B.
\end{equation}
Assuming the background expectation is known, the extracted signal yield is $S=N_{\rm obs}-B$.
In the Gaussian approximation, the statistical uncertainty on the signal yield is
\begin{equation}
\Delta S=\sqrt{S+B},
\end{equation}
and therefore the relative statistical uncertainty on the branching ratio is
\begin{equation}
\label{eq:ilc-br-precision}
\frac{\Delta {\rm Br}(H\to4b)}{{\rm Br}(H\to4b)}
=
\frac{\Delta S}{S}
=
\frac{\sqrt{S+B}}{S}.
\end{equation}
Using the BDTG-selected yields in Table~\ref{tab:ilc-bdt-yields}, we obtain at $300~{\rm fb}^{-1}$
\begin{equation}
\frac{\Delta {\rm Br}}{{\rm Br}}
=
\frac{\sqrt{32.86+3.35}}{32.86}
\simeq 0.183,
\end{equation}

corresponding to an accuracy of about $18.3\%$.
Since both signal and background scale linearly with luminosity, this uncertainty scales approximately as $1/\sqrt{L}$.

\begin{table}[htbp]
\centering
\begin{tabular}{|c|c|}
\hline\hline
Integrated luminosity & $\Delta{\rm Br}(H\to4b)/{\rm Br}(H\to4b)$ \\
\hline\hline
$300~{\rm fb}^{-1}$ & $18.3\%$ \\
$500~{\rm fb}^{-1}$ & $14.2\%$ \\
$1000~{\rm fb}^{-1}$ & $10.0\%$ \\
$3000~{\rm fb}^{-1}$ & $5.8\%$ \\
$4000~{\rm fb}^{-1}$ & $5.0\%$ \\
\hline\hline
\end{tabular}
\caption{\label{tab:ilc-br-precision}
Expected statistical precision on ${\rm Br}(H\to4b)$ obtained from Eq.~\eqref{eq:ilc-br-precision}, using the BDTG-selected ILC yields $S=32.86$ and $B=3.35$ at $300~{\rm fb}^{-1}$ and scaling with integrated luminosity.}
\end{table}

\begin{figure}[htb]
\includegraphics[width=0.60\textwidth]{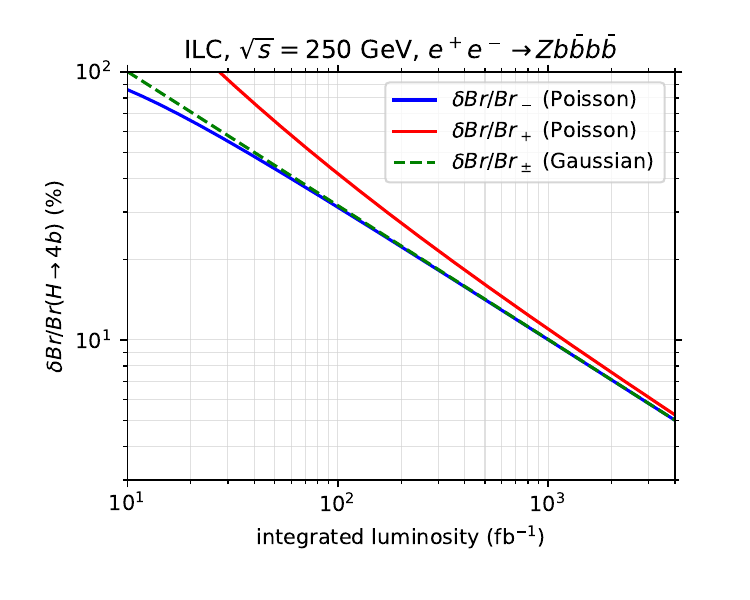}%
\vspace*{-0.5cm}
\caption{\label{fig:dbr_ilc}
Expected statistical accuracy of the ${\rm Br}(H\to4b)$ determination at the ILC as a function of integrated luminosity. The Gaussian estimate follows Eq.~\eqref{eq:ilc-br-precision}; the Poisson curves show the corresponding asymmetric small-counting uncertainties.}
\end{figure}

For low-statistics counting measurements, we use a Poisson treatment in addition to the Gaussian approximation.
The asymmetric Poisson uncertainty is estimated using the central Garwood interval for the total expected event count $S+B$~\cite{Garwood:1936}, implemented through the equivalent chi-square quantile representation and then converted to the relative uncertainty on ${\rm Br}(H\to4b)$.
At $300~{\rm fb}^{-1}$, this gives
$-18.2\%/+21.5\%$, compared with the Gaussian value of $18.3\%$.
This is why Fig.~\ref{fig:dbr_ilc} also shows the asymmetric Garwood-based Poisson uncertainty.
At higher luminosities the Poisson and Gaussian estimates converge, as expected once the number of selected events becomes large.

These results demonstrate that the ILC can do more than establish the existence of the $H\to4b$ decay mode.
Already at $300~{\rm fb}^{-1}$, the BDTG-selected sample gives a significance above $5\sigma$ for the setup used in this analysis.
At integrated luminosities of order $3-4~{\rm ab}^{-1}$, the branching ratio can be measured with a precision improving from about $6\%$ to $5\%$.
This is precisely the regime where the ILC becomes complementary to the HL-LHC: the LHC can provide evidence or discovery potential in a challenging hadronic environment, while the ILC can turn the same rare decay mode into a precision observable.


\section{Conclusions}
\label{sec:Conclusions}

The Higgs boson has now entered the precision era. The next step is not only to improve measurements of the dominant decay modes, but also to test rare channels that probe different components of the SM Higgs structure. In this paper we have proposed and studied the rare decay $H\to b\bar b b\bar b$ as a direct probe of Higgs interactions with bottom quarks and gauge bosons. Although this decay is rare, it is not negligible: its branching ratio is of order $10^{-3}$, and its internal structure contains physically distinct contributions that can be separated kinematically.

The dominant contribution comes from the $H\to b\bar b g\to b\bar b b\bar b$ topology, which accounts for about $68\%$ of the full
$H\to b\bar b b\bar b$ width. The $H\to ZZ^\ast\to b\bar b b\bar b$ contribution is also sizeable, at about $30\%$, while the loop-induced $H\to gg\to b\bar b b\bar b$ channel contributes at the percent level. We found ${\rm Br}(H\to4b)\simeq  1.66\times10^{-3}$. We also showed that interference among the leading contributions is destructive and reduces the full result relative to the incoherent sum of the separate channels. A complete amplitude-level treatment is therefore essential.

The kinematics of the decay provide the key to its observability. We identified a set of observables based on the ordered energies of the four $b$ quarks or jets in the Higgs rest frame, the invariant masses of two- and three-$b$ subsystems, and angular correlations among them. The two-dimensional and three-dimensional distributions show that the discrimination is genuinely multidimensional. No single variable carries all the useful information; the signal is characterised by correlated patterns among energies, invariant masses and angles. This is why a multivariate analysis is not an optional refinement, but the natural analysis strategy for this channel.

For the HL-LHC, we studied $pp\to WH\to Wb\bar b b\bar b$ at $\sqrt{s}=14$ TeV. The challenge is severe: after loose parton-level preselection, the total irreducible background is larger than the signal by a factor of about $1.6\times10^2$. After showering, hadronisation, Delphes detector simulation and $b$-tagging, we trained a boosted-decision-tree classifier using $20$ reconstructed kinematic variables. At $3000~{\rm fb}^{-1}$, the statistically optimal working point gives significance $\alpha=S/\sqrt{S+B}=3.49$ with $S/B\simeq3.1\%$. A tighter working point gives $\alpha\simeq3.1$, while improving the purity to $S/B\simeq5.2\%$. This is a more robust point once background modelling uncertainties are considered. These results demonstrate that the HL-LHC has realistic sensitivity to $H\to4b$, and that a combined high-luminosity ATLAS and CMS dataset could reach the level required for evidence, and possibly
discovery, of this rare decay.

At the same time, the LHC is not the ideal machine for precision studies of this mode. Even after a tight multivariate cut, thousands of background events survive. The LHC can make the decay visible, but the background environment makes a precise determination of the branching ratio and a detailed separation of the decay components extremely challenging.
-
The ILC changes the situation qualitatively. We studied $e^+e^-\to ZH\to Zb\bar b b\bar b$ at $\sqrt{s}=250$ GeV. Including ISR and beamstrahlung effects, the  signal cross section is about $0.370$ fb. The continuum $e^+e^-\to Zb\bar b b\bar b$ background is much smaller than at the LHC, and the reconstructed four-$b$ invariant mass remains  peaked around the Higgs mass after Delphes simulation. The same multivariate strategy gives a clean high-purity sample.  For $300~{\rm fb}^{-1}$, the BDTG-selected sample yields $\alpha=5.46$ after the optimal classifier cut, with $S/B\simeq9.8$. Thus, at the ILC this channel is not merely observable; it becomes measurable.

Using the selected event yields, the expected statistical uncertainty on the branching ratio is
\[
\frac{\Delta{\rm Br}(H\to4b)}{{\rm Br}(H\to4b)}
=
\frac{\sqrt{S+B}}{S}.
\]
 This gives an accuracy of about $18.3\%$ at $300~{\rm fb}^{-1}$ and reaches a precision improving from about $6\%$ to $5\%$ for integrated luminosities of order $3$--$4~{\rm ab}^{-1}$.

The analysis also establishes a baseline for searches for new physics in Higgs decays to four bottom quarks. Many BSM scenarios can enhance the $4b$ final state, for example through light intermediate resonances, modified Higgs couplings, or models with partial compositeness. The observables and multivariate strategy developed here can therefore be directly reinterpreted as tools for probing non-SM contributions to $H\to4b$-like final states. In this sense, the SM measurement proposed in this work is also a gateway to BSM searches in a difficult but highly motivated Higgs decay topology.

The main conclusion is therefore clear. The decay $H\to b\bar b b\bar b$ is a rare but realistic target for future Higgs studies. The HL-LHC can potentially establish its presence in associated production, while the ILC can measure it with high purity and  percent-level precision at high integrated luminosity. Together, these machines provide a realistic path from first observation to precision study of one of the rarest directly accessible Higgs decay modes involving bottom quarks.

\section*{Acknowledgments}

The authors acknowledge the use of the IRIDIS High Performance Computing Facility, and associated support services at the University of Southampton
which was very important to conduct this study.
AB acknowledges partial  support from the STFC grant ST/L000296/1 and  support from the Leverhulme Trust project MONDMag (RPG-2022-57). AB also thanks the NExT Institute for partial support.
The work of E.B. and V.B. was conducted under the state assignment of
Lomonosov Moscow State University. G.N. and S.R. acknowledge partial
support from grant IRN AP08053194 of the Ministry of Education and
Science of the Republic of Kazakhstan.
 
\bibliographystyle{unsrtnat} 
\bibliography{bib}

\end{document}